\documentclass[aps,prc,showpacs,twocolumn,superscriptaddress,nofootinbib]{revtex4-1}

\usepackage{graphicx}
\usepackage[dvips]{color}
\usepackage{colordvi}
\usepackage{bm}
\usepackage{amsmath}
\usepackage{amssymb}
\usepackage{verbatim}
\usepackage{feynmp}
\usepackage{multirow}
\usepackage{bigstrut}
\usepackage{array}
\usepackage{fix-cm}

\newcommand{\ad}[1]{a_{#1}}
\newcommand{\ac}[1]{a^{\dagger}_{#1}}

\newcommand{\bra}[1]{\langle #1 \vert}
\newcommand{\ket}[1]{\vert #1 \rangle}

\begin{document}
\begin{fmffile}{./diagfolder/gorkovdiag}

\title{{\it Ab-initio} self-consistent Gorkov-Green's function calculations of semi-magic nuclei \\ I. Formalism at second order with a two-nucleon interaction}

\author{V. Som\`a}
\email{vittorio.soma@cea.fr}
\affiliation{CEA-Saclay, IRFU/Service de Physique Nucl\'eaire, F-91191 Gif-sur-Yvette, France}

\author{T. Duguet}
\email{thomas.duguet@cea.fr}
\affiliation{CEA-Saclay, IRFU/Service de Physique Nucl\'eaire, F-91191 Gif-sur-Yvette, France}
\affiliation{National Superconducting Cyclotron Laboratory and Department of Physics and Astronomy,
Michigan State University, East Lansing, MI 48824, USA}

\author{C. Barbieri}
\email{C.Barbieri@surrey.ac.uk}
\affiliation{Department of Physics, University of Surrey, Guildford GU2 7XH, UK}

\date{\today}

\begin{abstract}
An ab-initio calculation scheme for finite nuclei based on self-consistent Green's functions in the Gorkov formalism is developed. It aims at describing properties of doubly-magic and semi-magic nuclei employing state-of-the-art microscopic nuclear interactions and explicitly treating pairing correlations through the breaking of $U(1)$ symmetry associated with particle number conservation. The present paper introduces the formalism, necessary to undertake applications at (self-consistent) second-order using two-nucleon interactions, in a detailed and self-contained fashion. First applications of such a scheme will be reported soon in a forthcoming publication. Future works will extend the present scheme to include three-nucleon interactions and implement more advanced truncation schemes.
\end{abstract}


\maketitle

%
%
\newcommand{\firstn}
{
\parbox{50mm}
{
\begin{fmfgraph*}(80,50)
\fmfcmd{%
style_def gorkov_n expr p =
draw_double p;
fill (harrow (p, .45));
fill (tarrow (p, .55))
enddef;}
\fmfleft{i} 
\fmfright{o}
\fmf{phantom,tension=10,label=$\phantom{aa}b$,l.s=right}{i,i1} 
\fmf{phantom,tension=10}{o,o1} 
\fmf{gorkov_n,left,tension=0.8,label=$c\phantom{aasssssiii}$,l.s=right,l.d=.26w}{o1,v1} 
\fmf{double,left,tension=0.8,label=$d\phantom{aasssssiii}$,l.s=right,l.d=.22w}{v1,o1} 
\fmf{dashes,label=$a\phantom{aaaaaaiii}$,l.s=right}{v1,i1}
\fmfdot{v1,i1}
\fmflabel{$\downarrow \omega'$}{o1}
\end{fmfgraph*} 
}
}

%
%
\newcommand{\firstnn}
{
\parbox{50mm}
{
\begin{fmfgraph*}(80,50)
\fmfcmd{%
style_def gorkov_n expr p =
draw_double p;
fill (harrow (p, .45));
fill (tarrow (p, .55))
enddef;}
\fmfleft{i} 
\fmfright{o}
\fmf{phantom,tension=10,label=$\phantom{aa}\bar{b}$,l.s=right}{i,i1} 
\fmf{phantom,tension=10}{o,o1} 
\fmf{gorkov_n,right,tension=0.8,label=$\bar{c}\phantom{aasssssiii}$,l.s=left,l.d=.26w}{v1,o1} 
\fmf{double,left,tension=0.8,label=$\bar{d}\phantom{aasssssiii}$,l.s=right,l.d=.22w}{v1,o1} 
\fmf{dashes,label=$\bar{a}\phantom{aaaaaaiii}$,l.s=right}{v1,i1}
\fmfdot{v1,i1}
\fmflabel{$\downarrow \omega'$}{o1}
\end{fmfgraph*} 
}
}

%
%
\newcommand{\firsta}
{
\parbox{50mm}
{
\begin{fmfgraph*}(80,50)
\fmfcmd{%
style_def gorkov_ain expr p =
draw_double p;
fill (harrow (reverse p, .45));
fill (harrow (p, .45))
enddef;
style_def gorkov_aout expr p =
draw_double p; 
fill (tarrow (reverse p, .55)); 
fill (tarrow (p, .55))
enddef;}
\fmfleft{i} 
\fmfright{o} 
\fmf{phantom,tension=10}{i,i1} 
\fmf{phantom,tension=10,label=$\bar{b}$,l.s=right}{o,o1} 
\fmf{gorkov_aout,left,tension=0.8,label=$\leftarrow \omega'$,l.s=left}{o1,i1} 
\fmf{dashes,label=$a\phantom{aaaaaaaaaaaaaiii}$,l.s=right}{o1,i1}
\fmflabel{$c$}{i1}
\fmflabel{$\bar{d}$}{o1}
\fmfdot{o1,i1}
\end{fmfgraph*} 
}
}

%
%
\newcommand{\firstaa}
{
\parbox{50mm}
{
\begin{fmfgraph*}(80,50)
\fmfcmd{%
style_def gorkov_ain expr p =
draw_double p;
fill (harrow (reverse p, .45));
fill (harrow (p, .45))
enddef;
style_def gorkov_aout expr p =
draw_double p; 
fill (tarrow (reverse p, .55)); 
fill (tarrow (p, .55))
enddef;}
\fmfleft{i} 
\fmfright{o} 
\fmf{phantom,tension=10}{i,i1} 
\fmf{phantom,tension=10,label=$d$,l.s=right}{o,o1} 
\fmf{gorkov_ain,right,tension=0.8,label=$\leftarrow \omega'$,l.s=right}{o1,i1} 
\fmf{dashes,label=$\bar{c}\phantom{aaaaaaaaaaaaaiii}$,l.s=right}{o1,i1}
\fmflabel{$\bar{a}$}{i1}
\fmflabel{$b$}{o1}
\fmfdot{o1,i1}
\end{fmfgraph*} 
}
}

%
%
\newcommand{\secondna}
{
\parbox{50mm}
{
\begin{fmfgraph*}(80,50)
\fmfcmd{%
style_def gorkov_n expr p =
draw_double p;
fill (harrow (p, .45));
fill (tarrow (p, .55))
enddef;}
\fmfleft{i1,i2} 
\fmfright{o1,o2} 
\fmf{gorkov_n,label=$\uparrow \omega'$,l.s=left}{i1,i2} 
\fmf{gorkov_n,left=.5,tension=.5,label=$\uparrow \omega''$,l.s=left}{o1,o2} 
\fmf{gorkov_n,left=.5,tension=.5,label=$\downarrow \omega'''$,l.s=left}{o2,o1} 
\fmf{dashes,label=$\phantom{a}d\phantom{aaaaaaaassaa}g\phantom{aaaaaa}$,l.s=right}{o1,i1}
\fmf{dashes,label=$\phantom{a}c\phantom{aaaaaaaassaa}f\phantom{aaaaaa}$,l.s=left}{o2,i2}
\fmflabel{$b$}{i1}
\fmflabel{$h$}{o1}
\fmflabel{$a$}{i2}
\fmflabel{$e$}{o2}
\fmfdot{o1,o2,i1,i2}
\end{fmfgraph*} 
}
}

%
%
\newcommand{\secondnapl}
{
\parbox{50mm}
{
\begin{fmfgraph*}(80,50)
\fmfcmd{%
style_def gorkov_n expr p =
draw_plain p;
fill (harrow (p, .45));
fill (tarrow (p, .55))
enddef;}
\fmfleft{i1,i2} 
\fmfright{o1,o2} 
\fmf{gorkov_n,label=$\uparrow \omega'$,l.s=left}{i1,i2} 
\fmf{gorkov_n,left=.5,tension=.5,label=$\uparrow \omega''$,l.s=left}{o1,o2} 
\fmf{gorkov_n,left=.5,tension=.5,label=$\downarrow \omega'''$,l.s=left}{o2,o1} 
\fmf{dashes,label=$\phantom{a}j\phantom{aaaaaaaassaa}g\phantom{aaaaaa}$,l.s=right}{o1,i1}
\fmf{dashes,label=$\phantom{a}i\phantom{aaaaaaaassaa}f\phantom{aaaaaa}$,l.s=left}{o2,i2}
\fmflabel{$d$}{i1}
\fmflabel{$h$}{o1}
\fmflabel{$c$}{i2}
\fmflabel{$e$}{o2}
\fmfdot{o1,o2,i1,i2}
\end{fmfgraph*} 
}
}

%
%
\newcommand{\secondnaold}
{
\parbox{50mm}
{
\begin{fmfgraph*}(80,50)
\fmfcmd{%
style_def gorkov_n expr p =
draw_double p;
fill (harrow (p, .45));
fill (tarrow (p, .55))
enddef;}
\fmfleft{i1,i2} 
\fmfright{o1,o2} 
\fmf{gorkov_n,label=$\uparrow \omega'$,l.s=left}{i1,i2} 
\fmf{gorkov_n,tension=0,label=$\uparrow \omega''$,l.s=left}{v1,v2} 
\fmf{gorkov_n,label=$\downarrow \omega'''$,l.s=left}{o2,o1} 
\fmf{dashes,label=$\phantom{aaa}i\phantom{aaassaa}f\phantom{aaaaaaa}$,l.s=right}{i2,v2}
\fmf{dashes,label=$\phantom{aaa}j\phantom{aadadsaa}g\phantom{aaaaaaaa}$,l.s=left}{i1,v1}
\fmf{phantom}{o1,v1}
\fmf{phantom}{o2,v2}
\fmf{double,left=1,tension=1}{v2,o2} 
\fmf{double,left=1,tension=1}{o1,v1} 
\fmflabel{$d$}{i1}
\fmflabel{$h$}{o1}
\fmflabel{$c$}{i2}
\fmflabel{$e$}{o2}
\fmfdot{v1,v2,i1,i2}
\end{fmfgraph*} 
}
}

%
%
\newcommand{\secondnb}
{
\parbox{50mm}
{
\begin{fmfgraph*}(80,50)
\fmfcmd{%
style_def gorkov_n expr p =
draw_double p;
fill (harrow (p, .45));
fill (tarrow (p, .55))
enddef;}
\fmfcmd{%
style_def gorkov_ain expr p =
draw_double p;
fill (harrow (reverse p, .45));
fill (harrow (p, .45))
enddef;
style_def gorkov_aout expr p =
draw_double p; 
fill (tarrow (reverse p, .55)); 
fill (tarrow (p, .55))
enddef;}
\fmfleft{i1,i2} 
\fmfright{o1,o2} 
\fmf{gorkov_n,label=$\uparrow \omega'$,l.s=left}{i1,i2} 
\fmf{gorkov_aout,left=.5,tension=.5,label=$\uparrow \omega''$,l.s=left}{o1,o2} 
\fmf{gorkov_ain,left=.5,tension=.5,label=$\downarrow \omega'''$,l.s=left}{o2,o1} 
\fmf{dashes,label=$\phantom{a}d\phantom{aaaaaaaassaa}\bar{h}\phantom{aaaaaa}$,l.s=right}{o1,i1}
\fmf{dashes,label=$\phantom{a}c\phantom{aaaaaaaassaa}f\phantom{aaaaaa}$,l.s=left}{o2,i2}
\fmflabel{$b$}{i1}
\fmflabel{$\bar{g}$}{o1}
\fmflabel{$a$}{i2}
\fmflabel{$e$}{o2}
\fmfdot{o1,o2,i1,i2}
\end{fmfgraph*} 
}
}

%
%
\newcommand{\secondnbnos}
{
\parbox{50mm}
{
\begin{fmfgraph*}(80,140)
\fmfcmd{%
style_def gorkov_n expr p =
draw_plain p;
fill (harrow (p, .45));
fill (tarrow (p, .55))
enddef;}
\fmfcmd{%
style_def gorkov_ain expr p =
draw_plain p;
fill (harrow (reverse p, .45));
fill (harrow (p, .45))
enddef;
style_def gorkov_aout expr p =
draw_plain p; 
fill (tarrow (reverse p, .55)); 
fill (tarrow (p, .55))
enddef;}
\fmfstraight
\fmfleft{i1,i2,i3,i4} 
\fmfright{o1,o2,o3,o4} 
\fmf{gorkov_n,label=$\uparrow \omega'$,l.s=left}{i2,i3} 
\fmf{gorkov_n,label=$\uparrow \omega$,l.s=left}{i1,i2} 
\fmf{gorkov_n,label=$\uparrow \omega$,l.s=left}{i3,i4} 
\fmf{phantom}{o1,o2} 
\fmf{phantom}{o3,o4} 
\fmf{gorkov_aout,left=.5,tension=.5,label=$\uparrow \omega''$,l.s=left}{o2,o3} 
\fmf{gorkov_ain,left=.5,tension=.5,label=$\downarrow \omega'''$,l.s=left}{o3,o2} 
\fmf{dashes,label=$\phantom{aaaaa}d\phantom{aaaaaassaa}\bar{h}\phantom{aaaaaa}$,l.s=right}{o2,i2}
\fmf{dashes,label=$\phantom{aaaaa}c\phantom{aaaaaassaa}f\phantom{aaaaaa}$,l.s=left}{o3,i3}
\fmflabel{$b$}{i1}
\fmflabel{$i$}{i2}
\fmflabel{$j$}{i3}
\fmflabel{$\bar{g}$}{o2}
\fmflabel{$a$}{i4}
\fmflabel{$e$}{o3}
\fmfdot{o2,o3,i2,i3}
\end{fmfgraph*} 
}
}

%
%
\newcommand{\secondnbnosdg}
{
\parbox{50mm}
{
\begin{fmfgraph*}(80,140)
\fmfcmd{%
style_def gorkov_n expr p =
draw_plain p;
fill (harrow (p, .45));
fill (tarrow (p, .55))
enddef;}
\fmfcmd{%
style_def gorkov_ain expr p =
draw_plain p;
fill (harrow (reverse p, .45));
fill (harrow (p, .45))
enddef;
style_def gorkov_aout expr p =
draw_plain p; 
fill (tarrow (reverse p, .55)); 
fill (tarrow (p, .55))
enddef;}
\fmfstraight
\fmfleft{i1,i2,i3,i4} 
\fmfright{o1,o2,o3,o4} 
\fmf{gorkov_n,l.s=left}{i2,i3} 
\fmf{gorkov_n,l.s=left}{i1,i2} 
\fmf{gorkov_n,l.s=left}{i3,i4} 
\fmf{phantom}{o1,o2} 
\fmf{phantom}{o3,o4} 
\fmf{gorkov_aout,left=.5,tension=.5}{o2,o3} 
\fmf{gorkov_ain,left=.5,tension=.5}{o3,o2} 
\fmf{dashes,label=$\phantom{aaaaa}d\phantom{aaaaaassaa}\bar{h}\phantom{aaaaaa}$,l.s=right}{o2,i2}
\fmf{dashes,label=$\phantom{aaaaa}c\phantom{aaaaaassaa}f\phantom{aaaaaa}$,l.s=left}{o3,i3}
\fmflabel{$b$}{i1}
\fmflabel{$i$}{i2}
\fmflabel{$j$}{i3}
\fmflabel{$\bar{g}$}{o2}
\fmflabel{$a$}{i4}
\fmflabel{$e$}{o3}
\fmfdot{o2,o3,i2,i3}
\end{fmfgraph*} 
}
}

%
%
\newcommand{\topeq}
{
\parbox{50mm}
{
\begin{fmfgraph*}(55,160)
\fmfcmd{%
style_def gorkov_n expr p =
draw_plain p;
fill (harrow (p, .45));
fill (tarrow (p, .55))
enddef;}
\fmfcmd{%
style_def gorkov_ain expr p =
draw_plain p;
fill (harrow (reverse p, .45));
fill (harrow (p, .45))
enddef;
style_def gorkov_aout expr p =
draw_plain p; 
fill (tarrow (reverse p, .55)); 
fill (tarrow (p, .55))
enddef;}
\fmfstraight
\fmfleft{i1,i2,i3,i4} 
\fmfright{o1,o2,o3,o4} 
\fmf{phantom,l.s=left}{i2,i3} 
\fmf{gorkov_ain,l.s=left}{i1,i2} 
\fmf{gorkov_ain,l.s=left}{i3,i4} 
\fmf{phantom}{o1,o2} 
\fmf{phantom}{o3,o4} 
\fmf{gorkov_ain,l.s=left}{o3,o2} 
\fmf{gorkov_aout,left=0.5,l.s=left}{o2,i2} 
\fmf{gorkov_aout,left=0.5,l.s=left}{o3,i3} 
\fmf{dashes,label=$\phantom{aaccaaa}\bar{d}\phantom{aaaaacccassaa}h\phantom{aaaaaa}$,l.s=right}{o2,i2}
\fmf{dashes,label=$\phantom{aaccaaa}c\phantom{aaaacaaccssaa}\bar{f}\phantom{aaaaaa}$,l.s=left}{o3,i3}
\fmflabel{$b$}{i1}
\fmflabel{$\bar{i}$}{i2}
\fmflabel{$j$}{i3}
\fmflabel{$g$}{o2}
\fmflabel{$\bar{a}$}{i4}
\fmflabel{$\bar{e}$}{o3}
\fmfdot{o2,o3,i2,i3}
\end{fmfgraph*} 
}
}

%
%
\newcommand{\topeqcmp}
{
\parbox{50mm}
{
\begin{fmfgraph*}(55,160)
\fmfcmd{%
style_def gorkov_n expr p =
draw_plain p;
fill (harrow (p, .45));
fill (tarrow (p, .55))
enddef;}
\fmfcmd{%
style_def gorkov_ain expr p =
draw_plain p;
fill (harrow (reverse p, .45));
fill (harrow (p, .45))
enddef;
style_def gorkov_aout expr p =
draw_plain p; 
fill (tarrow (reverse p, .55)); 
fill (tarrow (p, .55))
enddef;}
\fmfstraight
\fmfleft{i1,i2,i3,i4} 
\fmfright{o1,o2,o3,o4} 
\fmf{phantom,l.s=left}{i2,i3} 
\fmf{phantom,l.s=left}{i1,i2} 
\fmf{phantom,l.s=left}{i3,i4} 
\fmf{phantom}{o1,o2} 
\fmf{phantom}{o3,o4} 
\fmf{gorkov_ain,l.s=left,label=$\uparrow \omega'$,l.s=left}{o3,o2} 
\fmf{gorkov_aout,left=0.5,label=$\rightarrow \omega''$,l.s=left}{o2,i2} 
\fmf{gorkov_aout,left=0.5,label=$\leftarrow \omega'''$,l.s=left}{o3,i3} 
\fmf{dashes,label=$\phantom{aaccaaa}\bar{d}\phantom{aaaaacccassaa}h\phantom{aaaaaa}$,l.s=right}{o2,i2}
\fmf{dashes,label=$\phantom{aaccaaa}c\phantom{aaaacaaccssaa}\bar{f}\phantom{aaaaaa}$,l.s=left}{o3,i3}
\fmflabel{$g$}{o2}
\fmflabel{$\bar{e}$}{o3}
\fmfdot{o2,o3,i2,i3}
\end{fmfgraph*} 
}
}

%
%
\newcommand{\topeqqq}
{
\parbox{50mm}
{
\begin{fmfgraph*}(55,160)
\fmfcmd{%
style_def gorkov_n expr p =
draw_plain p;
fill (harrow (p, .45));
fill (tarrow (p, .55))
enddef;}
\fmfcmd{%
style_def gorkov_ain expr p =
draw_plain p;
fill (harrow (reverse p, .45));
fill (harrow (p, .45))
enddef;
style_def gorkov_aout expr p =
draw_plain p; 
fill (tarrow (reverse p, .55)); 
fill (tarrow (p, .55))
enddef;}
\fmfstraight
\fmfleft{i1,i2,i3,i4} 
\fmfright{o1,o2,o3,o4} 
\fmf{phantom,l.s=left}{i2,i3} 
\fmf{gorkov_ain,l.s=left}{o1,o2} 
\fmf{gorkov_ain,l.s=left}{o3,o4} 
\fmf{phantom}{o1,o2} 
\fmf{phantom}{o3,o4} 
\fmf{gorkov_ain,l.s=left}{i3,i2} 
\fmf{gorkov_aout,left=0.5,l.s=left}{o2,i2} 
\fmf{gorkov_aout,left=0.5,l.s=left}{o3,i3} 
\fmf{dashes,label=$\phantom{aaaaa}d\phantom{aaaaacasssaa}\bar{h}\phantom{aaaaaa}$,l.s=right}{o2,i2}
\fmf{dashes,label=$\phantom{aaaaa}\bar{c}\phantom{aaaascaassaa}f\phantom{aaaaaa}$,l.s=left}{o3,i3}
\fmflabel{$b$}{o1}
\fmflabel{$\bar{a}$}{o4}
\fmflabel{$\bar{j}$}{i3}
\fmflabel{$i$}{i2}
\fmflabel{$\bar{g}$}{o2}
\fmflabel{$e$}{o3}
\fmfdot{o2,o3,i2,i3}
\end{fmfgraph*} 
}
}

%
%
\newcommand{\topeqq}
{
\parbox{50mm}
{
\begin{fmfgraph*}(120,120)
\fmfcmd{%
style_def gorkov_n expr p =
draw_plain p;
fill (harrow (p, .45));
fill (tarrow (p, .55))
enddef;}
\fmfcmd{%
style_def gorkov_ain expr p =
draw_plain p;
fill (harrow (reverse p, .45));
fill (harrow (p, .45))
enddef;
style_def gorkov_aout expr p =
draw_plain p; 
fill (tarrow (reverse p, .55)); 
fill (tarrow (p, .55))
enddef;}
\fmfstraight
\fmfleft{i1,i2,i3,i4} 
\fmfright{o1,o2,o3,o4} 
\fmf{gorkov_ain,l.s=left}{i1,i2} 
\fmf{gorkov_ain,l.s=left}{o3,o4} 
\fmf{phantom}{i2,i3} 
\fmf{phantom}{o2,o3} 
\fmf{phantom}{i3,i4} 
\fmf{phantom}{o1,o2} 
\fmf{gorkov_ain,l.s=left}{v2,v1} 
\fmf{gorkov_aout,left=0.2}{i2,v1} 
\fmf{gorkov_aout,left=0.2}{o3,v2} 
\fmf{dashes,tension=0.0,label=$\phantom{aaaaa}d\phantom{aaaaaassaa}\bar{h}\phantom{aaaaaa}$,l.s=right}{i2,v1,o2}
\fmf{dashes,tension=0.0,label=$\phantom{aaaaa}c\phantom{aaaaaassaa}f\phantom{aaaaaa}$,l.s=left}{o3,v2,i3}
\fmfdot{o2,o1,i2,i1}
\end{fmfgraph*} 
}
}

%
%
\newcommand{\topneq}
{
\parbox{50mm}
{
\begin{fmfgraph*}(80,140)
\fmfcmd{%
style_def gorkov_n expr p =
draw_plain p;
fill (harrow (p, .45));
fill (tarrow (p, .55))
enddef;}
\fmfcmd{%
style_def gorkov_ain expr p =
draw_plain p;
fill (harrow (reverse p, .45));
fill (harrow (p, .45))
enddef;
style_def gorkov_aout expr p =
draw_plain p; 
fill (tarrow (reverse p, .55)); 
fill (tarrow (p, .55))
enddef;}
\fmfstraight
\fmfleft{i1,i2,i3,i4} 
\fmfright{o1,o2,o3,o4} 
\fmf{gorkov_aout,l.s=left}{i2,i3} 
\fmf{gorkov_ain,l.s=left}{i1,i2} 
\fmf{gorkov_ain,l.s=left}{i3,i4} 
\fmf{phantom}{o1,o2} 
\fmf{phantom}{o3,o4} 
\fmf{gorkov_aout,left=.5,tension=.5,l.s=left}{o2,o3} 
\fmf{gorkov_ain,left=.5,tension=.5,l.s=left}{o3,o2} 
\fmf{dashes,label=$\phantom{aaaaa}\bar{d}\phantom{aaaaaassaa}\bar{h}\phantom{aaaaaa}$,l.s=right}{o2,i2}
\fmf{dashes,label=$\phantom{aaaaa}c\phantom{aaaaaassaa}f\phantom{aaaaaa}$,l.s=left}{o3,i3}
\fmflabel{$b$}{i1}
\fmflabel{$\bar{i}$}{i2}
\fmflabel{$j$}{i3}
\fmflabel{$\bar{g}$}{o2}
\fmflabel{$\bar{a}$}{i4}
\fmflabel{$e$}{o3}
\fmfdot{o2,o3,i2,i3}
\end{fmfgraph*} 
}
}

%
%
\newcommand{\topneqnormal}
{
\parbox{50mm}
{
\begin{fmfgraph*}(80,140)
\fmfcmd{%
style_def gorkov_n expr p =
draw_plain p;
fill (harrow (p, .45));
fill (tarrow (p, .55))
enddef;}
\fmfcmd{%
style_def gorkov_ain expr p =
draw_plain p;
fill (harrow (reverse p, .45));
fill (harrow (p, .45))
enddef;
style_def gorkov_aout expr p =
draw_plain p; 
fill (tarrow (reverse p, .55)); 
fill (tarrow (p, .55))
enddef;}
\fmfstraight
\fmfleft{i1,i2,i3,i4} 
\fmfright{o1,o2,o3,o4} 
\fmf{fermion,l.s=left}{i2,i3} 
\fmf{fermion,l.s=left}{i1,i2} 
\fmf{fermion,l.s=left}{i3,i4} 
\fmf{phantom}{o1,o2} 
\fmf{phantom}{o3,o4} 
\fmf{fermion,left=.5,tension=.5,l.s=left}{o2,o3} 
\fmf{fermion,left=.5,tension=.5,l.s=left}{o3,o2} 
\fmf{dashes,label=$\phantom{aaaaa}f\phantom{aaaaaassadda}\phantom{aaaaaa}$,l.s=right}{o2,i2}
\fmf{dashes,label=$\phantom{aaaaa}e\phantom{aaaaaassadda}\phantom{aaaaaa}$,l.s=left}{o3,i3}
\fmflabel{$b$}{i1}
\fmflabel{$d$}{i2}
\fmflabel{$c$}{i3}
\fmflabel{$a$}{i4}
\fmfdot{o2,o3,i2,i3}
\end{fmfgraph*} 
}
}

%
%
\newcommand{\diagexex}
{
\parbox{50mm}
{
\begin{fmfgraph*}(60,120)
\fmfcmd{%
style_def gorkov_n expr p =
draw_plain p;
fill (harrow (p, .45));
fill (tarrow (p, .55))
enddef;}
\fmfcmd{%
style_def gorkov_ain expr p =
draw_plain p;
fill (harrow (reverse p, .45));
fill (harrow (p, .45))
enddef;
style_def gorkov_aout expr p =
draw_plain p; 
fill (tarrow (reverse p, .65)); 
fill (tarrow (p, .65))
enddef;}
\fmfstraight
\fmfleft{i1,i2,i3,i4} 
\fmfright{o1,o2,o3,o4} 
\fmf{gorkov_aout,l.s=left}{i2,o3} 
\fmf{gorkov_ain,l.s=left}{i1,i2} 
\fmf{gorkov_ain,l.s=left}{i3,i4} 
\fmf{phantom}{o1,o2} 
\fmf{phantom}{o3,o4} 
\fmf{gorkov_aout,l.s=left}{o2,i3} 
\fmf{gorkov_ain,l.s=left}{o3,o2} 
\fmf{dashes,label=$\phantom{aasaaa}d\phantom{aaaccassaassaa}h\phantom{aaaaaa}$,l.s=right}{o2,i2}
\fmf{dashes,label=$\phantom{aaasaa}c\phantom{aaassccaaassaa}\bar{f}\phantom{aaaaaa}$,l.s=left}{o3,i3}
\fmflabel{$b$}{i1}
\fmflabel{$i$}{i2}
\fmflabel{$j$}{i3}
\fmflabel{$g$}{o2}
\fmflabel{$\bar{a}$}{i4}
\fmflabel{$\bar{e}$}{o3}
\fmfdot{o2,o3,i2,i3}
\end{fmfgraph*} 
}
}

%
%
\newcommand{\secondnbold}
{
\parbox{50mm}
{
\begin{fmfgraph*}(80,50)
\fmfcmd{%
style_def gorkov_n expr p =
draw_double p;
fill (harrow (p, .45));
fill (tarrow (p, .55))
enddef;}
\fmfcmd{%
style_def gorkov_ain expr p =
draw_double p;
fill (harrow (reverse p, .45));
fill (harrow (p, .45))
enddef;
style_def gorkov_aout expr p =
draw_double p; 
fill (tarrow (reverse p, .55)); 
fill (tarrow (p, .55))
enddef;}
\fmfleft{i1,i2} 
\fmfright{o1,o2} 
\fmf{gorkov_n,label=$\uparrow \omega'$,l.s=left}{i1,i2} 
\fmf{gorkov_aout,tension=0,label=$\uparrow \omega''$,l.s=left}{v1,v2} 
\fmf{gorkov_ain,label=$\uparrow \omega'''$,l.s=right}{o1,o2} 
\fmf{dashes,label=$\phantom{aaa}c\phantom{aaassaa}f\phantom{aaaaaaa}$,l.s=right}{i2,v2}
\fmf{dashes,label=$\phantom{aaa}d\phantom{aadddddddaassaa}\bar{g}$,l.s=left}{i1,v1}
\fmf{dashes}{o1,v1}
\fmf{phantom}{o2,v2}
\fmf{double,left=1,tension=1}{v2,o2} 
\fmf{double,left=1,tension=1}{o1,v1} 
\fmflabel{$b$}{i1}
\fmflabel{$\bar{h}$}{o1}
\fmflabel{$a$}{i2}
\fmflabel{$e$}{o2}
\fmfdot{o1,v2,i1,i2}
\end{fmfgraph*} 
}
}

%
%
\newcommand{\secondnaa}
{
\parbox{50mm}
{
\begin{fmfgraph*}(80,50)
\fmfcmd{%
style_def gorkov_n expr p =
draw_double p;
fill (harrow (p, .45));
fill (tarrow (p, .55))
enddef;}
\fmfleft{i1,i2} 
\fmfright{o1,o2} 
\fmf{gorkov_n,label=$\uparrow \omega'$,l.s=right}{i2,i1} 
\fmf{gorkov_n,right=.5,tension=.5,label=$\downarrow \omega'''$,l.s=right}{o1,o2} 
\fmf{gorkov_n,right=.5,tension=.5,label=$\uparrow \omega''$,l.s=right}{o2,o1} 
\fmf{dashes,label=$\phantom{a}\bar{d}\phantom{aaaaaaaassaa}\bar{g}\phantom{aaaaaa}$,l.s=right}{o1,i1}
\fmf{dashes,label=$\phantom{a}\bar{c}\phantom{aaaaaaaassaa}\bar{f}\phantom{aaaaaa}$,l.s=left}{o2,i2}
\fmflabel{$\bar{b}$}{i1}
\fmflabel{$\bar{h}$}{o1}
\fmflabel{$\bar{a}$}{i2}
\fmflabel{$\bar{e}$}{o2}
\fmfdot{o1,o2,i1,i2}
\end{fmfgraph*} 
}
}

%
%
\newcommand{\secondnaaold}
{
\parbox{50mm}
{
\begin{fmfgraph*}(80,50)
\fmfcmd{%
style_def gorkov_n expr p =
draw_double p;
fill (harrow (p, .45));
fill (tarrow (p, .55))
enddef;}
\fmfcmd{%
style_def gorkov_ain expr p =
draw_double p;
fill (harrow (reverse p, .45));
fill (harrow (p, .45))
enddef;
style_def gorkov_aout expr p =
draw_double p; 
fill (tarrow (reverse p, .55)); 
fill (tarrow (p, .55))
enddef;}
\fmfleft{i1,i2} 
\fmfright{o1,o2} 
\fmf{gorkov_n,label=$\uparrow \omega'$,l.s=right}{i2,i1} 
\fmf{gorkov_n,tension=0,label=$\uparrow \omega''$,l.s=right}{v2,v1} 
\fmf{gorkov_n,label=$\downarrow \omega'''$,l.s=right}{o1,o2} 
\fmf{dashes,label=$\phantom{aaa}\bar{c}\phantom{aaassaa}\bar{f}\phantom{aaaaaaa}$,l.s=right}{i2,v2}
\fmf{dashes,label=$\phantom{aaa}\bar{d}\phantom{aaiassaa}\bar{g}\phantom{aaaaaaa}$,l.s=left}{i1,v1}
\fmf{phantom}{o1,v1}
\fmf{phantom}{o2,v2}
\fmf{double,left=1,tension=1}{v2,o2} 
\fmf{double,left=1,tension=1}{o1,v1} 
\fmflabel{$\bar{b}$}{i1}
\fmflabel{$\bar{h}$}{o1}
\fmflabel{$\bar{a}$}{i2}
\fmflabel{$\bar{e}$}{o2}
\fmfdot{v1,v2,i1,i2}
\end{fmfgraph*} 
}
}

%
%
\newcommand{\secondnbb}
{
\parbox{50mm}
{
\begin{fmfgraph*}(80,50)
\fmfcmd{%
style_def gorkov_n expr p =
draw_double p;
fill (harrow (p, .45));
fill (tarrow (p, .55))
enddef;}
\fmfcmd{%
style_def gorkov_ain expr p =
draw_double p;
fill (harrow (reverse p, .45));
fill (harrow (p, .45))
enddef;
style_def gorkov_aout expr p =
draw_double p; 
fill (tarrow (reverse p, .55)); 
fill (tarrow (p, .55))
enddef;}
\fmfleft{i1,i2} 
\fmfright{o1,o2} 
\fmf{gorkov_n,label=$\uparrow \omega'$,l.s=right}{i2,i1} 
\fmf{gorkov_ain,right=.5,tension=.5,label=$\downarrow \omega'''$,l.s=right}{o1,o2} 
\fmf{gorkov_aout,right=.5,tension=.5,label=$\uparrow \omega''$,l.s=right}{o2,o1} 
\fmf{dashes,label=$\phantom{a}\bar{d}\phantom{aaaaaaaassaa}\bar{g}\phantom{aaaaaa}$,l.s=right}{o1,i1}
\fmf{dashes,label=$\phantom{a}\bar{c}\phantom{aaaaaaaassaa}e\phantom{aaaaaa}$,l.s=left}{o2,i2}
\fmflabel{$\bar{b}$}{i1}
\fmflabel{$\bar{h}$}{o1}
\fmflabel{$\bar{a}$}{i2}
\fmflabel{$f$}{o2}
\fmfdot{o1,o2,i1,i2}
\end{fmfgraph*} 
}
}

%
%
\newcommand{\secondnbbold}
{
\parbox{50mm}
{
\begin{fmfgraph*}(80,50)
\fmfcmd{%
style_def gorkov_n expr p =
draw_double p;
fill (harrow (p, .45));
fill (tarrow (p, .55))
enddef;}
\fmfcmd{%
style_def gorkov_ain expr p =
draw_double p;
fill (harrow (reverse p, .45));
fill (harrow (p, .45))
enddef;
style_def gorkov_aout expr p =
draw_double p; 
fill (tarrow (reverse p, .55)); 
fill (tarrow (p, .55))
enddef;}
\fmfleft{i1,i2} 
\fmfright{o1,o2} 
\fmf{gorkov_n,label=$\uparrow \omega'$,l.s=right}{i2,i1} 
\fmf{gorkov_aout,tension=0,label=$\uparrow \omega''$,l.s=left}{v1,v2} 
\fmf{gorkov_ain,label=$\uparrow \omega'''$,l.s=right}{o1,o2} 
\fmf{dashes,label=$\phantom{aaa}\bar{c}\phantom{aaassaa}f\phantom{aaaaaaa}$,l.s=right}{i2,v2}
\fmf{dashes,label=$\phantom{aaa}\bar{d}\phantom{aadddddddaassaa}\bar{g}$,l.s=left}{i1,v1}
\fmf{dashes}{o1,v1}
\fmf{phantom}{o2,v2}
\fmf{double,left=1,tension=1}{v2,o2} 
\fmf{double,left=1,tension=1}{o1,v1} 
\fmflabel{$\bar{b}$}{i1}
\fmflabel{$\bar{h}$}{o1}
\fmflabel{$\bar{a}$}{i2}
\fmflabel{$e$}{o2}
\fmfdot{o1,v2,i1,i2}
\end{fmfgraph*} 
}
}

%
%
\newcommand{\secondaa}
{
\parbox{50mm}
{
\begin{fmfgraph*}(80,50)
\fmfcmd{%
style_def gorkov_ain expr p =
draw_double p;
fill (harrow (reverse p, .45));
fill (harrow (p, .45))
enddef;
style_def gorkov_n expr p =
draw_double p;
fill (harrow (p, .45));
fill (tarrow (p, .55))
enddef;
style_def gorkov_aout expr p =
draw_double p; 
fill (tarrow (reverse p, .55)); 
fill (tarrow (p, .55))
enddef;}
\fmfleft{i1,i2} 
\fmfright{o1,o2} 
\fmf{gorkov_aout,label=$\uparrow \omega'$,l.s=left}{i1,i2} 
\fmf{gorkov_n,left=.5,tension=.5,label=$\uparrow \omega''$,l.s=left}{o1,o2} 
\fmf{gorkov_n,left=.5,tension=.5,label=$\downarrow \omega'''$,l.s=left}{o2,o1} 
\fmf{dashes,label=$\phantom{a}\bar{d}\phantom{aaaaaaaassaa}h\phantom{aaaaaa}$,l.s=right}{o1,i1}
\fmf{dashes,label=$\phantom{a}c\phantom{aaaaaaaassaa}f\phantom{aaaaaa}$,l.s=left}{o2,i2}
\fmflabel{$\bar{b}$}{i1}
\fmflabel{$g$}{o1}
\fmflabel{$a$}{i2}
\fmflabel{$e$}{o2}
\fmfdot{o1,o2,i1,i2}
\end{fmfgraph*} 
}
}

%
%
\newcommand{\secondaasd}
{
\parbox{50mm}
{
\begin{fmfgraph*}(80,50)
\fmfcmd{%
style_def gorkov_ain expr p =
draw_double p;
fill (harrow (reverse p, .45));
fill (harrow (p, .45))
enddef;
style_def gorkov_n expr p =
draw_double p;
fill (harrow (p, .45));
fill (tarrow (p, .55))
enddef;
style_def gorkov_aout expr p =
draw_double p; 
fill (tarrow (reverse p, .55)); 
fill (tarrow (p, .55))
enddef;}
\fmfleft{i1,i2} 
\fmfright{o1,o2} 
\fmf{phantom}{i1,i2} 
\fmf{phantom}{o1,o2} 
\fmf{phantom}{v1,i1} 
\fmf{phantom}{v2,o2} 
\fmf{gorkov_n,left=.5,tension=0,label=$\uparrow \omega''$,l.s=right,l.d=.044w}{v1,v2} 
\fmf{gorkov_n,left=.5,tension=0,label=$\downarrow \omega'''$,l.s=left}{v2,v1} 
\fmf{dashes,label=$\phantom{aaaaa}h\phantom{aaa}\bar{b}$,l.s=right}{o1,v1}
\fmf{dashes,label=$c\phantom{aaaa}e\phantom{aaaa}$,l.s=left}{v2,i2}
\fmf{double}{i2,i1} 
\fmf{gorkov_aout,left=1,tension=0,label=$\leftarrow \omega'$,l.s=left}{o1,i1} 
\fmflabel{$g$}{v1}
\fmflabel{$f$}{v2}
\fmflabel{$\bar{d}$}{o1}
\fmflabel{$a$}{i2}
\fmfdot{i2,v2,v1,o1}
\end{fmfgraph*} 
}
}

%
%
\newcommand{\secondaaold}
{
\parbox{50mm}
{
\begin{fmfgraph*}(80,50)
\fmfcmd{%
style_def gorkov_n expr p =
draw_double p;
fill (harrow (p, .45));
fill (tarrow (p, .55))
enddef;}
\fmfcmd{%
style_def gorkov_ain expr p =
draw_double p;
fill (harrow (reverse p, .45));
fill (harrow (p, .45))
enddef;
style_def gorkov_aout expr p =
draw_double p; 
fill (tarrow (reverse p, .55)); 
fill (tarrow (p, .55))
enddef;}
\fmfleft{i1,i2} 
\fmfright{o1,o2} 
\fmf{phantom}{i1,i2} 
\fmf{phantom}{o1,o2} 
\fmf{gorkov_n,tension=0,label=$\uparrow\phantom{ii}\omega''$,l.s=right,l.d=-.13w}{v1,v4} 
\fmf{gorkov_n,tension=0,label=$\downarrow \omega'''$,l.s=left}{v3,v2} 
\fmf{dashes,label=$c\phantom{aaa}e\phantom{aaa}$,l.s=right}{i2,v4}
\fmf{dashes,label=$g\phantom{aiaaaaffaa}\bar{b}\phantom{aaaa}$,l.s=left}{v2,o1}
\fmf{dashes}{v2,v1}
\fmf{phantom}{v4,v3,o2}
\fmf{phantom}{i1,v1}
\fmf{double,right=1,tension=0,label=$f$,l.s=right}{v3,v4} 
\fmf{double,left=1,tension=0}{v2,v1} 
\fmf{double}{i2,i1} 
\fmf{gorkov_aout,left=1,tension=0,label=$\leftarrow \omega'$,l.s=left}{o1,i1} 
\fmflabel{$h$}{v1}
\fmflabel{$\bar{d}$}{o1}
\fmflabel{$a$}{i2}
\fmfdot{o1,v1,v4,i2}
\end{fmfgraph*} 
}
}

%
%
\newcommand{\secondab}
{
\parbox{50mm}
{
\begin{fmfgraph*}(80,50)
\fmfcmd{%
style_def gorkov_n expr p =
draw_double p;
fill (harrow (p, .45));
fill (tarrow (p, .55))
enddef;}
\fmfcmd{%
style_def gorkov_ain expr p =
draw_double p;
fill (harrow (reverse p, .45));
fill (harrow (p, .45))
enddef;
style_def gorkov_aout expr p =
draw_double p; 
fill (tarrow (reverse p, .55)); 
fill (tarrow (p, .55))
enddef;}
\fmfleft{i1,i2} 
\fmfright{o1,o2} 
\fmf{gorkov_aout,label=$\uparrow \omega'$,l.s=left}{i1,i2} 
\fmf{gorkov_aout,left=.5,tension=.5,label=$\uparrow \omega''$,l.s=left}{o1,o2} 
\fmf{gorkov_ain,left=.5,tension=.5,label=$\downarrow \omega'''$,l.s=left}{o2,o1} 
\fmf{dashes,label=$\phantom{a}\bar{d}\phantom{aaaaaaaassaa}\bar{g}\phantom{aaaaaa}$,l.s=right}{o1,i1}
\fmf{dashes,label=$\phantom{a}c\phantom{aaaaaaaassaa}f\phantom{aaaaaa}$,l.s=left}{o2,i2}
\fmflabel{$\bar{b}$}{i1}
\fmflabel{$\bar{h}$}{o1}
\fmflabel{$a$}{i2}
\fmflabel{$e$}{o2}
\fmfdot{o1,o2,i1,i2}
\end{fmfgraph*} 
}
}

%
%
\newcommand{\secondabsd}
{
\parbox{50mm}
{
\begin{fmfgraph*}(80,50)
\fmfcmd{%
style_def gorkov_n expr p =
draw_double p;
fill (harrow (p, .45));
fill (tarrow (p, .55))
enddef;}
\fmfcmd{%
style_def gorkov_ain expr p =
draw_double p;
fill (harrow (reverse p, .45));
fill (harrow (p, .45))
enddef;
style_def gorkov_aout expr p =
draw_double p; 
fill (tarrow (reverse p, .55)); 
fill (tarrow (p, .55))
enddef;}
\fmfleft{i1,i2} 
\fmfright{o1,o2} 
\fmf{phantom}{i1,i2} 
\fmf{phantom}{o1,o2} 
\fmf{phantom}{v1,i1} 
\fmf{phantom}{v2,o2} 
\fmf{gorkov_aout,left=.5,tension=0,label=$\uparrow \omega''$,l.s=right,l.d=.044w}{v1,v2} 
\fmf{gorkov_ain,left=.5,tension=0,label=$\downarrow \omega'''$,l.s=left}{v2,v1} 
\fmf{dashes,label=$\phantom{aaaaa}\bar{h}\phantom{aaa}\bar{b}$,l.s=right}{o1,v1}
\fmf{dashes,label=$c\phantom{aaa}f\phantom{aaaa}$,l.s=left}{v2,i2}
\fmf{double}{i2,i1} 
\fmf{gorkov_aout,left=1,tension=0,label=$\leftarrow \omega'$,l.s=left}{o1,i1} 
\fmflabel{$\bar{g}$}{v1}
\fmflabel{$e$}{v2}
\fmflabel{$\bar{d}$}{o1}
\fmflabel{$a$}{i2}
\fmfdot{i2,v2,v1,o1}
\end{fmfgraph*} 
}
}

%
%
\newcommand{\secondabold}
{
\parbox{50mm}
{
\begin{fmfgraph*}(80,50)
\fmfcmd{%
style_def gorkov_n expr p =
draw_double p;
fill (harrow (p, .45));
fill (tarrow (p, .55))
enddef;}
\fmfcmd{%
style_def gorkov_ain expr p =
draw_double p;
fill (harrow (reverse p, .45));
fill (harrow (p, .45))
enddef;
style_def gorkov_aout expr p =
draw_double p; 
fill (tarrow (reverse p, .55)); 
fill (tarrow (p, .55))
enddef;}
\fmfleft{i1,i2} 
\fmfright{o1,o2} 
\fmf{phantom}{i1,i2} 
\fmf{phantom}{o1,o2} 
\fmf{gorkov_aout,tension=0}{v1,v4} 
\fmf{gorkov_ain,tension=0,label=$\downarrow \omega'''$,l.s=right}{v2,v3} 
\fmf{dashes,label=$c\phantom{aaa}f\phantom{aaa}$,l.s=right}{i2,v4}
\fmf{dashes,label=$\phantom{aaiiaaa}\bar{h}\phantom{aiaa}\bar{b}\phantom{aaa}$,l.s=left}{v2,o1}
\fmf{phantom}{v4,v3,o2}
\fmf{phantom}{i1,v1,v2}
\fmf{double,right=1,tension=0,label=$e$,l.s=right}{v3,v4} 
\fmf{double,left=1,tension=0,label=$\nwarrow \omega''\phantom{aa}$,l.s=left,l.d=.02w,l.a=225}{v2,v1} 
\fmf{double}{i2,i1} 
\fmf{gorkov_aout,left=1,tension=0,label=$\leftarrow \omega'$,l.s=left}{o1,i1} 
\fmflabel{$\bar{g}$}{v2}
\fmflabel{$\bar{d}$}{o1}
\fmflabel{$a$}{i2}
\fmfdot{o1,v2,v4,i2}
\end{fmfgraph*} 
}
}

%
%
\newcommand{\secondaaa}
{
\parbox{50mm}
{
\begin{fmfgraph*}(80,50)
\fmfcmd{%
style_def gorkov_ain expr p =
draw_double p;
fill (harrow (reverse p, .45));
fill (harrow (p, .45))
enddef;
style_def gorkov_n expr p =
draw_double p;
fill (harrow (p, .45));
fill (tarrow (p, .55))
enddef;}
\fmfleft{i1,i2} 
\fmfright{o1,o2} 
\fmf{gorkov_ain,label=$\uparrow \omega'$,l.s=left}{i1,i2} 
\fmf{gorkov_n,left=.5,tension=.5,label=$\uparrow \omega''$,l.s=left}{o1,o2} 
\fmf{gorkov_n,left=.5,tension=.5,label=$\downarrow \omega'''$,l.s=left}{o2,o1} 
\fmf{dashes,label=$\phantom{a}d\phantom{aaaaaaaassaa}g\phantom{aaaaaa}$,l.s=right}{o1,i1}
\fmf{dashes,label=$\phantom{a}\bar{c}\phantom{aaaaaaaassaa}e\phantom{aaaaaa}$,l.s=left}{o2,i2}
\fmflabel{$b$}{i1}
\fmflabel{$h$}{o1}
\fmflabel{$\bar{a}$}{i2}
\fmflabel{$f$}{o2}
\fmfdot{o1,o2,i1,i2}
\end{fmfgraph*} 
}
}

%
%
\newcommand{\secondaaasd}
{
\parbox{50mm}
{
\begin{fmfgraph*}(80,50)
\fmfcmd{%
style_def gorkov_ain expr p =
draw_double p;
fill (harrow (reverse p, .45));
fill (harrow (p, .45))
enddef;
style_def gorkov_n expr p =
draw_double p;
fill (harrow (p, .45));
fill (tarrow (p, .55))
enddef;}
\fmfleft{i1,i2} 
\fmfright{o1,o2} 
\fmf{phantom}{i1,i2} 
\fmf{phantom}{o1,o2} 
\fmf{phantom}{v1,i1} 
\fmf{phantom}{v2,o2} 
\fmf{gorkov_n,left=.5,tension=0,label=$\uparrow \omega''$,l.s=left}{v1,v2} 
\fmf{gorkov_n,left=.5,tension=0,label=$\downarrow \phantom{ii} \omega'''$,l.s=left,l.d=-.13w}{v2,v1} 
\fmf{dashes,label=$\phantom{aaaaa}g\phantom{aaa}d$,l.s=right}{o1,v1}
\fmf{dashes,label=$\bar{a}\phantom{aaa}f\phantom{asaa}$,l.s=left}{v2,i2}
\fmf{double}{o2,o1} 
\fmf{gorkov_ain,right=1,tension=0}{o2,i2} 
\fmflabel{$h$}{v1}
\fmflabel{$e$}{v2}
\fmflabel{$b$}{o1}
\fmflabel{$\bar{c}$}{i2}
\fmflabel{$\uparrow \omega'''$}{o2}
\fmfdot{o1,v2,v1,i2}
\end{fmfgraph*} 
}
}

%
%
\newcommand{\secondaaaold}
{
\parbox{50mm}
{
\begin{fmfgraph*}(80,50)
\fmfcmd{%
style_def gorkov_n expr p =
draw_double p;
fill (harrow (p, .45));
fill (tarrow (p, .55))
enddef;}
\fmfcmd{%
style_def gorkov_ain expr p =
draw_double p;
fill (harrow (reverse p, .45));
fill (harrow (p, .45))
enddef;
style_def gorkov_aout expr p =
draw_double p; 
fill (tarrow (reverse p, .55)); 
fill (tarrow (p, .55))
enddef;}
\fmfleft{i1,i2} 
\fmfright{o1,o2} 
\fmf{phantom}{i1,i2} 
\fmf{phantom}{o1,o2} 
\fmf{gorkov_n,tension=0,label=$\uparrow \omega''$,l.s=left}{v1,v4} 
\fmf{gorkov_n,tension=0,label=$\downarrow \phantom{ii} \omega'''$,l.s=right,l.d=-.23w}{v3,v2} 
\fmf{dashes,label=$\bar{a}\phantom{aaa}f\phantom{aaa}$,l.s=right}{i2,v4}
\fmf{dashes,label=$h\phantom{aidddddddda}d\phantom{aaaa}$,l.s=left}{v2,o1}
\fmf{dashes}{v2,v1}
\fmf{phantom}{v4,v3,o2}
\fmf{phantom}{i1,v1}
\fmf{double,right=1,tension=0,label=$e$,l.s=right}{v3,v4} 
\fmf{double,left=1,tension=0}{v2,v1} 
\fmf{double}{o2,o1} 
\fmf{gorkov_ain,right=1,tension=0}{o2,i2} 
\fmflabel{$g$}{v1}
\fmflabel{$b$}{o1}
\fmflabel{$\bar{c}$}{i2}
\fmflabel{$\uparrow \omega''$}{o2}
\fmfdot{o1,v1,v4,i2}
\end{fmfgraph*} 
}
}

%
%
\newcommand{\secondabb}
{
\parbox{50mm}
{
\begin{fmfgraph*}(80,50)
\fmfcmd{%
style_def gorkov_n expr p =
draw_double p;
fill (harrow (p, .45));
fill (tarrow (p, .55))
enddef;}
\fmfcmd{%
style_def gorkov_ain expr p =
draw_double p;
fill (harrow (reverse p, .45));
fill (harrow (p, .45))
enddef;
style_def gorkov_aout expr p =
draw_double p; 
fill (tarrow (reverse p, .55)); 
fill (tarrow (p, .55))
enddef;}
\fmfleft{i1,i2} 
\fmfright{o1,o2} 
\fmf{gorkov_ain,label=$\uparrow \omega'$,l.s=left}{i1,i2} 
\fmf{gorkov_aout,left=.5,tension=.5,label=$\uparrow \omega''$,l.s=left}{o1,o2} 
\fmf{gorkov_ain,left=.5,tension=.5,label=$\downarrow \omega'''$,l.s=left}{o2,o1} 
\fmf{dashes,label=$\phantom{a}d\phantom{aaaaaaaassaa}\bar{h}\phantom{aaaaaa}$,l.s=right}{o1,i1}
\fmf{dashes,label=$\phantom{a}\bar{c}\phantom{aaaaaaaassaa}e\phantom{aaaaaa}$,l.s=left}{o2,i2}
\fmflabel{$b$}{i1}
\fmflabel{$\bar{g}$}{o1}
\fmflabel{$\bar{a}$}{i2}
\fmflabel{$f$}{o2}
\fmfdot{o1,o2,i1,i2}
\end{fmfgraph*} 
}
}

%
%
\newcommand{\secondabbsd}
{
\parbox{50mm}
{
\begin{fmfgraph*}(80,50)
\fmfcmd{%
style_def gorkov_n expr p =
draw_double p;
fill (harrow (p, .45));
fill (tarrow (p, .55))
enddef;}
\fmfcmd{%
style_def gorkov_ain expr p =
draw_double p;
fill (harrow (reverse p, .45));
fill (harrow (p, .45))
enddef;
style_def gorkov_aout expr p =
draw_double p; 
fill (tarrow (reverse p, .55)); 
fill (tarrow (p, .55))
enddef;}
\fmfleft{i1,i2} 
\fmfright{o1,o2} 
\fmf{phantom}{i1,i2} 
\fmf{phantom}{o1,o2} 
\fmf{phantom}{v1,i1} 
\fmf{phantom}{v2,o2} 
\fmf{gorkov_aout,left=.5,tension=0,label=$\uparrow \omega''$,l.s=left}{v1,v2} 
\fmf{gorkov_ain,left=.5,tension=0,label=$\downarrow \phantom{ii} \omega'''$,l.s=left,l.d=-.13w}{v2,v1} 
\fmf{dashes,label=$\phantom{aaaaa}\bar{g}\phantom{aaa}d$,l.s=right}{o1,v1}
\fmf{dashes,label=$\bar{a}\phantom{aaa}e\phantom{aaa}$,l.s=left}{v2,i2}
\fmf{double}{o2,o1} 
\fmf{gorkov_ain,right=1,tension=0}{o2,i2} 
\fmflabel{$\bar{h}$}{v1}
\fmflabel{$f$}{v2}
\fmflabel{$b$}{o1}
\fmflabel{$\bar{c}$}{i2}
\fmflabel{$\uparrow \omega'''$}{o2}
\fmfdot{o1,v2,v1,i2}
\end{fmfgraph*} 
}
}

%
%
\newcommand{\secondabbold}
{
\parbox{50mm}
{
\begin{fmfgraph*}(80,50)
\fmfcmd{%
style_def gorkov_n expr p =
draw_double p;
fill (harrow (p, .45));
fill (tarrow (p, .55))
enddef;}
\fmfcmd{%
style_def gorkov_ain expr p =
draw_double p;
fill (harrow (reverse p, .45));
fill (harrow (p, .45))
enddef;
style_def gorkov_aout expr p =
draw_double p; 
fill (tarrow (reverse p, .55)); 
fill (tarrow (p, .55))
enddef;}
\fmfleft{i1,i2} 
\fmfright{o1,o2} 
\fmf{phantom}{i1,i2} 
\fmf{phantom}{o1,o2} 
\fmf{gorkov_aout,tension=0,label=$\uparrow \omega''$,l.s=left}{v1,v4} 
\fmf{gorkov_ain,tension=0,label=$\downarrow \phantom{ii} \omega'''$,l.s=left,l.d=-.23w}{v2,v3} 
\fmf{dashes,label=$\bar{a}\phantom{aaa}e\phantom{aaa}$,l.s=right}{i2,v4}
\fmf{dashes,label=$\phantom{aaiiaaa}\bar{g}\phantom{aiaa}d\phantom{aaa}$,l.s=left}{v2,o1}
\fmf{phantom}{v4,v3,o2}
\fmf{phantom}{i1,v1,v2}
\fmf{double,right=1,tension=0,label=$f$,l.s=right}{v3,v4} 
\fmf{double,left=1,tension=0}{v2,v1} 
\fmf{double}{o2,o1} 
\fmf{gorkov_ain,right=1,tension=0}{o2,i2} 
\fmflabel{$\bar{h}$}{v2}
\fmflabel{$b$}{o1}
\fmflabel{$\bar{c}$}{i2}
\fmflabel{$\uparrow \omega''$}{o2}
\fmfdot{o1,v2,v4,i2}
\end{fmfgraph*} 
}
}

%
%
\newcommand{\nlfirstn}
{
\parbox{50mm}
{
\begin{fmfgraph}(80,50)
\fmfcmd{%
style_def gorkov_n expr p =
draw_double p;
fill (harrow (p, .45));
fill (tarrow (p, .55))
enddef;}
\fmfleft{i} 
\fmfright{o}
\fmf{phantom,tension=10,label=$\phantom{aa}b$,l.s=right}{i,i1} 
\fmf{phantom,tension=10}{o,o1} 
\fmf{gorkov_n,left,tension=0.8,label=$c\phantom{aasssssiii}$,l.s=right,l.d=.26w}{o1,v1} 
\fmf{double,left,tension=0.8,label=$d\phantom{aasssssiii}$,l.s=right,l.d=.22w}{v1,o1} 
\fmf{dashes,label=$a\phantom{aaaaaaiii}$,l.s=right}{v1,i1}
\fmfdot{v1,i1}
\fmflabel{$\downarrow \omega'$}{o1}
\end{fmfgraph} 
}
}

%
%
\newcommand{\nlfirstaaa}
{
\parbox{50mm}
{
\begin{fmfgraph}(80,50)
\fmfcmd{%
style_def gorkov_ain expr p =
draw_double p;
fill (harrow (reverse p, .45));
fill (harrow (p, .45))
enddef;
style_def gorkov_aout expr p =
draw_double p; 
fill (tarrow (reverse p, .55)); 
fill (tarrow (p, .55))
enddef;}
\fmfleft{i} 
\fmfright{o} 
\fmf{phantom,tension=10}{i,i1} 
\fmf{phantom,tension=10,label=$d$,l.s=right}{o,o1} 
\fmf{gorkov_aout,left,tension=0.8,label=$\leftarrow \omega'$,l.s=right}{o1,i1} 
\fmf{dashes,label=$\bar{c}\phantom{aaaaaaaaaaaaaiii}$,l.s=right}{o1,i1}
\fmflabel{$a$}{i1}
\fmflabel{$b$}{o1}
\fmfdot{o1,i1}
\end{fmfgraph} 
}
}

%
%
\newcommand{\nlfirstaa}
{
\parbox{50mm}
{
\begin{fmfgraph}(80,50)
\fmfcmd{%
style_def gorkov_ain expr p =
draw_double p;
fill (harrow (reverse p, .45));
fill (harrow (p, .45))
enddef;
style_def gorkov_aout expr p =
draw_double p; 
fill (tarrow (reverse p, .55)); 
fill (tarrow (p, .55))
enddef;}
\fmfleft{i} 
\fmfright{o} 
\fmf{phantom,tension=10}{i,i1} 
\fmf{phantom,tension=10,label=$d$,l.s=right}{o,o1} 
\fmf{gorkov_ain,right,tension=0.8,label=$\leftarrow \omega'$,l.s=right}{o1,i1} 
\fmf{dashes,label=$\bar{c}\phantom{aaaaaaaaaaaaaiii}$,l.s=right}{o1,i1}
\fmflabel{$a$}{i1}
\fmflabel{$b$}{o1}
\fmfdot{o1,i1}
\end{fmfgraph} 
}
}

%
%
\newcommand{\nlsecondnaold}
{
\parbox{50mm}
{
\begin{fmfgraph}(80,50)
\fmfcmd{%
style_def gorkov_n expr p =
draw_double p;
fill (harrow (p, .45));
fill (tarrow (p, .55))
enddef;}
\fmfleft{i1,i2} 
\fmfright{o1,o2} 
\fmf{gorkov_n,label=$\uparrow \omega'$,l.s=left}{i1,i2} 
\fmf{gorkov_n,tension=0,label=$\uparrow \omega''$,l.s=left}{v1,v2} 
\fmf{gorkov_n,label=$\downarrow \omega'''$,l.s=left}{o2,o1} 
\fmf{dashes,label=$\phantom{aaa}c\phantom{aaassaa}f\phantom{aaaaaaa}$,l.s=right}{i2,v2}
\fmf{dashes,label=$\phantom{aaa}d\phantom{aadadsaa}g\phantom{aaaaaaaa}$,l.s=left}{i1,v1}
\fmf{phantom}{o1,v1}
\fmf{phantom}{o2,v2}
\fmf{double,left=1,tension=1}{v2,o2} 
\fmf{double,left=1,tension=1}{o1,v1} 
\fmflabel{$b$}{i1}
\fmflabel{$h$}{o1}
\fmflabel{$a$}{i2}
\fmflabel{$e$}{o2}
\fmfdot{v1,v2,i1,i2}
\end{fmfgraph} 
}
}

%
%
\newcommand{\nlsecondnbold}
{
\parbox{50mm}
{
\begin{fmfgraph}(80,50)
\fmfcmd{%
style_def gorkov_n expr p =
draw_double p;
fill (harrow (p, .45));
fill (tarrow (p, .55))
enddef;}
\fmfcmd{%
style_def gorkov_ain expr p =
draw_double p;
fill (harrow (reverse p, .45));
fill (harrow (p, .45))
enddef;
style_def gorkov_aout expr p =
draw_double p; 
fill (tarrow (reverse p, .55)); 
fill (tarrow (p, .55))
enddef;}
\fmfleft{i1,i2} 
\fmfright{o1,o2} 
\fmf{gorkov_n,label=$\uparrow \omega'$,l.s=left}{i1,i2} 
\fmf{gorkov_aout,tension=0,label=$\uparrow \omega''$,l.s=left}{v1,v2} 
\fmf{gorkov_ain,label=$\uparrow \omega'''$,l.s=right}{o1,o2} 
\fmf{dashes,label=$\phantom{aaa}c\phantom{aaassaa}f\phantom{aaaaaaa}$,l.s=right}{i2,v2}
\fmf{dashes,label=$\phantom{aaa}d\phantom{aadddddddaassaa}\bar{g}$,l.s=left}{i1,v1}
\fmf{dashes}{o1,v1}
\fmf{phantom}{o2,v2}
\fmf{double,left=1,tension=1}{v2,o2} 
\fmf{double,left=1,tension=1}{o1,v1} 
\fmflabel{$b$}{i1}
\fmflabel{$\bar{h}$}{o1}
\fmflabel{$a$}{i2}
\fmflabel{$e$}{o2}
\fmfdot{o1,v2,i1,i2}
\end{fmfgraph} 
}
}

%
%
\newcommand{\nlsecondna}
{
\parbox{50mm}
{
\begin{fmfgraph}(80,50)
\fmfcmd{%
style_def gorkov_n expr p =
draw_double p;
fill (harrow (p, .45));
fill (tarrow (p, .55))
enddef;}
\fmfleft{i1,i2} 
\fmfright{o1,o2} 
\fmf{gorkov_n,label=$\uparrow \omega'$,l.s=left}{i1,i2} 
\fmf{gorkov_n,left=.5,tension=.5,label=$\uparrow \omega''$,l.s=left}{o1,o2} 
\fmf{gorkov_n,left=.5,tension=.5,label=$\downarrow \omega'''$,l.s=left}{o2,o1} 
\fmf{dashes,label=$\phantom{a}d\phantom{aaaaaassaa}g\phantom{aaaaaa}$,l.s=right}{o1,i1}
\fmf{dashes,label=$\phantom{a}c\phantom{aaaaaassaa}f\phantom{aaaaaa}$,l.s=left}{o2,i2}
\fmflabel{$b$}{i1}
\fmflabel{$h$}{o1}
\fmflabel{$a$}{i2}
\fmflabel{$e$}{o2}
\fmfdot{o1,o2,i1,i2}
\end{fmfgraph} 
}
}

%
%
\newcommand{\nlsecondnb}
{
\parbox{50mm}
{
\begin{fmfgraph}(80,50)
\fmfcmd{%
style_def gorkov_n expr p =
draw_double p;
fill (harrow (p, .45));
fill (tarrow (p, .55))
enddef;}
\fmfcmd{%
style_def gorkov_ain expr p =
draw_double p;
fill (harrow (reverse p, .45));
fill (harrow (p, .45))
enddef;
style_def gorkov_aout expr p =
draw_double p; 
fill (tarrow (reverse p, .55)); 
fill (tarrow (p, .55))
enddef;}
\fmfleft{i1,i2} 
\fmfright{o1,o2} 
\fmf{gorkov_n,label=$\uparrow \omega'$,l.s=left}{i1,i2} 
\fmf{gorkov_aout,left=.5,tension=.5,label=$\uparrow \omega''$,l.s=left}{o1,o2} 
\fmf{gorkov_ain,left=.5,tension=.5,label=$\downarrow \omega'''$,l.s=left}{o2,o1} 
\fmf{dashes,label=$\phantom{a}d\phantom{aaaaaassaa}g\phantom{aaaaaa}$,l.s=right}{o1,i1}
\fmf{dashes,label=$\phantom{a}c\phantom{aaaaaassaa}f\phantom{aaaaaa}$,l.s=left}{o2,i2}
\fmflabel{$b$}{i1}
\fmflabel{$h$}{o1}
\fmflabel{$a$}{i2}
\fmflabel{$e$}{o2}
\fmfdot{o1,o2,i1,i2}
\end{fmfgraph} 
}
}

%
%
\newcommand{\nlsecondnboldold}
{
\parbox{50mm}
{
\begin{fmfgraph}(80,50)
\fmfcmd{%
style_def gorkov_n expr p =
draw_double p;
fill (harrow (p, .45));
fill (tarrow (p, .55))
enddef;}
\fmfcmd{%
style_def gorkov_ain expr p =
draw_double p;
fill (harrow (reverse p, .45));
fill (harrow (p, .45))
enddef;
style_def gorkov_aout expr p =
draw_double p; 
fill (tarrow (reverse p, .55)); 
fill (tarrow (p, .55))
enddef;}
\fmfleft{i1,i2} 
\fmfright{o1,o2} 
\fmf{gorkov_n,label=$\uparrow \omega'$,l.s=left}{i1,i2} 
\fmf{gorkov_aout,tension=0,label=$\uparrow \omega''$,l.s=left}{v1,v2} 
\fmf{gorkov_ain,label=$\uparrow \omega'''$,l.s=right}{o1,o2} 
\fmf{dashes,label=$\phantom{aaa}c\phantom{aaassaa}f\phantom{aaaaaaa}$,l.s=right}{i2,v2}
\fmf{dashes,label=$\phantom{aaa}d\phantom{aadddddddaassaa}g$,l.s=left}{i1,v1}
\fmf{dashes}{o1,v1}
\fmf{phantom}{o2,v2}
\fmf{double,left=1,tension=1}{v2,o2} 
\fmf{double,left=1,tension=1}{o1,v1} 
\fmflabel{$b$}{i1}
\fmflabel{$h$}{o1}
\fmflabel{$a$}{i2}
\fmflabel{$e$}{o2}
\fmfdot{o1,v2,i1,i2}
\end{fmfgraph} 
}
}

%
%
\newcommand{\nlsecondaaanew}
{
\parbox{50mm}
{
\begin{fmfgraph}(80,50)
\fmfcmd{%
style_def gorkov_n expr p =
draw_double p;
fill (harrow (p, .45));
fill (tarrow (p, .55))
enddef;}
\fmfcmd{%
style_def gorkov_ain expr p =
draw_double p;
fill (harrow (reverse p, .45));
fill (harrow (p, .45))
enddef;
style_def gorkov_aout expr p =
draw_double p; 
fill (tarrow (reverse p, .55)); 
fill (tarrow (p, .55))
enddef;}
\fmfleft{i1,i2} 
\fmfright{o1,o2} 
\fmf{gorkov_ain,label=$\uparrow \omega'$,l.s=left}{i1,i2} 
\fmf{gorkov_n,left=.5,tension=.5,label=$\uparrow \omega''$,l.s=left}{o1,o2} 
\fmf{gorkov_n,left=.5,tension=.5,label=$\downarrow \omega'''$,l.s=left}{o2,o1} 
\fmf{dashes,label=$\phantom{a}d\phantom{aaaaaassaa}g\phantom{aaaaaa}$,l.s=right}{o1,i1}
\fmf{dashes,label=$\phantom{a}c\phantom{aaaaaassaa}f\phantom{aaaaaa}$,l.s=left}{o2,i2}
\fmflabel{$b$}{i1}
\fmflabel{$h$}{o1}
\fmflabel{$a$}{i2}
\fmflabel{$e$}{o2}
\fmfdot{o1,o2,i1,i2}
\end{fmfgraph} 
}
}

%
%
\newcommand{\nlsecondaaaold}
{
\parbox{50mm}
{
\begin{fmfgraph}(80,50)
\fmfcmd{%
style_def gorkov_n expr p =
draw_double p;
fill (harrow (p, .45));
fill (tarrow (p, .55))
enddef;}
\fmfcmd{%
style_def gorkov_ain expr p =
draw_double p;
fill (harrow (reverse p, .45));
fill (harrow (p, .45))
enddef;
style_def gorkov_aout expr p =
draw_double p; 
fill (tarrow (reverse p, .55)); 
fill (tarrow (p, .55))
enddef;}
\fmfleft{i1,i2} 
\fmfright{o1,o2} 
\fmf{phantom}{i1,i2} 
\fmf{phantom}{o1,o2} 
\fmf{gorkov_n,tension=0,label=$\uparrow \omega''$,l.s=left}{v1,v4} 
\fmf{gorkov_n,tension=0,label=$\downarrow \phantom{ii} \omega'''$,l.s=right,l.d=-.23w}{v3,v2} 
\fmf{dashes,label=$\bar{a}\phantom{aaa}f\phantom{aaa}$,l.s=right}{i2,v4}
\fmf{dashes,label=$h\phantom{aidddddddda}d\phantom{aaaa}$,l.s=left}{v2,o1}
\fmf{dashes}{v2,v1}
\fmf{phantom}{v4,v3,o2}
\fmf{phantom}{i1,v1}
\fmf{double,right=1,tension=0,label=$e$,l.s=right}{v3,v4} 
\fmf{double,left=1,tension=0}{v2,v1} 
\fmf{double}{o2,o1} 
\fmf{gorkov_ain,right=1,tension=0}{o2,i2} 
\fmflabel{$g$}{v1}
\fmflabel{$b$}{o1}
\fmflabel{$\bar{c}$}{i2}
\fmflabel{$\uparrow \omega''$}{o2}
\fmfdot{o1,v1,v4,i2}
\end{fmfgraph} 
}
}

%
%
\newcommand{\nlsecondabbold}
{
\parbox{50mm}
{
\begin{fmfgraph}(80,50)
\fmfcmd{%
style_def gorkov_n expr p =
draw_double p;
fill (harrow (p, .45));
fill (tarrow (p, .55))
enddef;}
\fmfcmd{%
style_def gorkov_ain expr p =
draw_double p;
fill (harrow (reverse p, .45));
fill (harrow (p, .45))
enddef;
style_def gorkov_aout expr p =
draw_double p; 
fill (tarrow (reverse p, .55)); 
fill (tarrow (p, .55))
enddef;}
\fmfleft{i1,i2} 
\fmfright{o1,o2} 
\fmf{phantom}{i1,i2} 
\fmf{phantom}{o1,o2} 
\fmf{gorkov_aout,tension=0,label=$\uparrow \omega''$,l.s=left}{v1,v4} 
\fmf{gorkov_ain,tension=0,label=$\downarrow \phantom{ii} \omega'''$,l.s=left,l.d=-.23w}{v2,v3} 
\fmf{dashes,label=$\bar{a}\phantom{aaa}e\phantom{aaa}$,l.s=right}{i2,v4}
\fmf{dashes,label=$\phantom{aaiiaaa}\bar{g}\phantom{aiaa}d\phantom{aaa}$,l.s=left}{v2,o1}
\fmf{phantom}{v4,v3,o2}
\fmf{phantom}{i1,v1,v2}
\fmf{double,right=1,tension=0,label=$f$,l.s=right}{v3,v4} 
\fmf{double,left=1,tension=0}{v2,v1} 
\fmf{double}{o2,o1} 
\fmf{gorkov_ain,right=1,tension=0}{o2,i2} 
\fmflabel{$\bar{h}$}{v2}
\fmflabel{$b$}{o1}
\fmflabel{$\bar{c}$}{i2}
\fmflabel{$\uparrow \omega''$}{o2}
\fmfdot{o1,v2,v4,i2}
\end{fmfgraph} 
}
}

%
%
\newcommand{\nlsecondaaa}
{
\parbox{50mm}
{
\begin{fmfgraph}(80,50)
\fmfcmd{%
style_def gorkov_ain expr p =
draw_double p;
fill (harrow (reverse p, .45));
fill (harrow (p, .45))
enddef;
style_def gorkov_n expr p =
draw_double p;
fill (harrow (p, .45));
fill (tarrow (p, .55))
enddef;}
\fmfleft{i1,i2} 
\fmfright{o1,o2} 
\fmf{phantom}{i1,i2} 
\fmf{phantom}{o1,o2} 
\fmf{phantom}{v1,i1} 
\fmf{phantom}{v2,o2} 
\fmf{gorkov_n,left=.5,tension=0,label=$\uparrow \omega''$,l.s=left}{v1,v2} 
\fmf{gorkov_n,left=.5,tension=0,label=$\downarrow \phantom{ii} \omega'''$,l.s=left,l.d=-.13w}{v2,v1} 
\fmf{dashes,label=$\phantom{aaaaa}g\phantom{aaa}d$,l.s=right}{o1,v1}
\fmf{dashes,label=$a\phantom{aaa}e\phantom{aaa}$,l.s=left}{v2,i2}
\fmf{double}{o2,o1} 
\fmf{gorkov_ain,right=1,tension=0}{o2,i2} 
\fmflabel{$h$}{v1}
\fmflabel{$f$}{v2}
\fmflabel{$b$}{o1}
\fmflabel{$\bar{c}$}{i2}
\fmflabel{$\uparrow \omega'''$}{o2}
\fmfdot{o1,v2,v1,i2}
\end{fmfgraph} 
}
}

%
%
\newcommand{\nlsecondabb}
{
\parbox{50mm}
{
\begin{fmfgraph}(80,50)
\fmfcmd{%
style_def gorkov_n expr p =
draw_double p;
fill (harrow (p, .45));
fill (tarrow (p, .55))
enddef;}
\fmfcmd{%
style_def gorkov_ain expr p =
draw_double p;
fill (harrow (reverse p, .45));
fill (harrow (p, .45))
enddef;
style_def gorkov_aout expr p =
draw_double p; 
fill (tarrow (reverse p, .55)); 
fill (tarrow (p, .55))
enddef;}
\fmfleft{i1,i2} 
\fmfright{o1,o2} 
\fmf{phantom}{i1,i2} 
\fmf{phantom}{o1,o2} 
\fmf{phantom}{v1,i1} 
\fmf{phantom}{v2,o2} 
\fmf{gorkov_aout,left=.5,tension=0,label=$\uparrow \omega''$,l.s=left}{v1,v2} 
\fmf{gorkov_ain,left=.5,tension=0,label=$\downarrow \phantom{ii} \omega'''$,l.s=left,l.d=-.13w}{v2,v1} 
\fmf{dashes,label=$\phantom{aaaaa}g\phantom{aaa}d$,l.s=right}{o1,v1}
\fmf{dashes,label=$a\phantom{aaa}e\phantom{aaa}$,l.s=left}{v2,i2}
\fmf{double}{o2,o1} 
\fmf{gorkov_ain,right=1,tension=0}{o2,i2} 
\fmflabel{$h$}{v1}
\fmflabel{$f$}{v2}
\fmflabel{$b$}{o1}
\fmflabel{$\bar{c}$}{i2}
\fmflabel{$\uparrow \omega'''$}{o2}
\fmfdot{o1,v2,v1,i2}
\end{fmfgraph} 
}
}

%
%
\newcommand{\nlsecondabbnew}
{
\parbox{50mm}
{
\begin{fmfgraph}(80,50)
\fmfcmd{%
style_def gorkov_n expr p =
draw_double p;
fill (harrow (p, .45));
fill (tarrow (p, .55))
enddef;}
\fmfcmd{%
style_def gorkov_ain expr p =
draw_double p;
fill (harrow (reverse p, .45));
fill (harrow (p, .45))
enddef;
style_def gorkov_aout expr p =
draw_double p; 
fill (tarrow (reverse p, .55)); 
fill (tarrow (p, .55))
enddef;}
\fmfleft{i1,i2} 
\fmfright{o1,o2} 
\fmf{gorkov_ain,label=$\uparrow \omega'$,l.s=left}{i1,i2} 
\fmf{gorkov_ain,left=.5,tension=.5,label=$\uparrow \omega''$,l.s=left}{o1,o2} 
\fmf{gorkov_aout,left=.5,tension=.5,label=$\downarrow \omega'''$,l.s=left}{o2,o1} 
\fmf{dashes,label=$\phantom{a}d\phantom{aaaaaassaa}g\phantom{aaaaaa}$,l.s=right}{o1,i1}
\fmf{dashes,label=$\phantom{a}c\phantom{aaaaaassaa}f\phantom{aaaaaa}$,l.s=left}{o2,i2}
\fmflabel{$b$}{i1}
\fmflabel{$h$}{o1}
\fmflabel{$a$}{i2}
\fmflabel{$e$}{o2}
\fmfdot{o1,o2,i1,i2}
\end{fmfgraph} 
}
}

%
%
\newcommand{\nlsecondabboldold}
{
\parbox{50mm}
{
\begin{fmfgraph}(80,50)
\fmfcmd{%
style_def gorkov_n expr p =
draw_double p;
fill (harrow (p, .45));
fill (tarrow (p, .55))
enddef;}
\fmfcmd{%
style_def gorkov_ain expr p =
draw_double p;
fill (harrow (reverse p, .45));
fill (harrow (p, .45))
enddef;
style_def gorkov_aout expr p =
draw_double p; 
fill (tarrow (reverse p, .55)); 
fill (tarrow (p, .55))
enddef;}
\fmfleft{i1,i2} 
\fmfright{o1,o2} 
\fmf{phantom}{i1,i2} 
\fmf{phantom}{o1,o2} 
\fmf{gorkov_aout,tension=0,label=$\uparrow \omega''$,l.s=left}{v1,v4} 
\fmf{gorkov_ain,tension=0,label=$\downarrow \phantom{ii} \omega'''$,l.s=left,l.d=-.23w}{v2,v3} 
\fmf{dashes,label=$a\phantom{aaa}e\phantom{aaa}$,l.s=right}{i2,v4}
\fmf{dashes,label=$\phantom{aaiiaaa}g\phantom{aiaa}d\phantom{aaa}$,l.s=left}{v2,o1}
\fmf{phantom}{v4,v3,o2}
\fmf{phantom}{i1,v1,v2}
\fmf{double,right=1,tension=0,label=$f$,l.s=right}{v3,v4} 
\fmf{double,left=1,tension=0}{v2,v1} 
\fmf{double}{o2,o1} 
\fmf{gorkov_ain,right=1,tension=0}{o2,i2} 
\fmflabel{$h$}{v2}
\fmflabel{$b$}{o1}
\fmflabel{$\bar{c}$}{i2}
\fmflabel{$\uparrow \omega''$}{o2}
\fmfdot{o1,v2,v4,i2}
\end{fmfgraph} 
}
}

%
%
\newcommand{\phifirst}
{
\parbox{50mm}
{
\begin{fmfgraph}(50,30)
\fmfcmd{%
style_def gorkov_n expr p =
draw_double p;
fill (harrow (p, .45));
fill (tarrow (p, .55))
enddef;}
\fmfcmd{%
style_def gorkov_ain expr p =
draw_double p;
fill (harrow (reverse p, .45));
fill (harrow (p, .45))
enddef;
style_def gorkov_aout expr p =
draw_double p; 
fill (tarrow (reverse p, .55)); 
fill (tarrow (p, .55))
enddef;}
\fmfleft{i} 
\fmfright{o} 
\fmf{gorkov_n,right,tension=1.2}{o,o} 
\fmf{gorkov_n,right,tension=1.2}{i,i} 
\fmf{dashes,label=$\bar{c}\phantom{aaaaaaaaaaaaaiii}$,l.s=right}{o,i}
\fmfdot{o,i}
\end{fmfgraph} 
}
}

%
%
\newcommand{\phifirstex}
{
\parbox{50mm}
{
\begin{fmfgraph}(80,30)
\fmfcmd{%
style_def gorkov_n expr p =
draw_double p;
fill (harrow (p, .45));
fill (tarrow (p, .55))
enddef;}
\fmfcmd{%
style_def gorkov_ain expr p =
draw_double p;
fill (harrow (reverse p, .45));
fill (harrow (p, .45))
enddef;
style_def gorkov_aout expr p =
draw_double p; 
fill (tarrow (reverse p, .55)); 
fill (tarrow (p, .55))
enddef;}
\fmfleft{i} 
\fmfright{o} 
\fmf{phantom,tension=10}{i,i1} 
\fmf{phantom,tension=10}{o,o1} 
\fmf{gorkov_ain,right,tension=1.8}{o1,i1} 
\fmf{gorkov_aout,right,tension=1.8}{i1,o1} 
\fmf{dashes,label=$\bar{c}\phantom{aaaaaaaaaaaaaiii}$,l.s=right}{o1,i1}
\fmflabel{$a$}{i1}
\fmflabel{$b$}{o1}
\fmfdot{o1,i1}
\end{fmfgraph} 
}
}

%
%
\newcommand{\philabn}
{
\parbox{50mm}
{
\begin{fmfgraph*}(60,40)
\fmfcmd{%
style_def gorkov_n expr p =
draw_double p;
fill (harrow (p, .70));
fill (tarrow (p, .80))
enddef;
style_def gorkov_nr expr p =
draw_double p;
fill (harrow (reverse p, .70));
fill (tarrow (reverse p, .80))
enddef;}
\fmfleft{i} 
\fmfright{o}
\fmf{gorkov_nr,right,tension=1.4,label=$\hspace{-.4cm}\downarrow \omega''$,l.s=right,l.d=.26w}{o,o} 
\fmf{gorkov_n,left,tension=1.4,label=$\downarrow \omega'$,l.s=right}{i,i} 
\fmf{dashes,label=$a\phantom{dddddddoi}c$,l.s=right}{o,i}
\fmfdot{i,o}
\fmflabel{$d$}{o}
\fmflabel{$b$}{i}
\end{fmfgraph*} 
}
}

%
%
\newcommand{\philaba}
{
\parbox{50mm}
{
\begin{fmfgraph*}(80,50)
\fmfcmd{%
style_def gorkov_ain expr p =
draw_double p;
fill (harrow (reverse p, .45));
fill (harrow (p, .45))
enddef;
style_def gorkov_aout expr p =
draw_double p; 
fill (tarrow (reverse p, .55)); 
fill (tarrow (p, .55))
enddef;}
\fmfleft{i} 
\fmfright{o} 
\fmf{phantom,tension=10}{i,i1} 
\fmf{phantom,tension=10,label=$\bar{b}$,l.s=right}{o,o1} 
\fmf{gorkov_aout,left,tension=0.8,label=$\leftarrow \omega'$,l.s=left}{o1,i1} 
\fmf{gorkov_ain,left,tension=0.8,label=$\rightarrow \omega'$,l.s=left}{i1,o1} 
\fmf{dashes,label=$a\phantom{aaaaaaaaaaaiii}$,l.s=right}{o1,i1}
\fmflabel{$c$}{i1}
\fmflabel{$\bar{d}$}{o1}
\fmfdot{o1,i1}
\end{fmfgraph*} 
}
}

%
%
\newcommand{\phisecondnn}
{
\parbox{50mm}
{
\begin{fmfgraph}(80,50)
\fmfcmd{%
style_def gorkov_n expr p =
draw_double p;
fill (harrow (p, .45));
fill (tarrow (p, .55))
enddef;}
\fmfcmd{%
style_def gorkov_ain expr p =
draw_double p;
fill (harrow (reverse p, .45));
fill (harrow (p, .45))
enddef;
style_def gorkov_aout expr p =
draw_double p; 
fill (tarrow (reverse p, .55)); 
fill (tarrow (p, .55))
enddef;}
\fmfleft{i1,i2} 
\fmfright{o1,o2} 
\fmf{gorkov_n,left=.5,tension=.5,label=$\uparrow \omega''$,l.s=left}{i1,i2} 
\fmf{gorkov_n,left=.5,tension=.5,label=$\downarrow \omega'''$,l.s=left}{i2,i1} 
\fmf{gorkov_n,left=.5,tension=.5,label=$\uparrow \omega''$,l.s=left}{o1,o2} 
\fmf{gorkov_n,left=.5,tension=.5,label=$\downarrow \omega'''$,l.s=left}{o2,o1} 
\fmf{dashes,label=$\phantom{a}d\phantom{aaaaaassaa}g\phantom{aaaaaa}$,l.s=right}{o1,i1}
\fmf{dashes,label=$\phantom{a}c\phantom{aaaaaassaa}f\phantom{aaaaaa}$,l.s=left}{o2,i2}
\fmfdot{o1,i1,o2,i2}
\end{fmfgraph} 
}
}

%
%
\newcommand{\phisecondna}
{
\parbox{50mm}
{
\begin{fmfgraph}(80,50)
\fmfcmd{%
style_def gorkov_n expr p =
draw_double p;
fill (harrow (p, .45));
fill (tarrow (p, .55))
enddef;}
\fmfcmd{%
style_def gorkov_ain expr p =
draw_double p;
fill (harrow (reverse p, .45));
fill (harrow (p, .45))
enddef;
style_def gorkov_aout expr p =
draw_double p; 
fill (tarrow (reverse p, .55)); 
fill (tarrow (p, .55))
enddef;}
\fmfleft{i1,i2} 
\fmfright{o1,o2} 
\fmf{gorkov_n,left=.5,tension=.5,label=$\uparrow \omega''$,l.s=left}{i1,i2} 
\fmf{gorkov_n,left=.5,tension=.5,label=$\downarrow \omega'''$,l.s=left}{i2,i1} 
\fmf{gorkov_ain,left=.5,tension=.5,label=$\uparrow \omega''$,l.s=left}{o1,o2} 
\fmf{gorkov_aout,left=.5,tension=.5,label=$\downarrow \omega'''$,l.s=left}{o2,o1} 
\fmf{dashes,label=$\phantom{a}d\phantom{aaaaaassaa}g\phantom{aaaaaa}$,l.s=right}{o1,i1}
\fmf{dashes,label=$\phantom{a}c\phantom{aaaaaassaa}f\phantom{aaaaaa}$,l.s=left}{o2,i2}
\fmfdot{o1,i1,o2,i2}
\end{fmfgraph} 
}
}

%
%
\newcommand{\phisecondaa}
{
\parbox{50mm}
{
\begin{fmfgraph}(80,50)
\fmfcmd{%
style_def gorkov_n expr p =
draw_double p;
fill (harrow (p, .45));
fill (tarrow (p, .55))
enddef;}
\fmfcmd{%
style_def gorkov_ain expr p =
draw_double p;
fill (harrow (reverse p, .45));
fill (harrow (p, .45))
enddef;
style_def gorkov_aout expr p =
draw_double p; 
fill (tarrow (reverse p, .55)); 
fill (tarrow (p, .55))
enddef;}
\fmfleft{i1,i2} 
\fmfright{o1,o2} 
\fmf{gorkov_ain,left=.5,tension=.5,label=$\uparrow \omega''$,l.s=left}{i1,i2} 
\fmf{gorkov_aout,left=.5,tension=.5,label=$\downarrow \omega'''$,l.s=left}{i2,i1} 
\fmf{gorkov_ain,left=.5,tension=.5,label=$\uparrow \omega''$,l.s=left}{o1,o2} 
\fmf{gorkov_aout,left=.5,tension=.5,label=$\downarrow \omega'''$,l.s=left}{o2,o1} 
\fmf{dashes,label=$\phantom{a}d\phantom{aaaaaassaa}g\phantom{aaaaaa}$,l.s=right}{o1,i1}
\fmf{dashes,label=$\phantom{a}c\phantom{aaaaaassaa}f\phantom{aaaaaa}$,l.s=left}{o2,i2}
\fmfdot{o1,i1,o2,i2}
\end{fmfgraph} 
}
}

%
%
\newcommand{\phithirda}
{
\parbox{50mm}
{
\begin{fmfgraph}(50,100)
\fmfcmd{%
style_def gorkov_na expr p =
draw_double p;
enddef;}
\fmfstraight
\fmfleft{i1,i2,i3} 
\fmfright{o1,o2,o3}
\fmf{gorkov_na,left=.35,tension=.5}{i1,i3} 
\fmf{gorkov_na,tension=0.0}{i2,i1} 
\fmf{gorkov_na,tension=0.0}{i3,i2} 
\fmf{gorkov_na,right=.35,tension=.5}{o1,o3} 
\fmf{gorkov_na,tension=0.0}{o2,o1} 
\fmf{gorkov_na,tension=0.0}{o3,o2} 
\fmf{dashes}{o1,i1}
\fmf{dashes}{o3,i3}
\fmf{dashes}{o2,i2}
\fmfdot{o1,i1,o2,i2,i3,o3}
\end{fmfgraph} 
}
}

%
%
\newcommand{\phithirdb}
{
\parbox{50mm}
{
\begin{fmfgraph}(80,100)
\fmfcmd{%
style_def gorkov_na expr p =
draw_double p;
enddef;}
\fmfstraight
\fmfleft{i1,i2,i3} 
\fmfright{o1,o2,o3}
\fmftop{vt}
\fmfbottom{vb}
\fmf{gorkov_na,left=.2,tension=.5}{i1,i3} 
\fmf{gorkov_na,left=.2,tension=.5}{i3,i1} 
\fmf{dashes}{i1,vb}
\fmf{dashes}{o3,i3}
\fmf{gorkov_na,left=.3,tension=.5}{o2,o3} 
\fmf{gorkov_na,left=.3,tension=.5}{o3,o2}
\fmf{phantom}{i1,vc,o3}
\fmffreeze
\fmf{gorkov_na,left=.3,tension=.5}{vb,vc} 
\fmf{gorkov_na,left=.3,tension=.5}{vc,vb} 
\fmf{phantom,left=.3,tension=.5}{vt,vc} 
\fmf{phantom,left=.3,tension=.5}{vc,vt} 
\fmf{dashes}{vc,o2}
\fmfdot{vc,o2,i1,vb,o3,i3}
\end{fmfgraph} 
}
}

%
%
\newcommand{\diagtma}
{
\parbox{50mm}
{
\begin{fmfgraph}(50,100)
\fmfcmd{%
style_def gorkov_n expr p =
draw_double p;
fill (harrow (p, .45));
fill (tarrow (p, .55))
enddef;}
\fmfcmd{%
style_def gorkov_ain expr p =
draw_double p;
fill (harrow (reverse p, .45));
fill (harrow (p, .45))
enddef;
style_def gorkov_aout expr p =
draw_double p; 
fill (tarrow (reverse p, .55)); 
fill (tarrow (p, .55))
enddef;}
\fmfstraight
\fmfleft{i1,i2,i3} 
\fmfright{o1,o2,o3} 
\fmf{double_arrow,tension=0.0}{i1,i2} 
\fmf{double_arrow,tension=0.0}{i2,i3} 
\fmf{double_arrow,left=.35,tension=.5}{o3,o1} 
\fmf{double_arrow,tension=0.0}{o1,o2} 
\fmf{double_arrow,tension=0.0}{o2,o3} 
\fmf{dashes}{o1,i1}
\fmf{dashes}{o3,i3}
\fmf{dashes}{o2,i2}
\fmfdot{i1,i2,i3,o1,o2,o3}
\end{fmfgraph} 
}
}

%
%
\newcommand{\diagrpa}
{
\parbox{50mm}
{
\begin{fmfgraph}(80,100)
\fmfcmd{%
style_def gorkov_n expr p =
draw_double p;
fill (harrow (p, .45));
fill (tarrow (p, .55))
enddef;}
\fmfcmd{%
style_def gorkov_ain expr p =
draw_double p;
fill (harrow (reverse p, .45));
fill (harrow (p, .45))
enddef;
style_def gorkov_aout expr p =
draw_double p; 
fill (tarrow (reverse p, .55)); 
fill (tarrow (p, .55))
enddef;}
\fmfstraight
\fmfleft{i1,i2,i3} 
\fmfright{o1,o2,o3}
\fmftop{vt}
\fmfbottom{vb}
\fmf{double_arrow,tension=.0}{i1,i3} 
\fmf{dashes}{i1,vb}
\fmf{dashes}{o3,i3}
\fmf{double_arrow,left=.3,tension=.5}{o2,o3} 
\fmf{double_arrow,left=.3,tension=.5}{o3,o2}
\fmf{phantom}{i1,vc,o3}
\fmffreeze
\fmf{double_arrow,left=.3,tension=.5}{vb,vc} 
\fmf{double_arrow,left=.3,tension=.5}{vc,vb} 
\fmf{phantom,left=.3,tension=.5}{vt,vc} 
\fmf{phantom,left=.3,tension=.5}{vc,vt} 
\fmf{dashes}{vc,o2}
\fmfdot{i1,i3,o2,o3,vb,vc}
\end{fmfgraph} 
}
}

%
%
\newcommand{\diaggoo}
{
\parbox{50mm}
{
\begin{fmfgraph*}(80,50)
\fmfcmd{%
style_def gorkov_n expr p =
draw_double p;
fill (harrow (p, .45));
fill (tarrow (p, .55))
enddef;}
\fmfleft{i1,i2}
\fmfright{o1,o1} 
\fmf{phantom}{o1,i1}
\fmf{phantom}{o2,i2}
\fmf{gorkov_n,label=$\uparrow \omega$,l.s=right,l.d=.17w}{i1,i2}
\fmflabel{$b$}{i1}
\fmflabel{$a$}{i2}
\end{fmfgraph*} 
}
}

%
%
\newcommand{\diaggtt}
{
\parbox{50mm}
{
\begin{fmfgraph*}(80,50)
\fmfcmd{%
style_def gorkov_n expr p =
draw_double p;
fill (harrow (p, .45));
fill (tarrow (p, .55))
enddef;}
\fmfleft{i1,i2} 
\fmfright{o1,o1} 
\fmf{phantom}{o1,i1}
\fmf{phantom}{o2,i2}
\fmf{gorkov_n,label=$\uparrow \omega$,l.s=left,l.d=.15w}{i2,i1}
\fmflabel{$\bar{b}$}{i1}
\fmflabel{$\bar{a}$}{i2}
\end{fmfgraph*} 
}
}

%
%
\newcommand{\diaggot}
{
\parbox{50mm}
{
\begin{fmfgraph*}(80,50)
\fmfcmd{%
style_def gorkov_aout expr p =
draw_double p; 
fill (tarrow (reverse p, .55)); 
fill (tarrow (p, .55))
enddef;}
\fmfleft{i1,i2}
\fmfright{o1,o1} 
\fmf{phantom}{o1,i1}
\fmf{phantom}{o2,i2}
\fmf{gorkov_aout,label=$\uparrow \omega$,l.s=right,l.d=.17w}{i1,i2}
\fmflabel{$\bar{b}$}{i1}
\fmflabel{$a$}{i2}
\end{fmfgraph*} 
}
}

%
%
\newcommand{\diaggotold} 
{
\parbox{50mm}
{
\begin{fmfgraph*}(80,50)
\fmfcmd{%
style_def gorkov_aout expr p =
draw_double p; 
fill (tarrow (reverse p, .55)); 
fill (tarrow (p, .55))
enddef;}
\fmfleft{i} 
\fmfright{o} 
\fmf{phantom,tension=10}{i,i1} 
\fmf{phantom,tension=10}{o,o1} 
\fmf{gorkov_aout,left,tension=0.8,label=$\leftarrow \omega$,l.s=left,l.d=.15w}{o1,i1} 
\fmf{phantom}{o1,i1}
\fmflabel{$a$}{i1}
\fmflabel{$\bar{b}$}{o1}
\end{fmfgraph*} 
}
}

%
%
\newcommand{\diaggto}
{
\parbox{50mm}
{
\begin{fmfgraph*}(80,50)
\fmfcmd{%
style_def gorkov_ain expr p =
draw_double p;
fill (harrow (reverse p, .45));
fill (harrow (p, .45))
enddef;}
\fmfleft{i1,i2}
\fmfright{o1,o1} 
\fmf{phantom}{o1,i1}
\fmf{phantom}{o2,i2}
\fmf{gorkov_ain,label=$\uparrow \omega$,l.s=right,l.d=.17w}{i1,i2}
\fmflabel{$b$}{i1}
\fmflabel{$\bar{a}$}{i2}
\end{fmfgraph*} 
}
}

%
%
\newcommand{\diaggtoold}
{
\parbox{50mm}
{
\begin{fmfgraph*}(80,50)
\fmfcmd{%
style_def gorkov_ain expr p =
draw_double p;
fill (harrow (reverse p, .45));
fill (harrow (p, .45))
enddef;}
\fmfleft{i} 
\fmfright{o} 
\fmf{phantom,tension=10}{i,i1} 
\fmf{phantom,tension=10}{o,o1} 
\fmf{gorkov_ain,right,tension=0.8,label=$\leftarrow \omega$,l.s=right,l.d=.15w}{o1,i1} 
\fmf{phantom}{o1,i1}
\fmflabel{$\bar{a}$}{i1}
\fmflabel{$b$}{o1}
\end{fmfgraph*} 
}
}

%
%
\newcommand{\diaggoozero}
{
\parbox{50mm}
{
\begin{fmfgraph*}(80,50)
\fmfcmd{%
style_def gorkov_nz expr p =
draw_plain p;
fill (harrow (p, .45));
fill (tarrow (p, .55))
enddef;}
\fmfleft{i1,i2}
\fmfright{o1,o1} 
\fmf{phantom}{o1,i1}
\fmf{phantom}{o2,i2}
\fmf{gorkov_nz,label=$\uparrow \omega$,l.s=right,l.d=.17w}{i1,i2}
\fmflabel{$b$}{i1}
\fmflabel{$a$}{i2}
\end{fmfgraph*} 
}
}

%
%
\newcommand{\diaggoozeronormal}
{
\parbox{50mm}
{
\begin{fmfgraph*}(80,50)
\fmfcmd{%
style_def gorkov_nz expr p =
draw_plain p;
fill (harrow (p, .45));
fill (tarrow (p, .55))
enddef;}
\fmfleft{i1,i2}
\fmfright{o1,o1} 
\fmf{phantom}{o1,i1}
\fmf{phantom}{o2,i2}
\fmf{fermion,label=$\uparrow \omega$,l.s=right,l.d=.17w}{i1,i2}
\fmflabel{$b$}{i1}
\fmflabel{$a$}{i2}
\end{fmfgraph*} 
}
}

%
%
\newcommand{\diaggttzero}
{
\parbox{50mm}
{
\begin{fmfgraph*}(80,50)
\fmfcmd{%
style_def gorkov_nz expr p =
draw_plain p;
fill (harrow (p, .45));
fill (tarrow (p, .55))
enddef;}
\fmfleft{i1,i2} 
\fmfright{o1,o1} 
\fmf{phantom}{o1,i1}
\fmf{phantom}{o2,i2}
\fmf{gorkov_nz,label=$\uparrow \omega$,l.s=left,l.d=.15w}{i2,i1}
\fmflabel{$\bar{b}$}{i1}
\fmflabel{$\bar{a}$}{i2}
\end{fmfgraph*} 
}
}

%
%
\newcommand{\diaggotzero}
{
\parbox{50mm}
{
\begin{fmfgraph*}(80,50)
\fmfcmd{%
style_def gorkov_aoutz expr p =
draw_plain p; 
fill (tarrow (reverse p, .55)); 
fill (tarrow (p, .55))
enddef;}
\fmfleft{i1,i2}
\fmfright{o1,o1} 
\fmf{phantom}{o1,i1}
\fmf{phantom}{o2,i2}
\fmf{gorkov_aoutz,label=$\uparrow \omega$,l.s=right,l.d=.17w}{i1,i2}
\fmflabel{$\bar{b}$}{i1}
\fmflabel{$a$}{i2}
\end{fmfgraph*} 
}
}

%
%
\newcommand{\diaggotzeroold}
{
\parbox{50mm}
{
\begin{fmfgraph*}(80,50)
\fmfcmd{%
style_def gorkov_aoutz expr p =
draw_plain p; 
fill (tarrow (reverse p, .55)); 
fill (tarrow (p, .55))
enddef;}
\fmfleft{i} 
\fmfright{o} 
\fmf{phantom,tension=10}{i,i1} 
\fmf{phantom,tension=10}{o,o1} 
\fmf{gorkov_aoutz,left,tension=0.8,label=$\leftarrow \omega$,l.s=left,l.d=.15w}{o1,i1} 
\fmf{phantom}{o1,i1}
\fmflabel{$a$}{i1}
\fmflabel{$\bar{b}$}{o1}
\end{fmfgraph*} 
}
}

%
%
\newcommand{\diaggtozero}
{
\parbox{50mm}
{
\begin{fmfgraph*}(80,50)
\fmfcmd{%
style_def gorkov_ainz expr p =
draw_plain p;
fill (harrow (reverse p, .45));
fill (harrow (p, .45))
enddef;}
\fmfleft{i1,i2}
\fmfright{o1,o1} 
\fmf{phantom}{o1,i1}
\fmf{phantom}{o2,i2}
\fmf{gorkov_ainz,label=$\uparrow \omega$,l.s=right,l.d=.17w}{i1,i2}
\fmflabel{$b$}{i1}
\fmflabel{$\bar{a}$}{i2}
\end{fmfgraph*} 
}
}

%
%
\newcommand{\diaggtozeroold}
{
\parbox{50mm}
{
\begin{fmfgraph*}(80,50)
\fmfcmd{%
style_def gorkov_ainz expr p =
draw_plain p;
fill (harrow (reverse p, .45));
fill (harrow (p, .45))
enddef;}
\fmfleft{i} 
\fmfright{o} 
\fmf{phantom,tension=10}{i,i1} 
\fmf{phantom,tension=10}{o,o1} 
\fmf{gorkov_ainz,right,tension=0.8,label=$\leftarrow \omega$,l.s=right,l.d=.15w}{o1,i1} 
\fmf{phantom}{o1,i1}
\fmflabel{$\bar{a}$}{i1}
\fmflabel{$b$}{o1}
\end{fmfgraph*} 
}
}

%
%
\newcommand{\diagv}
{
\parbox{50mm}
{
\begin{fmfgraph*}(55,30)
\fmfleft{i} 
\fmfright{o} 
\fmf{dashes,label=$\phantom{aaa}c\phantom{aaaaaassaa}d\phantom{aaa}$,l.s=right}{i,o} 
\fmfv{label=$a$,l.a=90,l.d=.17w}{i}
\fmfv{label=$b$,l.a=90,l.d=.17w}{o}
\fmfdot{i,o}
\end{fmfgraph*} 
}
}

%
%
\newcommand{\diagver}
{
\parbox{50mm}
{
\begin{fmfgraph*}(55,30)
\fmfstraight
\fmfleft{i1,i2,i3} 
\fmfright{o1,o2,o3} 
\fmf{dashes}{i2,o2} 
\fmf{fermion}{i1,i2}
\fmf{fermion}{i2,i3}
\fmf{fermion}{o1,o2}
\fmf{fermion}{o2,o3}
\fmfdot{i2,o2}
\fmflabel{$a$}{i3}
\fmflabel{$b$}{o3}
\fmflabel{$c$}{i1}
\fmflabel{$d$}{o1}
\end{fmfgraph*} 
}
}

%
%
\newcommand{\diagvew}
{
\parbox{50mm}
{
\begin{fmfgraph*}(55,30)
\fmfstraight
\fmfleft{i1,i2,i3} 
\fmfright{o1,o2,o3} 
\fmf{dashes}{i2,o2} 
\fmf{fermion}{i2,i1}
\fmf{fermion}{i2,i3}
\fmf{fermion}{o2,o1}
\fmf{fermion}{o2,o3}
\fmfdot{i2,o2}
\fmflabel{$a$}{i3}
\fmflabel{$b$}{o3}
\fmflabel{$c$}{i1}
\fmflabel{$d$}{o1}
\end{fmfgraph*} 
}
}

%
%
\newcommand{\diaggexa}
{
\parbox{50mm}
{
\begin{fmfgraph*}(55,60)
\fmfcmd{%
style_def gorkov_nz expr p =
draw_plain p;
fill (harrow (reverse p, .45));
fill (tarrow (reverse p, .55))
enddef;}
\fmfstraight
\fmfleft{i1,i2,i3} 
\fmfright{o} 
\fmf{dashes,label=$\phantom{a}c\phantom{aaaasaasaa}e\phantom{aaa}$,l.s=left}{i2,o} 
\fmf{fermion,label=$d$,l.s=left}{i1,i2}
\fmf{fermion}{i2,i3}
\fmfdot{i2,o}
\fmf{gorkov_nz,left,tension=1.2,label=$\downarrow \omega'$,l.s.=left}{o,o} 
\fmflabel{$\uparrow \omega \: \: b$}{i1}
\fmflabel{$\uparrow \omega \: \: a$}{i3}
\fmflabel{$f$}{o}
\end{fmfgraph*} 
}
}

%
%
\newcommand{\diaggexanormal}
{
\parbox{50mm}
{
\begin{fmfgraph*}(55,60)
\fmfcmd{%
style_def gorkov_nz expr p =
draw_plain p;
fill (harrow (reverse p, .45));
fill (tarrow (reverse p, .55))
enddef;}
\fmfstraight
\fmfleft{i1,i2,i3} 
\fmfright{o} 
\fmf{dashes,label=$\phantom{a}c\phantom{aaaaccsaasaa}\phantom{aaa}$,l.s=left}{i2,o} 
\fmf{fermion,label=$d$,l.s=left}{i1,i2}
\fmf{fermion}{i2,i3}
\fmfdot{i2,o}
\fmf{fermion,right,tension=1.2,l.s.=left}{o,o} 
\fmflabel{$\uparrow \omega \: \: b$}{i1}
\fmflabel{$\uparrow \omega \: \: a$}{i3}
\end{fmfgraph*} 
}
}

%
%
\newcommand{\diagconn}
{
\parbox{50mm}
{
\begin{fmfgraph*}(55,100)
\fmfcmd{%
style_def gorkov_n expr p =
draw_plain p;
fill (harrow (p, .45));
fill (tarrow (p, .55))
enddef;}
\fmfcmd{%
style_def gorkov_ain expr p =
draw_plain p;
fill (harrow (reverse p, .45));
fill (harrow (p, .45))
enddef;
style_def gorkov_aout expr p =
draw_plain p; 
fill (tarrow (reverse p, .55)); 
fill (tarrow (p, .55))
enddef;}
\fmfstraight
\fmfleft{i1,i2,i3} 
\fmfright{o1,o2,o3} 
\fmf{dashes,label=$\phantom{aa}\bar{e}\phantom{aasssccdddddiii}f$,l.s=right,l.d=.16w}{o2,i2} 
\fmf{gorkov_ain,l.s=left}{i1,i2}
\fmf{gorkov_ain}{o2,o3}
\fmfdot{i2,o2}
\fmf{gorkov_aout,left,tension=0.5,l.s.=left}{o2,i2} 
\fmflabel{$d$}{o2}
\fmflabel{$\bar{c}$}{i2}
\fmflabel{$\bar{a}$}{o3}
\fmflabel{$b$}{i1}
\end{fmfgraph*} 
}
}

%
%
\newcommand{\diagdisconn}
{
\parbox{50mm}
{
\begin{fmfgraph*}(70,40)
\fmfcmd{%
style_def gorkov_n expr p =
draw_plain p;
fill (harrow (p, .45));
fill (tarrow (p, .55))
enddef;}
\fmfcmd{%
style_def gorkov_ain expr p =
draw_plain p;
fill (harrow (reverse p, .45));
fill (harrow (p, .45))
enddef;
style_def gorkov_aout expr p =
draw_plain p; 
fill (tarrow (reverse p, .55)); 
fill (tarrow (p, .55))
enddef;}
\fmfstraight
\fmfleft{i1,i2,i3} 
\fmfright{o1,o2,o3} 
\fmf{dashes,label=$c\phantom{aasssssiii}\bar{d}$,l.s.=left}{o2,v1} 
\fmf{gorkov_ain,label=$\phantom{aaiii}f$,l.s.=left}{i1,i3}
\fmf{phantom}{v1,i2}
\fmfdot{v1,o2}
\fmf{gorkov_aout,left,tension=0.2,l.s.=left}{o2,v1} 
\fmf{gorkov_ain,left,tension=0.2,l.s.=left}{v1,o2} 
\fmflabel{$\bar{e}$}{o2}
\fmflabel{$b$}{i1}
\fmflabel{$\bar{a}$}{i3}
\end{fmfgraph*} 
}
}

%
%
\newcommand{\diaggexam}
{
\parbox{50mm}
{
\begin{fmfgraph*}(55,60)
\fmfcmd{%
style_def gorkov_nz expr p =
draw_plain p;
fill (harrow (p, .45));
fill (tarrow (p, .55))
enddef;}
\fmfstraight
\fmfright{i1,i2,i3} 
\fmfleft{o} 
\fmf{dashes,label=$\phantom{a}d\phantom{aaaasaa}f\phantom{aaa}$,l.s=left}{i2,o} 
\fmf{fermion}{i1,i2}
\fmf{fermion,label=$e$,l.s=left}{i2,i3}
\fmfdot{i2,o}
\fmf{gorkov_nz,left,tension=1.2,label=$\downarrow \omega'$,l.s.=right}{o,o} 
\fmflabel{$b \uparrow \omega$}{i1}
\fmflabel{$a \uparrow \omega$}{i3}
\fmflabel{$c$}{o}
\end{fmfgraph*} 
}
}

%
%
\newcommand{\diagseexam}
{
\parbox{50mm}
{
\begin{fmfgraph*}(55,60)
\fmfcmd{%
style_def gorkov_nz expr p =
draw_plain p;
fill (harrow (reverse p, .45));
fill (tarrow (reverse p, .55))
enddef;}
\fmfstraight
\fmfleft{i1,i2,i3} 
\fmfright{o} 
\fmf{dashes,label=$\phantom{av}c\phantom{sassvvasaa}e\phantom{aaa}$,l.s=left}{i2,o} 
\fmfdot{i2,o}
\fmf{gorkov_nz,left,tension=1.2,label=$\downarrow \omega'$,l.s.=left}{o,o}
\fmf{phantom,left,tension=1.2,label=$\phantom{v}d\phantom{sasssasaa}f$,l.s=right}{i1,i2}
\end{fmfgraph*} 
}
}

%
%
\newcommand{\diagseexamred}
{
\parbox{50mm}
{
\begin{fmfgraph*}(120,120)
\fmfcmd{%
style_def gorkov_nz expr p =
draw_plain p;
fill (harrow (reverse p, .45));
fill (tarrow (reverse p, .55))
enddef;}
\fmfstraight
\fmfleft{i1,i2,i3} 
\fmfright{o} 
\fmf{dashes,label=$\phantom{aaaaaaaaaaaauaaav}c\phantom{saassaa}i\phantom{aaaaaaaa}e\phantom{savsaa}g\phantom{aaaa}$,l.s=left}{i2,v1} 
\fmf{dashes,label=$d\phantom{svasaa}j\phantom{aaa}\phantom{aaav}f\phantom{svasdaa}h\phantom{aaaaaaaaaaaaaaa}$,l.s=right}{v2,o} 
\fmfdot{i2,o,v1,v2}
\fmf{gorkov_nz,left,tension=1.1,label=$\downarrow \omega'$,l.s.=left}{o,o}
\fmf{gorkov_nz,left,tension=0.8,label=$\leftarrow \omega''$,l.s.=left}{v1,v2}
\fmf{gorkov_nz,left,tension=0.8,label=$\rightarrow \omega'''$,l.s.=left}{v2,v1}
\fmf{phantom,left,tension=1.2}{i1,i2}
\end{fmfgraph*} 
}
}

%
%
\newcommand{\secondnboldold}
{
\parbox{50mm}
{
\begin{fmfgraph*}(80,50)
\fmfcmd{%
style_def gorkov_n expr p =
draw_double p;
fill (harrow (p, .45));
fill (tarrow (p, .55))
enddef;}
\fmfcmd{%
style_def gorkov_ain expr p =
draw_double p;
fill (harrow (reverse p, .45));
fill (harrow (p, .45))
enddef;
style_def gorkov_aout expr p =
draw_double p; 
fill (tarrow (reverse p, .55)); 
fill (tarrow (p, .55))
enddef;}
\fmfleft{i1,i2} 
\fmfright{o1,o2} 
\fmf{gorkov_n,label=$\uparrow \omega'$,l.s=left}{i1,i2} 
\fmf{gorkov_aout,tension=0,label=$\uparrow \omega''$,l.s=left}{v1,v2} 
\fmf{gorkov_ain,label=$\uparrow \omega'''$,l.s=right}{o1,o2} 
\fmf{dashes,label=$\phantom{aaa}c\phantom{aaassaa}f\phantom{aaaaaaa}$,l.s=right}{i2,v2}
\fmf{dashes,label=$\phantom{aaa}d\phantom{aadddddddaassaa}g$,l.s=left}{i1,v1}
\fmf{dashes}{o1,v1}
\fmf{phantom}{o2,v2}
\fmf{double,left=1,tension=1}{v2,o2} 
\fmf{double,left=1,tension=1}{o1,v1} 
\fmflabel{$b$}{i1}
\fmflabel{$\bar{h}$}{o1}
\fmflabel{$a$}{i2}
\fmflabel{$\bar{e}$}{o2}
\fmfdot{o1,v2,i1,i2}
\end{fmfgraph*} 
}
}

\newcommand{\mC}{\mathcal{C}}
\newcommand{\mD}{\mathcal{D}}
\newcommand{\mF}{\mathcal{F}}
\newcommand{\mM}{\mathcal{M}}
\newcommand{\mN}{\mathcal{N}}
\newcommand{\mO}{\mathcal{O}}
\newcommand{\mP}{\mathcal{P}}
\newcommand{\mQ}{\mathcal{Q}}
\newcommand{\mR}{\mathcal{R}}
\newcommand{\mS}{\mathcal{S}}
\newcommand{\mU}{\mathcal{U}}
\newcommand{\mV}{\mathcal{V}}
\newcommand{\mX}{\mathcal{X}}
\newcommand{\mY}{\mathcal{Y}}
\newcommand{\mW}{\mathcal{W}}
\newcommand{\mZ}{\mathcal{Z}}
\newcommand{\mA}{\mathcal{A}}
\newcommand{\mB}{\mathcal{B}}
\newcommand{\nA}{\mathbf{A}}
\newcommand{\nB}{\mathbf{B}}
\newcommand{\nF}{\mathbf{F}}
\newcommand{\nG}{\mathbf{G}}
\newcommand{\nX}{\mathbf{X}}
\newcommand{\nY}{\mathbf{Y}}
\newcommand{\nU}{\mathbf{U}}
\newcommand{\nSigma}{\mathbf{\Sigma}}

\newcommand{\hf}[0]{\textstyle \frac{1}{2} \displaystyle}
\newcommand{\hfb}[0]{ \frac{1}{2}}
\newcommand{\rmd}{{\rm d}}

\section{Introduction}

Over the last decade the reach of ab-initio nuclear structure calculations has extended up to the region of medium-mass systems. Despite the significant progress from both theoretical and computational points of view, methods as coupled-cluster (CC)~\cite{Hagen:2010gd}, in-medium similarity renormalization group (IMSRG)~\cite{Tsukiyama:2010rj} or Dyson self-consistent Green's function~\cite{Barbieri:2009nx} (Dyson-SCGF) are however currently limited to a few tens of doubly-closed shell nuclei. Neighboring nuclei with $\pm$1 or $\pm$2 nucleons can also be reached with particle attachment or removal formalisms~\cite{Dickhoff:2004xx,Jansen:2011cc}. While improving further the convergence of such existing schemes, it is essential to extend their reach and the intrinsic predictive character of ab-initio methods to truly open shell nuclei. One way of doing so involves the development of multi-reference schemes, such as e.g. multi-reference CC~\cite{MRCC} or valence-space shell-model based on microscopic inputs from the ab-initio calculation of a doubly-closed shell core nucleus of reference~\cite{tsukiyama11a,Holt:2010yb}. Alternatively, one may prefer to keep the simplicity of a single-reference method. This requires however, in any of the approaches mentioned above, to formulate the expansion scheme around a vacuum that can tackle Cooper pair instabilities, e.g. to build the correlated state starting from a Bogoliubov vacuum that already incorporates zeroth-order pairing correlations. The objective of the present work is to realize the latter program within the particular frame of SCGF theory.

As alluded to above, SCGF methods are being successfully applied to the study of nuclear systems. Over the last two decades considerable progress has been made in the development of suitable formalisms and computational algorithms both for finite nuclei and infinite nuclear matter \cite{Dickhoff:2004xx}. In infinite systems, bulk and single-particle properties are typically computed through the resummation of particle-particle (pp) and hole-hole (hh) ladder diagrams, i.e. in the self-consistent T-matrix approximation, that tackles short-range correlations induced by the hard-core of conventional nucleon-nucleon (NN) interactions. Results have been obtained at zero and finite temperature for both symmetric and pure neutron matter based on various conventional NN potentials \cite{Soma:2006zx, Rios:2008ef, Rios:2009gb}. Recently, microscopic three-nucleon (NNN) forces have been incorporated \cite{Soma:2008nn, Soma:2009pf}. There have been also attempts to take into account nucleonic superfluidity through the consistent treatment of anomalous propagators \cite{Bozek:1999rv, Bozek:2001nx}.

In finite systems the most advanced SCGF calculations feature the Faddeev random-phase approximation (FRPA) technique, which allows the simultaneous inclusion of pp, hh and ph excitations, together with interferences among them \cite{Barbieri:2000pg, Barbieri:2007Atoms}. By employing a G-matrix resummation of scattering diagrams not included in the chosen model space it is also possible to use interactions with strong repulsive cores \cite{Barbieri:2006sq}. An important characteristic of the FRPA expansion is that it is based on combining one- and many-body propagators, each one representing different experimental processes including nuclear excitations and transfer of one or two nucleons. The method has therefore been applied to a variety of problems including the quenching of spectroscopic factors \cite{Barbieri:2009ej}, anharmonic excitations \cite{Barbieri:2003ExO16}, two-nucleon knockout \cite{Barbieri:2004Shh,Middleton:2006O16pn}, and the derivation of optical potentials \cite{Barbieri:2005NAscatt,Waldecker:2011by}. At the moment, applications can access all doubly-closed shell nuclei up to $^{56}\mbox{Ni}$ together with neighboring systems with $\pm1$ or $\pm2$ nucleons~\cite{Barbieri:2009ej,Barbieri:2009nx}.

In the present work SCGF calculations of finite nuclei are implemented within the Gorkov scheme, allowing for an explicit treatment of nucleonic superfluidity. Suitable numerical techniques are developed in order to perform systematic studies of doubly-magic and semi-magic medium-mass nuclei as will be soon reported on in a forthcoming publication~\cite{Soma11b}, referred to thereafter as Paper II.

One of our goals is to be able to tackle various types of nuclear interactions, in particular chiral potentials based on effective field theory (EFT) \cite{Epelbaum:2008ga} and low-momentum potentials obtained through the further application of renormalization group (RG) techniques \cite{Bogner:2009bt}. There are also yet unanswered fundamental questions as to what microscopic processes are responsible for the superfluid character of open-shell nuclei~\cite{terasaki02,barranco04a,Duguet:2007be,Lesinski:2008cd,Hebeler:2009dy,Duguet:2009gc,hergert09b,Lesinski:2011rn,hergert11a,Idini:2011gm}. This is one among several long-term objectives of the project to provide a fully ab-initio answer to such questions. The present work eventually relates as well to the long-term development of so-called \textit{non-empirical} energy density functionals (EDFs) \cite{Lesinski:2008cd, Duguet:2009gc, Drut:2009ce}. There exist on-going efforts to construct nuclear EDFs starting from underlying nuclear interactions, with the main goal of improving the predictive power away from known data that is rather poor for existing phenomenological EDF parameterizations. The connection with NN and NNN interactions is typically obtained by means of density matrix expansion (DME) techniques~\cite{negele72} and many-body perturbation theory, which allow for the construction of schemes that can be systematically tested and improved order by order in the interaction \cite{Bogner:2008kj, Gebremariam:2009ff, Gebremariam:2010ni,Stoitsov:2010ha,Bogner:2011kp}. In this regard, recent developments and applications of low-momentum potentials \cite{Jurgenson:2007td, Hebeler:2010xb, Jurgenson:2009qs}, which exhibit a more perturbative nature than traditional nuclear interactions, are instrumental. Schemes towards non-empirical EDFs, however, are presently available only in their first stages. Their development necessitates a comparison with fully microscopic methods that can provide useful benchmarks over which EDFs parameterizations can be tested and improved. In this context, SCGF techniques represent a valid ab-initio method of reference. In particular, as the breaking and restoration of symmetries (e.g. translational, rotational, particle number, ...) is central to nuclear EDF methods, it is crucial to develop an approach that includes and exploits the same concept \cite{Duguet:2010jc}, which is the case of Gorkov-Green's function method regarding particle-number symmetry.

The present paper aims at providing a detailed account of Gorkov's formalism that is eventually applied in Paper II. Given that the present work is the first nuclear structure application of {\it ab-initio} Gorkov self-consistent Green's function method, it is relevant to provide a self-contained account of the formalism expressed in a discrete basis, which is suited to finite nuclear systems. The present formalism is further specified to second-order in the self-energy expansion and formulated in terms of NN interactions only. The extension to more advanced truncation schemes and to the inclusion of NNN forces is postponed to future works.

The paper is organized as follows. Section~\ref{sec_H} introduces the general form of the nuclear Hamiltonian employed in the present work while Sec.~\ref{sec_Gorkov} defines generic features of Gorkov's formalism. Section~\ref{sec_Gorkov_equation} discusses Gorkov's equation of motion under the form of an energy-dependent eigenvalue problem before computing normal and anomalous self-energies at second order and rewriting Gorkov's equation under the form of a more convenient energy-independent eigenvalue problem.
Further details regarding the extraction of observable are provided in Sec.~\ref{observables}, while Sec.~\ref{phi_derivable} discusses the conserving character of the employed truncation scheme. Finally, conclusions are given in Sec.~\ref{conclusions} followed by several appendices complementing the body of the paper with relevant technical details.

\section{Nuclear Hamiltonian}
\label{sec_H}

\subsection{Single-particle basis}

Let us first introduce the particular labeling of single-particle states that will be used throughout the work.
We consider a basis $\{a_a^{\dagger} \}$ of the one-body Hilbert space $\mathcal{H}_1$ that can be divided into two blocks according to the value (or more precisely to the sign) of an appropriate symmetry quantum number. To any state $a$ belonging to the first block, one can associate a single-particle state $\tilde{a}$ belonging to the second block and having the same quantum numbers as $a$, except for the one differentiating the two blocks. Typically there exists an anti-unitary transformation $\mathcal{T}$, leaving the Hamiltonian invariant and connecting, up to a phase $\eta_a$, state $a$ with state $\tilde{a}$. With that in mind one defines a basis $\{\bar{a}_a^{\dagger} \}$, partner of the initial one $\{a_a^{\dagger} \}$, through~\footnote{In the following, a {\em tilde} ($\tilde{~}$) always refer to the sole quantum numbers of the opposite block while {\em barred} ($\bar{~}$) quantities involve an additional phase factor $\eta_{a/\tilde{a}}$.}
\begin{equation}
\label{eq:gen_aad}
\bar{a}_{a}^{\dagger}(t) \equiv \eta_a a_{\tilde{a}}^{\dagger}(t)\, , \qquad
\bar{a}_{a}(t) \equiv \eta_a a_{\tilde{a}}(t) \: ,
\end{equation}
which corresponds to exchanging the state $a$ by its partner $\tilde{a}$ up to the phase $\eta_a$. By convention $\tilde{\tilde{a}}=a$ with $\eta_a \, \eta_{\tilde{a}}~=~-1$.

As discussed in Sec. \ref{appli0plus}, $\mathcal{T}$ will eventually be specified as the time-reversal transformation in our applications to even-even nuclei with $J^{\Pi} = 0^+$ ground states.

\subsection{Hamiltonian}
\label{sec:nuclearmbp}

Let us consider a finite system of $N$ fermions interacting via a two-body potential $V^{\text{NN}}$.
The corresponding Hamiltonian can be written as
\begin{eqnarray}
\label{eq:hamiltonian}
H_{\text{tot}} &\equiv& T + V^{\text{NN}}
\nonumber \\ \displaystyle &\equiv& \sum_{ab} T_{ab} \, a_a^{\dagger} a_b + \frac{1}{(2!)^2}\sum_{abcd} \bar{V}_{abcd} \, a_a^{\dagger} a_b^{\dagger} a_d a_c
\end{eqnarray}
where
\begin{equation}
\label{eq:kinme}
T_{ab} \equiv (  a | T |  b )
\end{equation}
is the matrix element of the kinetic energy operator $T$ and
\begin{eqnarray}
\label{eq:vanti2}
\bar{V}_{abcd} &\equiv& \langle ab | V^{\text{NN}} | cd \rangle
\\ \displaystyle &\equiv&
( \mbox{1:}a;\mbox{2:}b | V^{\text{NN}} | \mbox{1:}c;\mbox{2:}d ) -
( \mbox{1:}a;\mbox{2:}b | V^{\text{NN}} | \mbox{1:}d;\mbox{2:}c )\nonumber
\end{eqnarray}
is the antisymmetrized matrix element of $V^{\text{NN}}$ expanded in terms of direct-product states, denoted by $|1,2)$. Whenever a superscript or subscript index $\bar{x}$ appears in a matrix element \eqref{eq:kinme} or \eqref{eq:vanti2}, the associated annihilation (creation) operator is to be intended of the form \eqref{eq:gen_aad}, i.e. $\bar{a}_x$ ($\bar{a}_x^{\dagger}$). It follows that, e.g.,
\begin{subequations}
\label{barmatrixlements}
\begin{eqnarray}
\label{eq:kinme_bar}
T_{\bar{a}\bar{b}} &\equiv& \eta_a \, \eta_b \,(  \tilde{a} | T |  \tilde{b} ) \: ,
\\ \displaystyle
\nonumber \\ \displaystyle
\label{eq:vanti2_bar}
\bar{V}_{\bar{a}b\bar{c}d} &\equiv& \eta_a \, \eta_c \, \langle \tilde{a} b | V^{\text{NN}} |  \tilde{c} d \rangle \:,
\end{eqnarray}
\end{subequations}
etc. As discussed in Appendix \ref{sec:tris}, properties of operator $\mathcal{T}$ leads to the following useful relations
\begin{subequations}
\label{eq:TVbarstar}
\begin{eqnarray}
\label{eq:Tbarstar}
T_{\bar{a}\bar{b}} &=& T_{ab}^* \: ,
\\ \displaystyle
\nonumber \\ \displaystyle
\label{eq:Vbarstar}
\bar{V}_{\bar{a}\bar{b}\bar{c}\bar{d}} &=& \bar{V}_{abcd}^* \:.
\end{eqnarray}
\end{subequations}

\subsection{Centre-of-mass correction}

In the study of a N-body self-bound system, a separation can be made between the motion of its centre-of-mass and the motion of the nucleons relative to it. Specifically, the N-body Hamiltonian \eqref{eq:hamiltonian} can be divided into
\begin{equation}
\label{eq:H_cm-in}
H_{\text{tot}} \equiv H_{\text{cm}} + H_{\text{rel}} \: ,
\end{equation}
where $H_{\text{cm}} $ represents the centre-of-mass kinetic energy and the internal Hamiltonian $H_{\text{rel}}$ does not depend on centre-of-mass coordinates. Eigenfunctions of $H_{\text{tot}}$ can therefore be expressed as products of eigenfunctions of $H_{\text{cm}}$ and eigenfunctions of $H_{\text{rel}}$. Consequently, the energy is the sum of centre-mass energy $E_{\text{cm}}$ and internal energy $E_{\text{rel}}$
\begin{equation}
\label{eq:E_cm-in}
E_{\text{tot}} = E_{\text{cm}} + E_{\text{rel}} \: .
\end{equation}
Nuclei being self-bound objects, one is interested in the translationally invariant, internal Hamiltonian $H_{\text{rel}}$ and the corresponding energy $E_{\text{rel}}$. Subtracting the (known) centre-of-mass kinetic contribution from the total Hamiltonian, one indeed works with the internal Hamiltonian
\begin{equation}
\label{eq:H_in}
H_{\text{rel}} = H_{\text{tot}} - H_{\text{cm}} = T_{\text{rel}} + V^{\text{NN}} \, .
\end{equation}
The internal kinetic energy can be expressed either as a sum of one- and a two-body operators
\begin{equation}
\label{eq:T_in_a}
T_{\text{rel}}^{(a)} = \left( 1- \frac{1}{\hat{N}} \right ) \sum_i \frac{\mathbf{p}_i^2}{2M} - \frac{1}{\hat{N}} \sum_{i<j} \frac{\mathbf{p}_i \cdot \mathbf{p}_j}{M} \: ,
\end{equation}
or as a straight two-body operator
\begin{equation}
\label{eq:T_in_b}
T_{\text{rel}}^{(b)} = \frac{1}{\hat{N}} \sum_{i<j} \frac{(\mathbf{p}_i - \mathbf{p}_j )^2}{2 M} \: .
\end{equation}
Here $\mathbf{p}_i $ represents the momentum of the $i$-th nucleon, $M$ the nucleon mass and $\hat{N}$ is the particle number operator.
In theories that do not conserve particle number, $\hat{N}$ cannot be replaced by its eigenvalue such that expressions \eqref{eq:T_in_a} and \eqref{eq:T_in_b} are not equivalent. Considering a series expansion in $\hat{N}^{-1}$, it could be shown \cite{Hergert:2009na} that form \eqref{eq:T_in_a} displays the correct power counting and should be therefore employed in calculations over Fock space.

In the following, we consider Hamiltonian \eqref{eq:H_in} with choice \eqref{eq:T_in_a} at first order in $\hat{N}^{-1}$, i.e. $\hat{N}^{-1}$ is replaced by its average value $N^{-1}$. For simplicity, and unless otherwise stated, we denote in the following $H_{\text{rel}}$ by $H$ such that $T$ actually embodies the one-body part of $T_{\text{rel}}^{(a)}$ (first term in Eq. \eqref{eq:T_in_a}) and such that $V^{\text{NN}}$ incorporates the two-body part of $T_{\text{rel}}^{(a)}$ (second term in Eq. \eqref{eq:T_in_a}).

\section{Gorkov formalism}
\label{sec_Gorkov}

\subsection{Standard propagator and superfluid systems}

Let us consider the N-body ground-state  $|  \Psi_0^N \rangle$ solution of
\begin{equation}
\label{eq:eigenh}
H \, |  \Psi_k^{N} \rangle = E_k^{N} \, |  \Psi_k^{N} \rangle
\end{equation}
with the lowest eigenvalue $E_0^N$.
The fundamental object of Green's function theory is the one-body propagator defined as
\begin{equation}
\label{eq:normalg}
i \, \mathcal{G}_{ab}^{(N,N)}(t,t') \equiv
\langle \Psi_0^N | T \left \{
a_{a}(t) a_{b}^{\dagger}(t')
\right\}
| \Psi_0^N \rangle \: ,
\end{equation}
where the operator $T$ orders $a$ and $a^{\dagger}$ according to their time argument (larger times to the left) and where annihilation and creation operators are in the Heisenberg representation
\begin{subequations}
\label{eq:HS_reps_standard}
\begin{eqnarray}
a_{a}(t) =& a_{a}^{(H)}(t) &\equiv \exp[i H t] \, a_{a}  \exp[-i H t] \: ,
\\
a_{a}^{\dagger}(t) =& \left[a_{a}^{(H)}(t) \right]^{\dagger} &\equiv \exp[i H t] \, a_{a}^{\dagger} \exp[-i H t] \: .
\end{eqnarray}
\end{subequations}
The knowledge of $\mathcal{G}$ enables the computation of expectation values of all one-body operators plus the two-body ground-state energy, i.e. the expectation value of the Hamiltonian if only two-body forces are considered. One can define two-, three-, ..., $N$-body propagators in a similar way, in order to evaluate up to $N$-nucleon observables.

Green's functions' equations of motion take the form of a set of $N$ coupled integro-differential equations, each of them involving $(i\!-\!1)$-, $i$- and $(i\!+\!1)$-body propagators. In order to compute the one-body propagator, one can as well derive a perturbative expansion that translates into an infinite series of diagrams. Both approaches provide systematic ways of approximating the exact solution. The connection between the diagrammatic expansion and the equation of motion for $\mathcal{G}$ leads to the definition of the (irreducible) self-energy $\tilde{\Sigma}$ and the derivation of Dyson's equation
\begin{equation}
 \mathcal{G}_{ab}^{(N,N)}  \displaystyle
= \mathcal{G}^{(N,N) \, (0)}_{ab}
+ \sum_{cd} \mathcal{G}^{(N,N) \, (0)}_{ac} \, \tilde{\Sigma}_{cd}  \,
\mathcal{G}_{db}^{(N,N)} \: ,
\end{equation}
where $\mathcal{G}^{(0)}$ is the one-body propagator of the unperturbed system associated with a one-body Hamiltonian $H_0$ of choice.

The scheme is in principle exact, i.e. if one can compute the perturbative expansion up to infinite order. Approximations are introduced by including only a certain class or subset of terms in the computation of the self-energy. Such a subset is chosen according to a hierarchy between the various types of diagrams whose rationale depends on the system under consideration. The validity of the standard perturbative expansion, however, is not always guaranteed. In particular, nuclear interactions inducing strong pairing correlations between constituents of the many-body system make the usual expansion inappropriate for the large majority of nuclei. The breakdown of the perturbative expansion is signaled by the appearance of (Cooper) instabilities, which occur when summing up certain classes of diagrams and point to the necessity of developing an alternative diagrammatic method~\cite{mehta1,henley,balian}.

\subsection{Auxiliary many-body problem}

In the presence of Cooper instabilities, one can develop an alternative expansion method accounting in a controlled fashion for the appearance and destruction of condensed nucleonic pairs.

Instead of targeting the actual ground state $|  \Psi_0^{N} \rangle$ of the system, one considers a symmetry breaking state $|  \Psi_0 \rangle$, i.e. a wave packet, defined as a superposition of actual ground states of $(N\!-\!2)$-, $N$-, $(N\!+\!2)$-, ... particle systems, i.e.
\begin{equation}
\label{eq:psi0}
| \Psi_0 \rangle \equiv \sum_{N}^{\mbox{\footnotesize{even}}} c_N \, | \Psi_0^N \rangle \, ,
\end{equation}
where $c_N$ denote unknown complex coefficients. The sum over even particle numbers is said to respect the (even) number-parity quantum number. Together with such a state, one considers the grand-canonical-like potential $\Omega \equiv H - \mu N$, with $\mu$ the chemical potential, in place of $H$ \footnote{Let us remark that the analogy with a grand-canonical ensemble holds only on a formal level, as here eigenstates of \protect$\Omega$ are pure states $| \protect\psi_0^N \protect\rangle$, \protect$| \protect\psi_0^{N \pm 2} \protect\rangle$, ... while the \protect\textit{admixture} or symmetry breaking state \protect$| \protect\Psi_0 \protect\rangle$ is a pure state as well. In other words, one is not introducing a statistical density operator to describe the system.}.
The state $| \Psi_0 \rangle$ is chosen to minimize
\begin{equation}
\label{eq:Omega0}
\Omega_0 =  \langle  \Psi_0 | \Omega |  \Psi_0 \rangle
\end{equation}
under the constraint
\begin{equation}
\label{eq:N0}
N = \langle  \Psi_0 |  N |  \Psi_0 \rangle \, ,
\end{equation}
i.e. it is not an eigenstate of the particle number operator but it has a fixed number of particles on average.
Equation \eqref{eq:Omega0}, together with the normalization condition\begin{equation}
\label{eq:norm_cn}
\langle  \Psi_0 |  \Psi_0 \rangle = \sum_{N}^{\mbox{\footnotesize{even}}} |c_N|^2 = 1 \, ,
\end{equation}
determines the set of coefficients $c_N$, while Eq. \eqref{eq:N0} fixes the chemical potential $\mu$.

By targeting $|  \Psi_0 \rangle$, the initial problem that aimed at describing the many-body system with $N$ nucleons is replaced with an auxiliary problem, whose solution approximates the initial one. The validity of such an approximation resides in the degeneracy characterizing the ground state of the system. The presence of a condensate (ideally) implies that pairs of nucleons can be added or removed from the ground-state of the system with the same energy cost, independently of $N$. Such an hypothesis translates into the fact that the binding energies of the systems with $N, N\!\pm\!2, N\!\pm\!4, ...$ particles differ by $2\mu$; i.e. the idealized situation considered here corresponds to the ansatz that \textit{all} ground states obtained from the system with $N$ nucleons by removing or adding pairs of particles are degenerate eigenstates of $\Omega$ such that their binding energies fulfill
\begin{equation}
\label{eq:2mu}
... \approx E_0^{N+2}-E_0^N \approx E_0^{N}-E_0^{N-2} \approx ... \approx 2 \mu \, ,
\end{equation}
with $\mu$ \textit{independent} of $N$.
If the assumption is valid, the energy obtained by solving the auxiliary many-body problem provides the energy of the initial problem as
\begin{equation}
\label{eq:eneapp}
\Omega_0 = \sum_{N'} |c_{N'}|^2 \Omega_0^{N'}
 \approx E_0^{N} - \mu N \: ,
\end{equation}
which follows from Eqs. \eqref{eq:Omega0}, \eqref{eq:norm_cn} and \eqref{eq:2mu}.

\subsection{Gorkov propagators}
\label{sec:gorkov_prop}

In order to access all one-body information contained in $| \Psi_0 \rangle$, one must generalize the one-body propagator defined in Eq.~\eqref{eq:normalg} by introducing additional objects that account for the formation and destruction of pairs. One thus defines a set of four Green's functions, known as Gorkov propagators \cite{Gorkov:1958}, through
\begin{subequations}
\label{eq:gg}
\begin{equation}
\label{eq:gg11}
i \, G^{11}_{ab}(t,t') \equiv
\langle \Psi_0 | T \left \{
a_{a}(t) a_{b}^{\dagger}(t')
\right\}
| \Psi_0 \rangle \: ,
\end{equation}
\begin{equation}
\label{eq:gg12}
i \, G^{12}_{ab}(t,t') \equiv
\langle \Psi_0  | T \left \{
a_{a}(t) \bar{a}_{b}(t')
\right\}
| \Psi_0 \rangle \: ,
\end{equation}
\begin{equation}
\label{eq:gg21}
i \, G^{21}_{ab}(t,t') \equiv
\langle \Psi_0  | T \left \{
\bar{a}_{a}^{\dagger}(t) a_{b}^{\dagger}(t')
\right\}
| \Psi_0 \rangle \: ,
\end{equation}
\begin{equation}
\label{eq:gg22}
i \, G^{22}_{ab}(t,t') \equiv
\langle \Psi_0  | T \left \{
\bar{a}_{a}^{\dagger}(t) \bar{a}_{b}(t')
\right\}
| \Psi_0 \rangle \: ,
\end{equation}
\end{subequations}
where single-particle operators associated with the partner basis are as defined in Eq. \eqref{eq:gen_aad} and where the modified Heisenberg representation is introduced through
\begin{subequations}
\label{eq:HS_reps}
\begin{eqnarray}
a_{a}(t) =& a_{a}^{(\Omega)}(t) &\equiv \exp[i \Omega t] \, a_{a} \, \exp[-i \Omega t] \: ,
\\
a_{a}^{\dagger}(t) =& \left[a_{a}^{(\Omega)}(t) \right]^{\dagger} &\equiv \exp[i \Omega t] \, a_{a}^{\dagger} \, \exp[-i \Omega t] \: .
\end{eqnarray}
\end{subequations}
Besides the time dependence and quantum numbers $a$ and $b$ identifying single-particle states, Gorkov propagators $G_{ab}^{g_1g_2}$ carry two labels $g_1$ and $g_2$ that span Gorkov's space. When $g_1=1$ $(g_1=2)$ a particle is annihilated in the block of $a$ (created in the block of $\tilde{a}$) and vice versa for $g_2$; i.e. $g_2=1$ $(g_2=2)$ corresponds to a second particle created in the block of $b$ (annihilated in the block of $\tilde{b}$). Green's functions $G^{11}$ and $G^{22}$ are called \textit{normal} propagators while off-diagonal ones, $G^{12}$ and $G^{21}$, are denoted as \textit{anomalous}  propagators.

\subsection{Nambu's matrix formalism}

Gorkov's propagators can be conveniently grouped into a matrix representation, introduced by Nambu \cite{Nambu:1960tm}.
First one defines the two-component vector
\begin{subequations}
\label{eq:A_nambu}
\begin{equation}
\nA_a(t) \equiv
\left(
\begin{tabular}{c}
$a_a(t)$ \\  $\bar{a}_a^{\dagger}(t)$
\end{tabular}
\right) \: ,
\end{equation}
and its self adjoint
\begin{equation}
\nA_a^{\dagger}(t)  =
\left(
\begin{tabular}{cc}
$a_a^{\dagger}(t)$  & $\bar{a}_a(t)$
\end{tabular}
\right) \: ,
\end{equation}
\end{subequations}
denoting generalized annihilation and creation operators. Their components fulfill the anti-commutation relations
\begin{equation}
\label{eq:commAA}
\left \{ A^{g_1}_a(t), A^{g_2 \; \dagger}_b(t) \right \} =
 \, \delta_{g_1 g_2} \,  \delta_{ab} \: ,
\end{equation}
where the extra label labels the rows (columns) of the annihilation (creation) vector operator. One can then write the four propagators \eqref{eq:gg} in the matrix form
\begin{eqnarray}
\label{eq:gnambu}
i \, \nG_{ab}(t,t') &\equiv&
\langle \Psi_0 | T \left \{
\nA_{a}(t) \nA_{b}^{\dagger}(t')
\right\}
| \Psi_0 \rangle
\nonumber \\ \displaystyle
 &=&
i \, \left(
\begin{tabular}{cc}
$G^{11}_{ab}(t,t')$ & $G^{12}_{ab}(t,t')$ \\
& \\
$G^{21}_{ab}(t,t')$ & $G^{22}_{ab}(t,t')$
\end{tabular}
\right) \: ,
\end{eqnarray}
where the time ordering operator acts separately on each element of the Gorkov's matrix $\nA \nA^\dagger$.
In general, any object $O^{g_1g_2}_{ab}$ defined in Gorkov's space can be put into such a matrix form
\begin{equation}
\mathbf{O}_{ab} (t,t') \equiv
\left(
\begin{tabular}{cc}
$O^{11}_{ab}(t,t')$ & $O^{12}_{ab}(t,t')$ \\
& \\
$O^{21}_{ab}(t,t')$ & $O^{22}_{ab}(t,t')$
\end{tabular}
\right) \: ,
\end{equation}
with $g_1$ and $g_2$ labeling respectively the rows and the columns of the matrix.

\subsection{Energy representation}

For most applications it is convenient to transform the propagators from time to energy representation. In systems at equilibrium governed by a time-independent Hamiltonian, one-body Green's functions depend only on the difference of their two time arguments, i.e. $\nG_{ab}(t,t') = \nG_{ab}(t-t')$. Gorkov propagators in the energy domain are thus obtained through the Fourier transformation
\begin{equation}
\label{eq:FT_Gorkov}
\nG_{ab}(\omega) = \int_{-\infty}^{+\infty} d (t-t') \, e^{i \omega(t-t')} \, \nG_{ab}(t-t') \: .
\end{equation}
The energy representation is more suitable to analyzing the physical content of single-particle propagators, as will become clear in the following.

\subsection{Gorkov's equations}

In the standard case, the derivation of the equations of motion and the formulation of a diagrammatic expansion for the one-body propagator lead to defining the irreducible self-energy and Dyson's equation, through which the propagator of the interacting system can actually be computed.
One proceeds similarly in the Gorkov formalism. The first step consists in separating the Hamiltonian into an ``unperturbed'' one-body part and an interacting part. This is conveniently achieved by introducing an auxiliary, one-body Hermitian potential $U$ taking the general form
\begin{equation}
\label{eq:u}
U \equiv
\sum_{ab} \left[
U_{ab} \, a^{\dagger}_a a_b - U_{ab} \, \bar{a}_a \bar{a}_b^{\dagger} + \tilde{U}_{ab} \, a_a^{\dagger} \bar{a}_b^{\dagger} +  \tilde{U}^{\dagger}_{ab} \, \bar{a}_a a_b \right]\: ,
\end{equation}
and by defining
\begin{equation}
\Omega = \underbrace{T +  U - \mu N}_{\displaystyle \equiv \Omega_U} + \underbrace{V^{\text{NN}} - U}_{\displaystyle \equiv \Omega_I} \: .
\end{equation}
The one-body grand potential $\Omega_U$ defines unperturbed Gorkov propagators $\nG^{(0)}$ in energy representation through
\begin{equation}
\label{eq:g0}
\left[ \nG^{(0)}(\omega) \right]^{-1} \equiv \omega - \Omega_U \: .
\end{equation}
The choice of $\nG^{(0)}$ corresponds to selecting an appropriate unperturbed ground state which acts as a reference vacuum for the application of Wick's theorem, and is crucial for the convergence of the perturbative series. In particular, one cannot expand the interacting superfluid ground-state $| \Psi_0 \rangle$ around a non-superfluid unperturbed state, i.e. unperturbed propagators must already contain the basic features characterizing the interacting ones.

The requirement that the unperturbed ground state is superfluid translates into a choice of a $\Omega_U$ that breaks particle number, as is evident from the form of the auxiliary potential \eqref{eq:u}. Applying Wick's theorem in the derivation of the perturbative expansion, anomalous contractions naturally appear, and are afterwards identified with anomalous Gorkov propagators.

Once the unperturbed ground state is defined, one writes down the perturbative series for the interacting propagator $\nG$ and defines normal and anomalous one-line irreducible self-energies. Self-consistency is obtained by computing self-energy diagrams in terms of fully dressed propagators $\nG$ and by only retaining {\it skeleton} diagrams, i.e. diagrams with no self-energy insertions (see App.~\ref{diag_rules}). Working in the energy representation, the four irreducible self-energies read
\begin{equation}
\tilde{\mathbf{\Sigma}}_{ab} (\omega) \equiv
\left(
\begin{tabular}{cc}
$\tilde{\Sigma}^{11}_{ab}(\omega)$ & $\tilde{\Sigma}^{12}_{ab}(\omega)$ \\
& \\
$\tilde{\Sigma}^{21}_{ab}(\omega)$ & $\tilde{\Sigma}^{22}_{ab}(\omega)$
\end{tabular}
\right) \: ,
\end{equation}
and can be divided into a proper part and a contribution coming from the auxiliary potential, i.e.
\begin{equation}
\label{eq:self_tilde}
\tilde{\nSigma}_{ab}(\omega) \equiv \nSigma_{ab} (\omega) - \nU_{ab} \, .
\end{equation}
Eventually, standard Dyson's equation is generalized as set of coupled equations involving the two types of propagators and self-energies. These are known as Gorkov's equations \cite{Gorkov:1958} and read, in Nambu's notation,
\begin{equation}
\label{eq:gorkov}
\mathbf{G}_{ab}(\omega)  \displaystyle
= \mathbf{G}^{(0)}_{ab}(\omega)
+ \sum_{cd} \mathbf{G}^{(0)}_{ac}(\omega) \, \tilde{\mathbf{\Sigma}}_{cd}(\omega) \,
\mathbf{G}_{db}(\omega)  \: . \:\:\:\:\:
\end{equation}
As Dyson's equation in the standard case, Gorkov's equations represent an expansion of interacting or \textit{dressed} one-body normal and anomalous Green's functions in terms of unperturbed ones.
If the method is self-consistent, the final result does not depend on the choice of the auxiliary potential, which disappears from the equations once the propagators are dressed with the corresponding self-energies. From a practical point of view it is useful to track where the auxiliary potential enters and how its cancelation is eventually worked out. This point is addressed in Section \ref{sec:calc_prop}, where the solution of Gorkov's equations is discussed. In particular, and since such a solution is to be found through an iterative procedure, one is still interested in choosing a good auxiliary potential as a starting point.

Let us further remark that, as the auxiliary potential \eqref{eq:u} has a one-body character, i.e. it acts as a mean field, the search for the ground state of $\Omega_U$ corresponds to solving a Bogoliubov-like problem, as becomes evident when writing the unperturbed grand potential in its Nambu's form
\begin{equation}
\label{eq:omega_u}
[\Omega_U]_{ab} =
\left(
\begin{tabular}{cc}
$T_{ab} + U_{ab}-\mu \, \delta_{ab} $ & $\tilde{U}_{ab}$ \\
$\tilde{U}_{ab}^{\dagger}$ & $ -T_{ab} -U_{ab}+\mu \, \delta_{ab}$
\end{tabular}
\right) \: .
\end{equation}
In fact a convenient choice for $\Omega_U$ is constituted by $\Omega_{HFB}$, i.e. one first solves the Hartree-Fock-Bogoliubov (HFB) problem and then uses the resulting propagators $\mathbf{G}^{HFB}_{ab}$ as the unperturbed ones. Notice that the self-energy corresponding to this solution, $\mathbf{\Sigma}^{HFB}$, eventually differs from the first-order self-energy $\mathbf{\Sigma}^{(1)}$ as soon as higher orders are included in the calculation because of the associated self-consistent dressing of the one-body propagators.

\subsection{Lehmann representation}
\label{sec:lehmann}

Let us consider a complete set of normalized eigenstates of $\Omega$ with no definite particle number
\begin{equation}
\label{eq:kapp}
\Omega | \Psi_{k} \rangle = \Omega_{k} | \Psi_{k} \rangle \: ,
\end{equation}
and spanning the Fock space ${\cal F}$. Inserting the corresponding completeness relation, $G^{11}(t,t')$ becomes
\begin{widetext}
\begin{eqnarray}
\label{eq:leh11f}
 G^{11}_{ab}(t,t') &=&   \displaystyle
-i \theta(t-t')  \sum_k \langle \Psi_0 |  a_a | \Psi_k \rangle
\langle \Psi_k |  a_{b}^{\dagger}  | \Psi_0 \rangle \, e^{i[\Omega_0 - \Omega_k](t-t')}
+i \theta(t'-t)  \sum_k
\langle \Psi_0 |  a_{b}^{\dagger}   | \Psi_k \rangle
\langle \Psi_k |  a_a  | \Psi_0 \rangle \, e^{-i[\Omega_0 - \Omega_k](t-t')}  \,  . \:\:\:\:\: \nonumber
\end{eqnarray}
\end{widetext}
Using the integral representation of the theta function and reading out the Fourier transform, one obtains the propagator in energy representation under the form
\begin{eqnarray}
\label{eq:leh11capp}
G^{11}_{ab}(\omega) &=&
\sum_k \frac{\langle \Psi_0 | a_a | \Psi_k \rangle\langle
\Psi_k |  a_{b}^{\dagger} |
\Psi_0 \rangle}{\omega - [\Omega_k - \Omega_0] + i\eta}
\nonumber \\ &+& \displaystyle
\sum_k \frac{\langle \Psi_0 | a_{b}^{\dagger} | \Psi_k \rangle\langle
\Psi_k |  a_a |
\Psi_0 \rangle}{\omega + [\Omega_k - \Omega_0] - i\eta} \: .
\end{eqnarray}
One can proceed similarly for the other three Gorkov-Green's functions and obtain the following set of Lehmann representations
\begin{subequations}
\label{eq:leh}
\begin{equation}
\label{eq:leh11}
G^{11}_{ab} (\omega) =  \sum_{k} \left\{
\frac{\mU_{a}^{k} \,\mU_{b}^{k*}}
{\omega-\omega_{k} + i \eta}
+ \frac{\bar{\mV}_{a}^{k*} \, {\bar{\mV}_{b}^{k}}}{\omega+\omega_{k} - i \eta} \right\} \: ,
\end{equation}
\begin{equation}
\label{eq:leh12}
G^{12}_{ab} (\omega) =   \sum_{k}
\left\{
\frac{\mU_{a}^{k} \,\mV_{b}^{k*}}
{\omega-\omega_{k} + i \eta} + \frac{\bar{\mV}_{a}^{k*} \, {\bar{\mU}_{b}^{k}}}{\omega+\omega_{k} - i \eta}
\right\}  \, ,
\end{equation}
\begin{equation}
\label{eq:leh21}
G^{21}_{ab} (\omega) =  \sum_{k}
\left\{
\frac{\mV_{a}^{k} \,\mU_{b}^{k*}}
{\omega-\omega_{k}  + i \eta}
+ \frac{\bar{\mU}_{a}^{k*} \, {\bar{\mV}_{b}^{k}}}{\omega+\omega_{k} - i \eta}
\right\} \, ,
\end{equation}
\begin{equation}
\label{eq:leh22}
G^{22}_{ab} (\omega) =   \sum_{k} \left\{
\frac{\mV_{a}^{k} \,\mV_{b}^{k*}}
{\omega-\omega_{k} + i \eta}
+  \frac{\bar{\mU}_{a}^{k*} \, {\bar{\mU}_{b}^{k}}}{\omega+\omega_{k} - i \eta} \right\} \: .
\end{equation}
\end{subequations}
with Gorkov's spectroscopic amplitudes defined as
\begin{subequations}
\label{eq:specg}
\begin{eqnarray}
\label{eq:specXg}
{\mU_{a}^{k*}} &\equiv&\langle \Psi_{k} | a_{a}^{\dagger} | \Psi_0 \rangle \, ,
\\
\label{eq:specYg}
{\mV_{a}^{k*}} &\equiv& \langle \Psi_{k} | \bar{a}_{a} | \Psi_0 \rangle \, ,
\end{eqnarray}
\end{subequations}
and
\begin{subequations}
\label{eq:specgb}
\begin{eqnarray}
\label{eq:specXgb}
\bar{\mU}_{a}^{k*} &\equiv&\langle \Psi_{k} | \bar{a}_{a}^{\dagger} | \Psi_0 \rangle \, ,
\\
\label{eq:specYgb}
\bar{\mV}_{a}^{k*} &\equiv& \langle \Psi_{k} | a_{a} | \Psi_0 \rangle \, ,
\end{eqnarray}
\end{subequations}
from which follows that\footnote{Similarly to Eq.~\eqref{barmatrixlements}, we may equivalently write Eqs.~\eqref{eq:Gamp} as ${\bar{\mU}_{a}^{k}} = +\mU_{\bar{a}}^{k}$ and ${\bar{\mV}_{a}^{k}} = - \mV_{\bar{a}}^{k}$.}
\begin{subequations}
\label{eq:Gamp}
\begin{eqnarray}
\label{eq:GampU}
{\bar{\mU}_{a}^{k}} &=& +\eta_a \, \mU_{\tilde{a}}^{k} \: ,
\\
\label{eq:GampV}
{\bar{\mV}_{a}^{k}} &=& - \eta_a \, \mV_{\tilde{a}}^{k} \: .
\end{eqnarray}
\end{subequations}
The poles of the propagators\footnote{As discussed later on, eigensolutions of Gorkov's equations come in pairs $(\omega_k,-\omega_k)$ such that one should only sum on positive solutions in Eq.~\eqref{eq:leh}.} are given by $\omega_{k} \equiv \Omega_k - \Omega_0$. The relation of such poles to separation energies between the N-body ground state and eigenstates of the $N\pm1$ systems is polluted by the breaking of particle number symmetry and is less transparent than for standard Dyson-Green's function. Still, the structure of the propagator naturally suggests to approximate one-nucleon separation energies as
\begin{subequations}
\label{eq:epmk}
\begin{eqnarray}
E_k^{+} \equiv  \mu + \omega_k &=&  \langle  \Psi_k | H |  \Psi_k \rangle - \langle  \Psi_0 | H |  \Psi_0 \rangle  \nonumber \\
&&- \mu \left[\langle  \Psi_k | N |  \Psi_k \rangle - (N+1)\right] \,\,\,, \label{eq:epmk1} \\
E_k^{-}  \equiv \mu - \omega_k &=&  \langle  \Psi_0 | H |  \Psi_0 \rangle  - \langle  \Psi_k | H |  \Psi_k \rangle \nonumber \\
&&+ \mu \left[\langle  \Psi_k | N |  \Psi_k \rangle - (N-1)\right] \,\,\,, \label{eq:epmk2}
\end{eqnarray}
\end{subequations}
where the error associated with the difference between the average number of particles in state $|  \Psi_k \rangle$ and the targeted particle number $N\pm1$ is taken care of by the last term in Eqs.~\eqref{eq:epmk1} and~\eqref{eq:epmk2}.

It is useful to introduce a Nambu representation for the Lehmann form of the propagators by defining the two-component vectors
\begin{subequations}
\label{eq:XYdef}
\begin{eqnarray}
\nX_a^{k \dagger} &\equiv& \langle \Psi_{k} | \nA_{a}^{\dagger} | \Psi_0 \rangle
= \left(
\begin{tabular}{cc}
$\mU_{a}^{k*}$  & $\mV_{a}^{k*}$
\end{tabular}
\right) ,
\\ \displaystyle
\nY^k_a &\equiv& \langle \Psi_{k} | \nA_{a} | \Psi_0 \rangle =
\left(
\begin{tabular}{cc}
$\bar{\mV}_{a}^{k*}$  \\
$\bar{\mU}_{a}^{k*}$
\end{tabular}
\right)
\, ,
\end{eqnarray}
\end{subequations}
where $\nA$ and $\nA^{\dagger}$ have been introduced in Eq. \eqref{eq:A_nambu}, and by writing
\begin{equation}
\label{eq:naleh}
\mathbf{G}_{ab} (\omega) =  \sum_{k} \left\{
\frac{\nX_a^{k} \, \nX_b^{k \dagger}}
{\omega-\omega_{k} + i \eta}
+ \frac{\nY_a^{k} \, \nY_b^{k \dagger}}{\omega+\omega_{k} - i \eta} \right\} \: .
\end{equation}
Note that vectors~(\ref{eq:XYdef}) contain equivalent physics information and are transformed into each other by
\begin{equation}
\nX_a^{k} =
\left( \begin{tabular}{cc}
0 & -1  \\
1 &  0
\end{tabular}  \right)
 {\nY_{\bar{a}}^{k}}^{ *} \: .
\end{equation}

\subsection{Symmetry properties}
\label{sec:symmetries}

The four Gorkov propagators and self-energies are not independent from each other and can be related through certain symmetry operations.
Starting from the definition of Gorkov Green's functions \eqref{eq:gg} and their Fourier transforms \eqref{eq:FT_Gorkov}, one can first prove that
\begin{subequations}
\label{eq:symG}
\begin{eqnarray}
\label{eq:symG11}
G^{22}_{ab} (\omega)  &=& - \, \eta_a \eta_b \, G^{11}_{\tilde{b}\tilde{a}} (-\omega) = - \, G^{11}_{\bar{b}\bar{a}} (-\omega) \: , \\
\label{eq:symG12}
G^{12}_{ab} (\omega)  &=& + \, \eta_a \eta_b \, G^{12}_{\tilde{b}\tilde{a}} (-\omega) = + \, G^{12}_{\bar{b}\bar{a}} (-\omega)  \: , \\
\label{eq:symG21}
G^{21}_{ab} (\omega)  &=& + \, \eta_a \eta_b \, G^{21}_{\tilde{b}\tilde{a}} (-\omega) = + \, G^{21}_{\bar{b}\bar{a}} (-\omega)  \: .
\end{eqnarray}
\end{subequations}
Result \eqref{eq:symG} is easily derived from Lehmann representation \eqref{eq:leh}, together with properties \eqref{eq:Gamp}, as
\begin{subequations}
\begin{eqnarray}
G^{22}_{ab} (\omega) &=&  \sum_{k} \left\{
\frac{\mV_{a}^{k} \,\mV_{b}^{k*}}
{\omega-\omega_{k} + i \eta} +
\frac{\bar{\mU}_{a}^{k*} \, {\bar{\mU}_{b}^{k}}}{\omega+\omega_{k} - i \eta} \right\}
\nonumber \\ \displaystyle
&=&  \sum_{k} \left\{
- \frac{\bar{\mV}_{\bar{a}}^{k} \,\bar{\mV}_{\bar{b}}^{k*}}
{-\omega+\omega_{k} - i \eta}
-  \frac{\mU_{\bar{a}}^{k*} \, {\mU_{\bar{b}}^{k}}}{-\omega-\omega_{k} + i \eta} \right\}\nonumber \\ \displaystyle
&=&  - G^{11}_{\bar{b}\bar{a}} (-\omega) \: ,
\end{eqnarray}
and
\begin{eqnarray}
G^{21}_{ab} (\omega) &=&  \sum_{k} \left\{
\frac{\mV_{a}^{k} \,\mU_{b}^{k*}}
{\omega-\omega_{k} + i \eta} +
\frac{\bar{\mU}_{a}^{k*} \, {\bar{\mV}_{b}^{k}}}{\omega+\omega_{k} - i \eta} \right\}
\nonumber \\ \displaystyle
&=&  \sum_{k} \left\{
- \frac{\bar{\mV}_{\bar{a}}^{k} \,(-\bar{\mU}_{\bar{b}}^{k*})}
{-\omega+\omega_{k} - i \eta}
-  \frac{\mU_{\bar{a}}^{k*} \, (-\mV_{\bar{b}}^{k}) }{-\omega-\omega_{k} + i \eta} \right\}\nonumber \\ \displaystyle
&=&  G^{21}_{\bar{b}\bar{a}} (-\omega)  \: .
\end{eqnarray}
\end{subequations}
By separating the real and imaginary parts of the poles in Eq.~\eqref{eq:naleh} the Gorkov propagator splits into its hermitian
and antihermitian components
\begin{equation}
\mathbf{G}_{ab} (\omega) =
\left( \begin{tabular}{cc}
              $A_{ab}(\omega)$     &            $C_{ab} (\omega)$ \\
$C^\dagger_{ab}(\omega)$     &  $\tilde{A}_{ab} (\omega)$
\end{tabular} \right)
+
i \left( \begin{tabular}{cc}
              $B_{ab}(\omega)$     &            $D_{ab} (\omega)$ \\
$D^\dagger_{ab}(\omega)$     &  $\tilde{B}_{ab} (\omega)$
\end{tabular} \right)
\end{equation}
where $A_{ab} (\omega)$ and $B_{ab} (\omega)$ are hermitian matrices in the one-body Hilbert space ${\cal H}_{1}$. Note that because of the presence of an antihermitian component
$G^{11}_{ab} (\omega) \neq  \left[G^{11}_{ba}(\omega) \right]^*$ and
$G^{21}_{ab} (\omega) \neq \left[G^{12}_{\bar{a}\bar{b}}(\omega) \right]^*$.
From \eqref{eq:symG11} and \eqref{eq:symG12} it follows that
\begin{eqnarray}
\tilde{A}_{ab} (\omega)  &=& - A_{\bar{b}\bar{a}} (-\omega)  \: , \nonumber \\
\tilde{B}_{ab} (\omega)  &=& - B_{\bar{b}\bar{a}} (-\omega)  \: ,  \nonumber \\
C_{ab} (\omega)           &=& + C_{\bar{b}\bar{a}} (-\omega)  \: , \nonumber \\
D_{ab} (\omega)           &=& + D_{\bar{b}\bar{a}} (-\omega)  \: . \nonumber
\end{eqnarray}

Similar symmetry properties are valid for normal and anomalous self-energies. Starting from Gorkov's equation \eqref{eq:gorkov} and making use of relations \eqref{eq:symG} one can prove that the equivalence between Gorkov's equation \eqref{eq:gorkov} and its conjugate \eqref{eq:gorkov_conj}
requires
\begin{subequations}
\label{eq:symS}
\begin{eqnarray}
\label{eq:symS11}
\Sigma^{22}_{ab} (\omega)  &=& - \, \eta_a \eta_b \, \Sigma^{11}_{\tilde{b}\tilde{a}} (-\omega) = - \, \Sigma^{11}_{\bar{b}\bar{a}} (-\omega) \: , \\
\label{eq:symS12}
\Sigma^{12}_{ab} (\omega)  &=& + \, \eta_a \eta_b \, \Sigma^{12}_{\tilde{b}\tilde{a}} (-\omega) = + \, \Sigma^{12}_{\bar{b}\bar{a}} (-\omega)  \: , \\
\label{eq:symS21}
\Sigma^{21}_{ab} (\omega)  &=& + \, \eta_a \eta_b \, \Sigma^{21}_{\tilde{b}\tilde{a}} (-\omega) = + \, \Sigma^{21}_{\bar{b}\bar{a}} (-\omega)  \: .
\end{eqnarray}
\end{subequations}
Such properties are general and should be required from any truncation scheme used to compute self-energies. At first order, they are confirmed by the explicit evaluation of normal and anomalous diagrams in Eq. \eqref{eq:lambda_h}. At second order one can check that they are indeed fulfilled by expressions \eqref{eq:leh_self11} and \eqref{eq:leh_self}.

\subsection{Spectroscopic content of Gorkov propagators}

Let us now discuss quantities that are useful to analyze the spectroscopic content of Gorkov propagators.
First, one defines generalized spectroscopic factors through the 2x2 Nambu matrix
\label{eq:SF}
\begin{eqnarray}
\mathbf{F}_{k} &\equiv& \sum_a \langle \Psi_{0} | \nA_{a} | \Psi_{k} \rangle \, \langle \Psi_{k} | \nA_{a}^{\dagger} | \Psi_0 \rangle \\
&=&  \sum_{a} \nX_{a}^{k} \, \nX_{a}^{k \dagger}
\: , \nonumber
\end{eqnarray}
which is independent of the one-body basis used and whose normal components generalize traditional spectroscopic factors for addition and removal of a nucleon
\begin{subequations}
\label{eq:specfuv}
\begin{eqnarray}
\label{eq:specfU}
\mF_k^{+} \equiv F_k^{11} &=& \sum_a \left| \langle \Psi_k | a_{a}^{\dagger} | \Psi_0 \rangle \right|^2 \\
&=& \sum_a \left|  \mU_{a}^{k} \right|^2 \, , \nonumber
\\
\label{eq:specfV}
\mF_k^{-} \equiv F_k^{22}  &=& \sum_a \left| \langle \Psi_{k} | a_{a} | \Psi_0 \rangle \right|^2 \\
&=& \sum_a \left| \mV_{a}^{k} \right|^2 \, . \nonumber
\end{eqnarray}
\end{subequations}
As states $| \Psi_0 \rangle$ and $| \Psi_k \rangle$ do not carry a definite particle number, such spectroscopic factors do not possess the sharp physical interpretation of the usual ones. Still, and although $\mF_k^+$ ($\mF_k^-$) contains contributions from the addition (removal) of a nucleon to (from) systems characterized by different particle numbers, the dominating contribution remains associated with the addition (removal) to (from) the actual targeted ground-state $| \Psi^N_0 \rangle$.

Next is Gorkov's one-nucleon spectral function $\mathbf{S}(\omega)$ summing one-nucleon {\it addition} $\mathbf{S}^+(\omega)$ and {\it removal} $\mathbf{S}^-(\omega)$ components. Such spectral functions are not only (energy-dependent) 2x2 matrices in Nambu space but also matrices on the one-body Hilbert space ${\cal H}_{1}$. They are extracted from the imaginary part of Gorkov's propagators through
\begin{subequations}
\label{eq:sfunc}
\begin{eqnarray}
\label{eq:sfunc2}
\mathbf{S}_{ab}^+ (\omega) &\equiv&  -\frac{1}{\pi} \, \mbox{Im} \, \mathbf{G}_{ab} (\omega) \nonumber \\
&=&  \sum_{k} \nX_a^{k} \, \nX_b^{k \dagger} \,
\delta(\omega-\omega_{k})
 \quad \hbox{for } \omega > 0  \; , \quad
\end{eqnarray}
\begin{eqnarray}
\label{eq:sfunc1}
\mathbf{S}_{ab}^- (\omega) &\equiv& + \frac{1}{\pi} \, \mbox{Im} \, \mathbf{G}_{ab} (\omega)
\nonumber \\
&=&  \sum_{k} \nY_a^{k} \, \nY_b^{k \dagger} \,
\delta(\omega+\omega_{k})
 \quad \hbox{for }  \omega  < 0 \; , \quad
\end{eqnarray}
\end{subequations}
where only $\omega_k = \pm (E^\pm_k-\mu) \geq 0$ contribute to the sum. Just as for spectroscopic factors, the normal components of Gorkov's spectral functions, e.g.,
\begin{subequations}
\label{eq:sfunc_11}
\begin{eqnarray}
\label{eq:sfunc2_11}
{S}_{ab}^{p} (E) &=& {S}_{ab}^{+ \, 11} (E\!-\!\mu) =  \sum_{k} \mU_{a}^{k} \,\mU_{b}^{k*} \,\, \delta(E-E^+_{k}) \: , \qquad
\end{eqnarray}
\begin{eqnarray}
\label{eq:sfunc1_11}
{S}_{ab}^{h} (E) &=& {S}_{ab}^{- \, 11} (E\!-\!\mu) =  \sum_{k} \bar{\mV}_{a}^{k*} \, {\bar{\mV}_{b}^{k}} \, \, \delta(E-E^-_{k}) \: , \qquad
\end{eqnarray}
\end{subequations}
generalize standard particle and hole spectral functions. The normal one-body density matrix can be extracted by integrating the normal part of the removal spectral function ${S}_{ab}^{- \, 11} $,
\begin{subequations}
\label{eq:allbdm}
\begin{eqnarray}
\label{eq:nobdm}
\rho_{ab} &\equiv& \langle \Psi_0 | a_b^{\dagger} a_a | \Psi_0 \rangle  \\
&=&  \int_{-\infty}^{0} d \omega \, {S}_{ab}^{- \, 11} (\omega) \\
&=& \sum_{k} \bar{\mV}_{b}^k \, \bar{\mV}_{a}^{k*} \: , \nonumber
\end{eqnarray}
whereas the anomalous density matrix is obtained as
\begin{eqnarray}
\label{eq:aobdm}
\tilde{\rho}_{ab} &\equiv& \langle \Psi_0 | \bar{a}_b a_a | \Psi_0 \rangle  \\
&=&  \int_{-\infty}^{0} d \omega \, {S}_{ab}^{- \, 12} (\omega) \\
&=& \sum_{k} {\bar{\mU}_{b}^k} \, \bar{\mV}_{a}^{k*}   \: . \nonumber
\end{eqnarray}
\end{subequations}
Information contained in the spectral function can be characterized by computing its various moments, i.e.
\begin{eqnarray}
\label{eq:sfunc2_mom}
\mathbf{M}^{(n)}_{ab} &\equiv&  \int_{-\infty}^{\infty} d \omega \, \omega^n \, \mathbf{S}_{ab} (\omega) \: ,
\end{eqnarray}
which are (energy-independent) matrices in Nambu space and on ${\cal H}_1$. Making use of anti-commutation relation \eqref{eq:commAA} one can derive a sum rule for the generalized spectral function that directly relates to its zeroth moment\footnote{As discussed in Sec.~\ref{sec_espe}, the first moment $\mathbf{M}^{(1)}$ of Gorkov spectral function gives access to effective single-particle energies~\cite{Duguet:2011a}.}, i.e.
\begin{eqnarray}
\label{eq:sumruleSF}
{M}^{(0) \, g_1 g_2}_{ab}& = & \langle \Psi_0 | \left \{ A^{g_1}_a, A^{g_2 \, \dagger}_b \right \} | \Psi_0 \rangle
 =
\delta_{g_1 g_2} \delta_{ab} \: . \label{M0sumrule}
\end{eqnarray}
The usual sum rule associated with the normal part of the spectral function is recovered from Eq.~\eqref{M0sumrule} as
\begin{equation}
\label{eq:sumruleSF_normal}
\int_{-\infty}^{+\infty} d \omega \left [ {S}_{ab}^{- \, 11}(\omega)  + {S}_{ab}^{+ \, 11}(\omega)  \right ] = \delta_{ab} \, ,
\end{equation}
showing that diagonal matrix elements ${S}_{aa}^{11}(\omega)$ are nothing but probability distribution functions associated with the probability to remove/add a nucleon from/to the ground state from/on a given single-particle state $a$ and leave the residual system with a missing energy $\omega$. Equation~\eqref{eq:sumruleSF_normal} simply states that such a probability integrate to 1 when scanning missing energies from $-\infty$ to $+\infty$. A new sum rule associated with anomalous spectral functions can also be deduced from Eq.~\eqref{M0sumrule} as
\begin{equation}
\label{eq:sumruleSF_anomalous}
\int_{-\infty}^{+\infty} d \omega \left [ {S}_{ab}^{- \, 12}(\omega)  + {S}_{ab}^{+ \, 12}(\omega)  \right ] = 0 \: .
\end{equation}

Last but not least, one introduces the spectral strength distribution (SSD) through ${\cal \mathbf{Sp}} (E) \equiv \text{Tr}_{{\cal H}_1} \left[\mathbf{S} (E\!-\!\mu)\right] $, which reads as
\begin{eqnarray}
\label{eq:SSD}
{\cal \mathbf{Sp}} (E) &=& \sum_k \mathbf{F}_{k}^+ \, \delta(E-E^+_k)  + \mathbf{F}_{k}^- \, \delta(E-E^-_k) \: .
\end{eqnarray}
The SSD is a 2x2 matrix of energy-dependent functions and is independent of the single-particle basis used to compute it. Its normal part ${\cal S}p (E) \equiv \text{Sp}^{11}(E)$ reads as
\begin{eqnarray}
\label{eq:SSDan}
{\cal S}p (E) &=& \sum_k  \mF_k^{+} \, \delta(E-E^+_k) + \mF_k^{-} \, \delta(E-E^-_k) \: ,
\end{eqnarray}
and provides the probability to leave the system  with relative energy $E$ by adding/removing a nucleon to/from ground state $| \Psi_0 \rangle$.

\section{Gorkov's equations}
\label{sec_Gorkov_equation}

We now proceed to a form of Gorkov's equations allowing for a direct numerical implementation.

\subsection{Energy-dependent eigenvalue problem}
\label{sec:calc_prop}

Let us first transform Eq.~\eqref{eq:gorkov} into an eigenvalue equation for amplitudes $\mU^k$ and $\mV^k$, along with a normalization condition for those amplitudes. Multiplying Gorkov's equation \eqref{eq:gorkov} by $(\omega - \omega_k)$, the pole at $\omega=+\omega_k$ is extracted by taking the limit $\omega \rightarrow \omega_k$, such that substituting Lehmann representation \eqref{eq:naleh} for $\nG$ and operator form~\eqref{eq:g0} for $\nG^{(0)}$, one obtains
\begin{equation}
\left.
\nX_{a}^{k} \, {\nX_{b}^{k \dagger}} = \sum_{cd} \left( \omega - \Omega_U \right)^{-1}_{ac} \tilde{\nSigma}_{cd}(\omega) \, \nX_{d}^{k} \, {\nX_{b}^{k \dagger}}
\right|_{+\omega_k}
 \: . \nonumber
\end{equation}
Multiplying both sides by $\left( \omega - \Omega_U \right)_{ea}$ and summing over $a$ yields
\begin{equation}
\label{eq:omu_sigma}
\left.
\sum_{a} \left( \omega - \Omega_U \right)_{ea} \, \nX_{a}^{k}
\right|_{+\omega_k}
 =
\left.
\sum_{d}  \tilde{\nSigma}_{ed}(\omega) \, \nX_{d}^{k}
\right|_{+\omega_k}
\: , \nonumber
\end{equation}
such that Eqs.~\eqref{eq:self_tilde} and \eqref{eq:omega_u} finally allows writing the matrix eigenvalue equation
\begin{widetext}
\begin{equation}
\label{eq:eigen_uv_bar}
\sum_b
\left.
\left(
\begin{tabular}{cc}
$T_{ab} - \mu \, \delta_{ab} + \Sigma^{11}_{ab}(\omega)$ & $\Sigma^{12}_{ab}(\omega)$ \\
$\Sigma^{21}_{ab}(\omega)$ & $-T_{ab} + \mu \, \delta_{ab} + \Sigma^{22}_{ab}(\omega)$
\end{tabular}
\right)
\right|_{+\omega_k}
\left(
\begin{tabular}{c}
${\mU_{b}^{k}}$  \\
${\mV_{b}^{k}}$
\end{tabular}
\right)=+\omega_k \,
\left(
\begin{tabular}{c}
${\mU_{a}^{k}}$  \\
${\mV_{a}^{k}}$
\end{tabular}
\right) \: .
\end{equation}
whose solutions are amplitudes  $(\mU^k, \mV^k)$ and associated pole energy $\omega_k$. Equivalently, computing the residue at  $\omega = - \omega_k$ leads to
\begin{equation}
\label{eq:eigen_uv_star}
\sum_b
\left.
\left(
\begin{tabular}{cc}
$T_{ab} - \mu \, \delta_{ab} + \Sigma^{11}_{ab}(\omega)$ & $\Sigma^{12}_{ab}(\omega)$ \\
$\Sigma^{21}_{ab}(\omega)$ & $-T_{ab} + \mu \, \delta_{ab} + \Sigma^{22}_{ab}(\omega)$
\end{tabular}
\right)
\right|_{-\omega_k}
\left(
\begin{tabular}{c}
$\bar{\mV}_{b}^{k*}$  \\
$\bar{\mU}_{b}^{k*}$
\end{tabular}
\right) = -\omega_k \,
\left(
\begin{tabular}{c}
$\bar{\mV}_{a}^{k*}$  \\
$\bar{\mU}_{a}^{k*}$
\end{tabular}
\right) \: .
\end{equation}
\end{widetext}
Notice that the latter relationship can be also obtained from the conjugate of Eq. \eqref{eq:eigen_uv_bar} by using properties of Gorkov amplitudes and self-energies. Equations \eqref{eq:eigen_uv_bar} or \eqref{eq:eigen_uv_star} and their solutions are independent of auxiliary potential $U$, which canceled out. This leaves proper self-energy contributions only, which eventually act as energy-dependent potentials. The self-energies depend in turn on amplitudes $\mU^k$ and $\mV^k$ such that Eqs.~\eqref{eq:eigen_uv_bar} or~\eqref{eq:eigen_uv_star} must be solved iteratively. At each iteration the chemical potential $\mu$ must be fixed such that Eq. \eqref{eq:N0} is fulfilled, which translates into the necessity for amplitude $\mV$ to satisfy
\begin{equation}
\label{eq:NV}
N = \sum_a \, \rho_{aa} = \sum_{a,k} \left| {\mV_{a}^k} \right|^2 \: ,
\end{equation}
where $\rho_{ab}$
is the (normal) one-body density matrix \eqref{eq:nobdm}.

As demonstrated in App.~\ref{app_normalization}, the spectroscopic amplitudes solution of Eq.~\eqref{eq:eigen_uv_bar} or~\eqref{eq:eigen_uv_star} fulfill normalization conditions
\begin{subequations}
\begin{eqnarray}
\label{eq:norm_x}
\sum_a  \left| \nX_{a}^{k} \right|^2
&=& 1 + \sum_{ab} \, \nX_{a}^{k \dagger} \,
 \left. \frac{\partial  \nSigma_{ab}(\omega)}{\partial \omega}\right|_{+\omega_k}  {\nX_{b}^{k}} \: ,
\\
\label{eq:norm_y}
\sum_a  \left| \nY_{a}^{k} \right|^2
&=& 1 + \sum_{ab} \, {\nY_{a}^{k \dagger}} \,
 \left. \frac{\partial  \nSigma_{ab}(\omega)}{\partial \omega}\right|_{-\omega_k} \nY_{b}^{k} \: ,
\end{eqnarray}
\end{subequations}
where only the proper self-energy appears because of the energy independence of the auxiliary potential.

\subsection{First-order self-energies}

\begin{figure}[h]
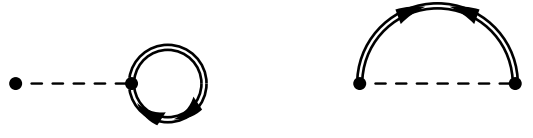

\begin{tabular}{cc}
& \\
\hspace{-.5cm} \nlfirstn &
\hspace{-1cm} \nlfirstaa \\
\end{tabular}
\vspace{-.2cm}
\caption{First-order normal $\Sigma^{11 \, (1)}$ (left) and anomalous $\Sigma^{21 \, (1)}$ (right) self-energy diagrams. Double lines denote self-consistent normal (two arrows in the same direction) and anomalous (two arrows in opposite directions) propagators while dashed lines embody antisymmetrized matrix elements of the NN interaction.
}
\label{fig:first}
\end{figure}

In Fig. \ref{fig:first}, first-order diagrams contributing to normal and anomalous self-energies are displayed. Diagrammatic rules appropriate to the computation of Gorkov's propagators and for the evaluation of self-energy diagrams are discussed in App. \ref{sec:diagrules}, while the $\Phi$-derivability of the presently used truncation scheme is addressed in Sec.~\ref{phi_derivable}.

The four first-order self-energies diagrams are computed in Eqs. \eqref{eq:self111},  \eqref{eq:self221}, \eqref{eq:self121} and \eqref{eq:self211}, and read
\begin{subequations}
\label{eq:lambda_h}
\begin{eqnarray}
\Sigma^{11 \, (1)}_{ab} &=& +\sum_{cd} \bar{V}_{acbd} \, \rho_{dc}
\equiv +\Lambda_{ab}
= + \Lambda_{ab}^{\dagger}
 \\  \displaystyle
\Sigma^{22 \, (1)}_{ab} &=&  -\sum_{cd} \bar{V}_{\bar{b}d\bar{a}c} \, \rho_{cd}^*
= - \Lambda_{\bar{a}\bar{b}}^*  \, ,
 \\  \displaystyle
\label{eq:h_tilde}
\Sigma^{12 \, (1)}_{ab} &=&   \frac{1}{2}  \sum_{cd} \bar{V}_{a\bar{b}c\bar{d}} \, \tilde{\rho}_{cd}
\equiv + \tilde{h}_{ab} \: ,
 \\  \displaystyle
\label{eq:h_tilde_dagger}
\Sigma^{21 \, (1)}_{ab} &=& \frac{1}{2}  \sum_{cd} \bar{V}_{b\bar{a}c\bar{d}}^* \,  \tilde{\rho}_{cd}^* = + \tilde{h}_{ab}^{\dagger}
\: ,
\end{eqnarray}
\end{subequations}
where the normal ($\rho_{ab}$) and anomalous ($\tilde{\rho}_{ab}$) density matrices have been defined in Eqs. \eqref{eq:allbdm}.

\subsection{HFB limit}
\label{sec_HFB_limit}

Neglecting higher-order contributions to the self-energy, Eqs. \eqref{eq:eigen_uv_bar} and \eqref{eq:lambda_h} combine to give
\begin{widetext}
\begin{equation}
\label{eq:eigen_hfb_ns}
\sum_b
\left(
\begin{tabular}{cc}
$T_{ab} + \Lambda_{ab} - \mu \, \delta_{ab} $ & $\tilde{h}_{ab}$ \\
$\tilde{h}_{ab}^{\dagger}$ & $-T_{\bar{a}\bar{b}}^*  - \Lambda_{\bar{a}\bar{b}}^* + \mu \, \delta_{\bar{a}\bar{b}}$
\end{tabular}
\right)
\left(
\begin{tabular}{c}
$\mU_{b}^{k}$  \\
$\mV_{b}^{k}$
\end{tabular}
\right)
= \omega_k \,
\left(
\begin{tabular}{c}
$\mU_{a}^{k}$  \\
$\mV_{a}^{k}$
\end{tabular}
\right)
\: ,
\end{equation}
\end{widetext}
which is nothing but the HFB eigenvalue problem in the case where time-reversal invariance is not assumed. In such a limit, $\mU^k$ and $\mV^k$ define the unitary Bogoliubov transformation~\cite{ring80a} according to
\begin{subequations}
\label{bogo1}
\begin{eqnarray}
\label{eq:bogo_trans}
a_a &=& \sum_{k} \mU_a^k \, \beta_k + \bar{\mV}_a^{k*} \, \beta_k^{\dagger}  \, , \\
a_a^{\dagger} &=& \sum_{k} \mU_a^{k*} \, \beta_k^{\dagger} + \bar{\mV}_a^{k} \, \beta_k  \,  .
\end{eqnarray}
\end{subequations}
Moreover, normalization condition \eqref{eq:norm_y} reduces in this case to the well-known HFB identity
\begin{equation}
\label{eq:norm_hfb}
\sum_a  \left| \nY_{a}^{k} \right|^2 = \sum_a  \left| \mU_{a}^{k} \right|^2 + \sum_a  \left| \mV_{a}^{k} \right|^2
= 1  \: .
\end{equation}
Let us now stress that, despite the energy independence of first-order self energies, some fragmentation of the single-particle strength is already accounted for at the HFB level such that one deals with quasi-particle degrees of freedom. In particular one can deduce from Eq. \eqref{eq:norm_hfb} that (generalized) spectroscopic factors defined in Eq. \eqref{eq:specfuv} are already smaller than one. Such a fragmentation is an established consequence of static pairing correlations that are explicitly treated at the HFB level through particle number symmetry breaking.

Finally, let us underline again that, whenever higher orders are to be included in the calculation, first-order self-energies~(\ref{eq:lambda_h}) are self-consistently modified (in particular through the further fragmentation of the quasi-particle strength) such that they do not correspond anymore to standard Hartree-Fock and Bogoliubov potentials, in spite of their energy independence. They actually correspond to the energy-independent part of the (dynamically) {\it correlated} self-energy.

\subsection{Second-order self-energies}

Let us now discuss second-order contributions to normal and anomalous (irreducible) self-energies.
\begin{figure}[h]
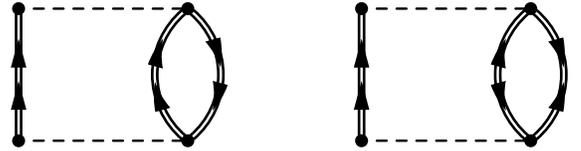

\begin{tabular}{cc}
& \\
\hspace{-.8cm} \nlsecondna &
\hspace{-.8cm} \nlsecondnb \\
& \\
\end{tabular}
\caption{Second-order normal self-energies $\Sigma^{11 \, (2')}$ (left) and $\Sigma^{11 \, (2'')}$ (right).
See Fig. \ref{fig:first} for conventions.}
\label{fig:secondn}
\end{figure}
\begin{figure}[h]
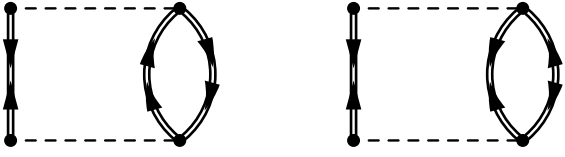

\begin{tabular}{cc}
& \\
\hspace{-.8cm} \nlsecondaaanew &
\hspace{-.8cm} \nlsecondabbnew \\
& \\
\end{tabular}
\caption{Second-order anomalous self-energies $\Sigma^{21 \, (2')}$ (left) and $\Sigma^{21 \, (2'')}$ (right).
See Fig. \ref{fig:first} for conventions.}
\label{fig:seconda}
\end{figure}
In Figs. \ref{fig:secondn} and \ref{fig:seconda} the four types of normal and anomalous self-energies are depicted. The evaluation of all second-order diagrams is performed in App. \ref{sec:diagrules}. Before addressing their expressions, let us introduce useful quantities
\begin{subequations}
\label{eq:mpr}
\begin{eqnarray}
\mM^{k_1k_2k_3}_{a} &\equiv& \sum_{ijk} \bar{V}_{akij} \,\mU_{i}^{k_1} \mU_{j}^{k_2} \bar{\mV}_{k}^{k_3} \, ,
\\ \displaystyle
\mP^{k_1k_2k_3}_{a} &\equiv& \sum_{ijk} \bar{V}_{a\bar{k}i\bar{j}} \, \mU_{i}^{k_1} \mV_{k}^{k_2} \bar{\mU}_{j}^{k_3}
= \mM^{k_1k_3k_2}_{a} \, , \,\,\,\,
\\ \displaystyle
\mR^{k_1k_2k_3}_{a} &\equiv& \sum_{ijk} \bar{V}_{a\bar{k}\bar{i}j} \,  \mV_{k}^{k_1} \mU_{j}^{k_2} \bar{\mU}_{i}^{k_3}
= \mM^{k_3k_2k_1}_{a} \,,\,\,\,\,
\end{eqnarray}
\end{subequations}
and
\begin{subequations}
\label{eq:mpr_bis}
\begin{eqnarray}
\mN^{k_1k_2k_3}_{a} &\equiv& \sum_{ijk} \bar{V}_{akij} \, \mV_{i}^{k_1} \mV_{j}^{k_2} \bar{\mU}_{k}^{k_3}  \, ,
\\ \displaystyle
\mQ^{k_1k_2k_3}_{a} &\equiv& \sum_{ijk} \bar{V}_{a\bar{k}i\bar{j}} \,\mV_{i}^{k_1} \mU_{k}^{k_2} \bar{\mV}_{j}^{k_3}
= \mN^{k_1k_3k_2}_{a} \, , \,\,\,\,
\\ \displaystyle
\mS^{k_1k_2k_3}_{a} &\equiv& \sum_{ijk} \bar{V}_{a\bar{k}\bar{i}j} \,  \mU_{k}^{k_1} \mV_{j}^{k_2} \bar{\mV}_{i}^{k_3}
= \mN^{k_3k_2k_1}_{a} \,,\,\,\,\,
\end{eqnarray}
\end{subequations}
in terms of which second-order self-energies are expressed below.
Using relations \eqref{eq:Gamp} one shows that
\begin{subequations}
\label{eq:mpn_bar}
\begin{eqnarray}
\bar{\mM}^{k_1k_2k_3}_{a} &=&  \eta_a \, {\mM}^{k_1k_2k_3}_{\tilde{a}} \, ,
\\ \displaystyle
\bar{\mP}^{k_1k_2k_3}_{a} &=&  \eta_a \, {\mP}^{k_1k_2k_3}_{\tilde{a}} \, ,
\\ \displaystyle
\bar{\mR}^{k_1k_2k_3}_{a} &=&  \eta_a \, {\mR}^{k_1k_2k_3}_{\tilde{a}} \, ,
\end{eqnarray}
\end{subequations}
and
\begin{subequations}
\label{eq:nqs_bar}
\begin{eqnarray}
\label{eq:n_bar}
\bar{\mN}^{k_1k_2k_3}_{a} &=&  -\eta_a \, {\mN}^{k_1k_2k_3}_{\tilde{a}} \, ,
\\ \displaystyle
\bar{\mQ}^{k_1k_2k_3}_{a} &=&  -\eta_a \, {\mQ}^{k_1k_2k_3}_{\tilde{a}} \, ,
\\ \displaystyle
\bar{\mS}^{k_1k_2k_3}_{a} &=&  -\eta_a \, {\mS}^{k_1k_2k_3}_{\tilde{a}} \, .
\end{eqnarray}
\end{subequations}
Given that $\mP$ and $\mR$ can be obtained from $\mM$ through odd permutations of indices $\{ k_1, k_2, k_3 \}$ and taking into account the symmetries of interaction matrix elements, one can prove that such quantities display the properties
\begin{subequations}
\label{eq:pro_m}
\begin{eqnarray}
\label{eq:pro_m1}
\sum_{k_1 k_2 k_3}  {\mM^{k_1k_2k_3}_{a}} \, \, {\mM^{k_1k_2k_3}_{b}}^{*}
&=& + \sum_{k_1 k_2 k_3}  {\mP^{k_1k_2k_3}_{a}} \, \, {\mP^{k_1k_2k_3}_{b}}^{*} \nonumber \\
&=& + \sum_{k_1 k_2 k_3}  {\mR^{k_1k_2k_3}_{a}} \, \, {\mR^{k_1k_2k_3}_{b}}^{*} \: ,  \nonumber \\
\end{eqnarray}
and
\begin{eqnarray}
\label{eq:pro_m2}
\sum_{k_1 k_2 k_3}  {\mM^{k_1k_2k_3}_{a}} \, \, {\mP^{k_1k_2k_3}_{b}}^{*}
&=& + \sum_{k_1 k_2 k_3}  {\mM^{k_1k_2k_3}_{a}} \, \, {\mR^{k_1k_2k_3}_{b}}^{*} \nonumber \\
&=& + \sum_{k_1 k_2 k_3}  {\mP^{k_1k_2k_3}_{a}} \, \, {\mM^{k_1k_2k_3}_{b}}^{*} \nonumber \\
&=& - \sum_{k_1 k_2 k_3}  {\mP^{k_1k_2k_3}_{a}} \, \, {\mR^{k_1k_2k_3}_{b}}^{*} \nonumber \\
&=& + \sum_{k_1 k_2 k_3}  {\mR^{k_1k_2k_3}_{a}} \, \, {\mM^{k_1k_2k_3}_{b}}^{*} \nonumber \\
&=& - \sum_{k_1 k_2 k_3}  {\mR^{k_1k_2k_3}_{a}} \, \, {\mP^{k_1k_2k_3}_{b}}^{*}  \: . \nonumber \\
\end{eqnarray}
\end{subequations}
Similarly, for $\mN$, $\mQ$ and $\mS$ one has
\begin{subequations}
\label{eq:pro_n}
\begin{eqnarray}
\label{eq:pro_n1}
\sum_{k_1 k_2 k_3}  {\mN^{k_1k_2k_3}_{a}}^{*} \, \, \mN^{k_1k_2k_3}_{b}
&=& + \sum_{k_1 k_2 k_3}  {\mQ^{k_1k_2k_3}_{a}}^{*} \, \, \mQ^{k_1k_2k_3}_{b} \nonumber \\
&=& + \sum_{k_1 k_2 k_3}  {\mS^{k_1k_2k_3}_{a}}^{*} \, \, \mS^{k_1k_2k_3}_{b} \: , \nonumber \\
\end{eqnarray}
and
\begin{eqnarray}
\label{eq:pro_n2}
\sum_{k_1 k_2 k_3}  {\mN^{k_1k_2k_3}_{a}}^{*} \, \, \mQ^{k_1k_2k_3}_{b}
&=& + \sum_{k_1 k_2 k_3}  {\mN^{k_1k_2k_3}_{a}}^{*} \, \, \mS^{k_1k_2k_3}_{b}  \nonumber \\
&=& + \sum_{k_1 k_2 k_3}  {\mQ^{k_1k_2k_3}_{a}}^{*} \, \, \mN^{k_1k_2k_3}_{b} \nonumber \\
&=& - \sum_{k_1 k_2 k_3}  {\mQ^{k_1k_2k_3}_{a}}^{*} \, \, \mS^{k_1k_2k_3}_{b} \nonumber \\
&=& + \sum_{k_1 k_2 k_3}  {\mS^{k_1k_2k_3}_{a}}^{*} \, \, \mN^{k_1k_2k_3}_{b} \nonumber \\
&=& - \sum_{k_1 k_2 k_3}  {\mS^{k_1k_2k_3}_{a}}^{*} \, \, \mQ^{k_1k_2k_3}_{b}  \: . \nonumber \\
\end{eqnarray}
\end{subequations}
Analogous properties can be derived for terms mixing $\{ \mM, \mP, \mR \}$ and $\{ \mN, \mQ, \mS \}$.

Let us now consider $\Sigma^{11}$, whose second-order contributions, evaluated in Eqs. \eqref{eq:self112a} and \eqref{eq:self112b}, can be written as
\begin{widetext}
\begin{eqnarray}
\label{eq:selftwoa}
\Sigma_{ab}^{11 \, (2')} (\omega) &=& \frac{1}{2}
\sum_{k_1k_2k_3}  \left\{
\frac{{\mM^{k_1k_2k_3}_{a}} \, ({\mM^{k_1k_2k_3}_{b}})^{*}}{\omega-E_{k_1 k_2 k_3} + i \eta}
+  \frac{(\bar{\mN}^{k_1k_2k_3}_{a})^{*} \, \bar{\mN}^{k_1k_2k_3}_{b}}{\omega+E_{k_1 k_2 k_3} - i \eta}
\right\} \\
\label{eq:selftwob}
\Sigma_{ab}^{11 \, (2'')} (\omega) &=&
- \sum_{k_1k_2k_3}  \left\{
\frac{{\mM^{k_1k_2k_3}_{a}} \, ({\mP^{k_1k_2k_3}_{b}})^{*}}{\omega-E_{k_1 k_2 k_3} + i \eta}
+  \frac{(\bar{\mN}^{k_1k_2k_3}_{a})^{*} \, \bar{\mQ}^{k_1k_2k_3}_{b}}{\omega+E_{k_1 k_2 k_3} - i \eta}
\right\} \, ,
\end{eqnarray}
where the notation $E_{k_1 k_2 k_3} \equiv \omega_{k_1} + \omega_{k_2} + \omega_{k_3}$ has been introduced.
Summing the two terms and using properties \eqref{eq:pro_m} and \eqref{eq:pro_n} one obtains
\begin{eqnarray}
\label{eq:leh_self11}
\Sigma_{ab}^{11 \, (2'+2'')} (\omega) &=&
\sum_{k_1k_2k_3}  \left\{
\frac{{\mC^{k_1k_2k_3}_{a}}
\, ({\mC^{k_1k_2k_3}_{b}})^{*}}{\omega-E_{k_1 k_2 k_3} + i \eta}
+ \frac{({\bar{\mD}^{k_1k_2k_3}_{a}})^{*}
\, \bar{\mD}^{k_1k_2k_3}_{b}}{\omega+E_{k_1 k_2 k_3} - i \eta}
\right\} \: ,
\end{eqnarray}
where
\begin{subequations}
\begin{eqnarray}
\mC^{k_1k_2k_3}_{a} &\equiv& \frac{1}{\sqrt{6}} \left [ {\mM^{k_1k_2k_3}_{a}} - {\mP^{k_1k_2k_3}_{a}} - {\mR^{k_1k_2k_3}_{a}} \right ] \: ,
\\
\mD^{k_1k_2k_3}_{a} &\equiv& \frac{1}{\sqrt{6}} \left [ {\mN^{k_1k_2k_3}_{a}} - {\mQ^{k_1k_2k_3}_{a}} - {\mS^{k_1k_2k_3}_{a}} \right ] \: .
\end{eqnarray}
\end{subequations}
Notice that from Eqs. \eqref{eq:mpn_bar} and \eqref{eq:nqs_bar} follow $\bar{\mC}^{k_1k_2k_3}_{a} = +\eta_a \, \mC^{k_1k_2k_3}_{\tilde{a}}$ and
$\bar{\mD}^{k_1k_2k_3}_{a} = -\eta_a \, \mD^{k_1k_2k_3}_{\tilde{a}}$.
All other second-order self-energies computed in Section \ref{sec:self_eva} can be written similarly according to
\begin{subequations}
\label{eq:leh_self}
\begin{eqnarray}
\label{eq:leh_self12}
\Sigma_{ab}^{12 \, (2'+2'')} (\omega) &=&
\sum_{k_1k_2k_3}  \left\{
\frac{{\mC^{k_1k_2k_3}_{a}}
\, ( {\mD}^{k_1k_2k_3}_{b}  )^* }{\omega-E_{k_1 k_2 k_3} + i \eta}
+ \frac{({\bar{\mD}^{k_1k_2k_3}_{a}})^{*}
\, {\bar{\mC}^{k_1k_2k_3}_{b}}}{\omega+E_{k_1 k_2 k_3} - i \eta}
\right\} \: ,
\\ \displaystyle
\label{eq:leh_self21}
\Sigma_{ab}^{21 \, (2'+2'')} (\omega) &=&
\sum_{k_1k_2k_3}  \left\{
\frac{{\mD^{k_1k_2k_3}_{a}}
\, ({\mC^{k_1k_2k_3}_{b}})^{*}}{\omega-E_{k_1 k_2 k_3} + i \eta}
+ \frac{({\bar{\mC}^{k_1k_2k_3}_{a}})^*
\, \bar{\mD}^{k_1k_2k_3}_{b}}{\omega+E_{k_1 k_2 k_3} - i \eta}
\right\} \: ,
 \\ \displaystyle
\label{eq:leh_self22}
\Sigma_{ab}^{22 \, (2'+2'')} (\omega) &=&
\sum_{k_1k_2k_3}  \left\{
\frac{{\mD^{k_1k_2k_3}_{a}}
\, (\mD^{k_1k_2k_3}_{b})^*}{\omega-E_{k_1 k_2 k_3} + i \eta}
+ \frac{({\bar{\mC}^{k_1k_2k_3}_{a}})^*
\, {\bar{\mC}^{k_1k_2k_3}_{b}}}{\omega+E_{k_1 k_2 k_3} - i \eta}
\right\} \: .
\end{eqnarray}
\end{subequations}

\subsection{Energy-independent eigenvalue problem}
\label{sec:matrix_rep}

Defining quantities $\mW$ and $\mZ$ through
\begin{subequations}
\label{eq:wz}
\begin{eqnarray}
(\omega_k-E_{k_1 k_2 k_3}) \, \mW^{k_1k_2k_3}_{k} &\equiv&
\sum_a \left[ ({\mC^{k_1k_2k_3}_{a}})^{*} \, \mU^k_a + ({\mD^{k_1k_2k_3}_{a}})^{*} \, \mV^k_a \right] \: ,
\\
(\omega_k+E_{k_1 k_2 k_3}) \, \mZ^{k_1k_2k_3}_{k} &\equiv&
\sum_a \left[  \bar{\mD}^{k_1k_2k_3}_{a} \, \mU^k_a + {\bar{\mC}^{k_1k_2k_3}_{a}} \, \mV^k_a \right] \: ,
\end{eqnarray}
\end{subequations}
Gorkov's equations \eqref{eq:eigen_uv_bar} computed at second-order can be rewritten as
\begin{subequations}
\label{eq:gorkov_premat}
\begin{eqnarray}
\omega_k \, \mU^k_a &=& \sum_b \left [ (T_{ab} - \mu \, \delta_{ab} + \Lambda_{ab}) \, \mU^k_b + \tilde{h}_{ab} \, \mV^k_b \right]
+ \sum_{k_1k_2k_3} \left [ {\mC^{k_1k_2k_3}_{a}} \, \mW^{k_1k_2k_3}_{k}
+ ({\bar{\mD}^{k_1k_2k_3}_{a}})^{*} \, \mZ^{k_1k_2k_3}_{k} \right]  \: ,
\\
\omega_k \, \mV^k_a &=& \sum_b \left [\tilde{h}_{ab}^{\dagger} \, \mU^k_b  -(T_{ab} - \mu \, \delta_{ab} + \Lambda_{\bar{a}\bar{b}}^*) \, \mV^k_b\right]
+ \sum_{k_1k_2k_3} \left [{\mD^{k_1k_2k_3}_{a}} \, \mW^{k_1k_2k_3}_{k}
+ ({\bar{\mC}^{k_1k_2k_3}_{a}})^{*} \, \mZ^{k_1k_2k_3}_{k} \right]  \: .
\end{eqnarray}
\end{subequations}
The four relations above provide a set of coupled equations for unknowns $\mU$, $\mV$, $\mW$ and $\mZ$ that can be displayed in a matrix form
\begin{equation}
\label{eq:matrix_uvwz}
\omega_k
\left(
\begin{array}{c}
\mU  \\ \mV \\ \mW \\ \mZ
\end{array}
\right)_k
=
\left(
\begin{array}{cccc}
T-\mu+\Lambda &  \tilde{h} &\quad  \mC \ &\quad  \bar{\mD}^{*}  \\
\tilde{h}^{\dagger} & -T+\mu-\bar{\Lambda}^* &\quad  \mD &\quad  \bar{\mC}^* \\
\mC^{\dagger} & \mD^{\dagger}  &\quad  E  &\quad  0 \\
\bar{\mD}^{T} & \bar{\mC}^{T}  &\quad  0 &\; \; -E
\end{array}
\right)
\left(
\begin{array}{c}
\mU  \\ \mV \\ \mW \\ \mZ
\end{array}
\right)_k
\equiv \Xi
\left(
\begin{array}{c}
\mU  \\ \mV \\ \mW \\ \mZ
\end{array}
\right)_k
\: ,
\end{equation}
\end{widetext}
where $\Xi$ is an energy-independent Hermitian matrix. The diagonalization of $\Xi$ is equivalent to solving Gorkov's equation. Such a transformation is made possible by the explicit energy dependence embodied in the Lehmann representation, i.e. the known pole structure of the propagators--and consequently of second-order self-energies--is used to recast Gorkov's equations under the form of an energy-independent eigenvalue problem whose eigenvalues and eigenvectors yield the complete set of poles of Gorkov-Green's functions. The solution of such an eigenvalue problem has to be found self-consistently while satisfying Eq. \eqref{eq:NV}.

A normalization condition for the column vectors in Eq. \eqref{eq:matrix_uvwz} is obtained for each solution $k$ by inserting second-order self-energies \eqref{eq:leh_self11} and \eqref{eq:leh_self} into Eq. \eqref{eq:norm_x} (or equivalently into Eq. \eqref{eq:norm_y}). One obtains
\begin{equation}
\sum_a  \left[ \left| \mU_{a}^{k} \right|^2 + \phantom{\int} \hspace{-.4cm} \left| \mV_{a}^{k} \right|^2 \right ]+
\sum_{k_1 k_2 k_3} \left [ \left| \mW^{k_1k_2k_3}_k \right|^2 +
\left| \mZ^{k_1k_2k_3}_k \right|^2 \right] =1 \: .
\end{equation}
The fact that $\Xi$ is Hermitian implies that eigenvalues $\omega_k$ are real. Moreover, similarly to the HFB problem~\cite{ring80a}, solutions come in pairs with opposite sign, i.e. for any solution $\{ \mU^k,\mV^k,\mW_k,\mZ_k,\omega_k \}$ there exists another solution $\{ \bar{\mV}^{k *},\bar{\mU}^{k *},\bar{\mZ}_k^*,\bar{\mW}_k^*,-\omega_k \}$. This can be checked either by substituting $\omega$ with $-\omega$ in the steps that led to Eq. \eqref{eq:matrix_uvwz} or by re-deriving Eq. \eqref{eq:matrix_uvwz} starting from Eq. \eqref{eq:eigen_uv_star} instead of Eq. \eqref{eq:eigen_uv_bar}.

Let us discuss in some detail the structure of $\Xi$. The upper-left block
\begin{equation}
\Xi^{(1)} \equiv
\left(
\begin{array}{cc}
T-\mu+\Lambda &  \tilde{h}  \\
\tilde{h}^{\dagger} & -T+\mu-\bar{\Lambda}^*
\end{array}
\right)
\end{equation}
represents the ``mean-field'' sector. If second-order self-energies are zero, $\Xi=\Xi^{(1)}=\Xi^{HFB}$ and one recovers the HFB eigenvalue problem of fixed dimensionality (twice the size of the single-particle basis) for amplitudes $\mU$ and $\mV$ discussed in Sec.~\ref{sec_HFB_limit}.

The upper-right
\begin{equation}
\Xi^{(2)} \equiv
\left(
\begin{array}{cc}
\mC &  \bar{\mD}^*  \\
\mD & \bar{\mC}^*
\end{array}
\right)
\end{equation}
and lower-left ${\Xi^{(2)}}^{\dagger}$ blocks contain second-order couplings between one quasi-particle and three-quasi-particle configurations. Such couplings further fragment the single-particle strength as compared to the pure HFB approximation. As a matter of fact, following the iterative process leading to a self-consistent solution of Gorkov's equations, the dimension of $\Xi^{(2)}$ grows, i.e. a larger number of poles is generated in Gorkov-Green's functions at each iteration. A propagator with an initial number of poles $N_p^0=2 \, N_{b}$ generates at first iteration a second-order self-energy with approximately $(N_b)^3$ poles, which reflects into a matrix $\Xi^{(2)}$ of dimension $N_p^0+2(N_b)^3$. After $n$ iterations the propagator and the second-order self-energy will contain respectively $\mathcal{O}((N_b)^{3^n})$ and $\mathcal{O}((N_b)^{3^{(n+1)}})$ poles, and the dimension of $\Xi^{(2)}$ will be $\mathcal{O}((N_b)^{3^{(n+1)}})$. This exponential growth of the number of poles seems to prevent the achievement of  convergence in an actual calculation. In practice, one limits the growth of the number of poles by Krylov projection techniques \cite{Schirmer:1989BlkLanc,Dewulf:1997Bagel}, as discussed in Paper II, while ensuring the convergence of the calculation.

\subsection{Application to $J^{\Pi} = 0^+$ states}

The results obtained so far are general and valid for any choice of single-particle basis $\{a_a^{\dagger} \}$ and even-number parity state $| \Psi_0 \rangle$. If the target system, however, possesses specific symmetries, one can exploit them to simplify the set of equations. The first applications of the scheme developed in the present paper will be dedicated to studying the ground-state of even-even semi-magic nuclei, i.e. states characterized by angular momentum and parity $J^{\Pi} = 0^+$. Appendix~\ref{appli0plus} specifies the set of equations provided above to the particular case of such $J^{\Pi} = 0^+$ states.

\section{Quantities of interest}
\label{observables}

\subsection{Binding energy}

The energy sum rule first derived by Galitskii \cite{Galitskii:1958} and formalized by Koltun \cite{Koltun:1972} expresses the expectation value of the Hamiltonian in terms of the one-body propagator. It is one of the appealing features of Green's functions theory, since the energy of the system, a two-body observable, can be computed exactly from a one-body quantity.

The purpose of this subsection is to derive the analogous of the Koltun sum rule in the more general context of Gorkov-Green's functions. Let us first recall that the equation of motion of annihilation operators defined in their Heisenberg representation through Eq. \eqref{eq:HS_reps} reads
\begin{equation}
\label{eq:da}
i \frac{d a_a (t)}{dt} = [\Omega,a_a(t)] \: ,
\end{equation}
and that the normal Gorkov propagator is defined at equal times through
\begin{equation}
\label{eq:gtimezero_Gkv}
\langle \Psi_0 | a_b^{\dagger}(0) a_a(t) | \Psi_0 \rangle \Big{|}_{t=0} \equiv
\frac{1}{2 \pi i} \int_{C \uparrow} d \omega \, {G}^{11}_{ab} (\omega) \, .
\end{equation}
From the definition of the Fourier transform one can then derive
\begin{equation}
\label{eq:gderint_Gkv}
\left. \frac{d \, {G}^{11}_{ab}(t)}{dt} \right|_{t=0^-} =
\frac{1}{2 \pi i} \int_{C \uparrow} d \omega \, \omega \, {G}^{11}_{ab} (\omega) \: .
\end{equation}
Also, it is useful for the following to compute the three commutators
\begin{subequations}
\label{eq:commut}
\begin{eqnarray}
\sum_a a_a^{\dagger} \, [T,a_a] &=& - T \: ,
\\ \displaystyle
\sum_a a_a^{\dagger} \, [V^{\text{NN}},a_a] &=& - 2 V^{\text{NN}} \: ,
\\ \displaystyle
\sum_a a_a^{\dagger} \, [N,a_a] &=& - N \: .
\end{eqnarray}
\end{subequations}
Let us now write $\Omega_0 = \langle \Psi_0 | \Omega |  \Psi_0 \rangle$ as
\begin{eqnarray}
\Omega_0 &=& \frac{1}{2}   \underbrace{\langle \Psi_0 | T - \mu N |  \Psi_0 \rangle}_{\displaystyle\equiv A_0} +
 \frac{1}{2} \underbrace{\langle \Psi_0 | T + 2 V^{\text{NN}} - \mu N |  \Psi_0 \rangle}_{\displaystyle\equiv B_0} \: . \nonumber
\end{eqnarray}
Using Eq. \eqref{eq:gtimezero_Gkv} one has
\begin{eqnarray}
A_0 &=&
 \sum_{ab} \left[ T_{ab} - \mu \, \delta_{ab}  \right] \langle \Psi_0 | a_a^{\dagger} a_b |  \Psi_0 \rangle
\nonumber \\ \displaystyle &=&
\sum_{ab} \left[ T_{ab} - \mu \, \delta_{ab}  \right]
\frac{1}{2 \pi i} \int_{C \uparrow} d \omega \, {G}^{11}_{ba} (\omega) \, ,
\end{eqnarray}
while using Eqs. \eqref{eq:da}, \eqref{eq:gderint_Gkv} and \eqref{eq:commut}, one can also write
\begin{eqnarray}
B_0 &=&
-  \sum_a \langle \Psi_0 | a_a^{\dagger} [\Omega, a_a] |  \Psi_0 \rangle
\nonumber \\ \displaystyle &=& \nonumber
- i \,  \sum_a \langle \Psi_0 | a_a^{\dagger}(0) \left.
\frac{d a_a(t)}{dt}\right|_{t=0} |  \Psi_0 \rangle
\nonumber \\ \displaystyle &=&
 \sum_a \frac{1}{2 \pi i} \int_{C \uparrow} d \omega \, \omega \, {G}^{11}_{aa} (\omega) \: .
\end{eqnarray}
Hence one obtains the generalized Koltun sum rule
\begin{equation}
\label{eq:koltun_gorkov}
\Omega_0 =
\sum_{ab} \frac{1}{4 \pi i} \int_{C \uparrow} d \omega \, G^{11}_{ba} (\omega) \left[ T_{ab}
-\mu \, \delta_{ab} + \omega \, \delta_{ab} \right] \, ,
\end{equation}
where the normal Gorkov propagator $G^{11}$ appears instead of the Dyson one.

\subsection{Effective single-particle energies}
\label{sec_espe}

In Ref.~\cite{Duguet:2011a}, an extensive discussion about effective single-particle energies (ESPE) in doubly-closed shell nuclei was proposed. Results were based on the definition of ESPE going back to Baranger~\cite{baranger70a} and the fact that eigenstates of the nuclear Hamiltonian are also eigenstates of the particle number operator. Such a definition and its associated properties need to be revisited in the context of Gorkov-Green's function were particle-number, as a good symmetry, is lost, i.e. for methods formulated over Fock space rather than over the Hilbert space associated with a definite number of particles.

As in Refs.~\cite{baranger70a,Duguet:2011a}, ESPE are naturally computed as eigenvalues of the so-called centroid matrix, which in the present context is nothing but the normal part of the first moment of the spectral function introduced in Eq.~\eqref{eq:sfunc2_mom}
\begin{eqnarray}
\label{eq:SSDa}
M^{(1)\, 11}_{ab} &\equiv& \sum_k   \mU_{a}^{k} \,\mU_{b}^{k*} \, E^+_k + \bar{\mV}_{a}^{k*} \, {\bar{\mV}_{b}^{k}} \, E^-_k \: .
\end{eqnarray}
By definition of $M^{(1)\, 11}$, the computation (or extraction) of ESPEs requires the full spectroscopic strength, i.e. the complete set of separation energies and spectroscopic amplitudes from both one-nucleon stripping {\it and} pickup processes. This is particularly critical as one moves away from doubly closed-shell nuclei as the low-lying strength becomes more and more fragmented, e.g. by pairing correlations, into both the additional and the removal channels. This is precisely the focus of the presently developed theoretical method to access the complete spectroscopic one-nucleon addition and removal strength in open-shell nuclei from which ESPEs can be extracted.

Let us now derive a sum rule for $M^{(1)\, 11}$ that complements the one provided for $M^{(0)\, 11}$ in Eq.~\eqref{eq:sumruleSF} and that provides ESPEs with a transparent physical meaning. Considering the first term on the right-hand side of Eq. \eqref{eq:SSDa}, substituting the definition of spectroscopic amplitudes and one-nucleon additional energy, one finds
\begin{eqnarray}
\sum_{k} \mU_{a}^{k} \,\mU_{b}^{k*} E_{k}^{+} &=&
 \sum_{k}
  \bra{\Psi_0} \ad {a} \ket{\Psi_k} \, \bra{\Psi_k} \ac {b} \ket{\Psi_0}
 (\mu + \omega_k) \nonumber \\
&=&
 \sum_{k}
 \bra{\Psi_0}  \ad {a}  \Omega_k \ket{\Psi_k}
 \bra{\Psi_k} \ac {b} \ket{\Psi_0}
\nonumber   \\
 &-& \sum_{k}
 \bra{\Psi_0}  \Omega_0 \ad {a} \ket{\Psi_k}
 \bra{\Psi_k} \ac {b} \ket{\Psi_0}
\nonumber   \\
&+&  \sum_{k}  \mu \,
 \bra{\Psi_0}  \ad {a} \ket{\Psi_k}
 \bra{\Psi_k} \ac {b} \ket{\Psi_0}
\nonumber   \\
 &=&
 \bra{\Psi_0}  [ \ad {a} , \Omega ] \ac {b} \ket{\Psi_0}
 + \mu \, \bra{\Psi_0}  \ad {a}  \ac {b} \ket{\Psi_0}
 \, , \qquad
 \label{eq:seplus}
\end{eqnarray}
where a completeness relation over ${\cal F}$ was removed. Similarly, one obtains
\begin{eqnarray}
\sum_{k} \bar{\mV}_{a}^{k*} \, {\bar{\mV}_{b}^{k}} \, E_{k}^{-} &=&
 \bra{\Psi_0}  \ac {b} [ \ad {a} , \Omega ]  \ket{\Psi_0} + \mu \, \bra{\Psi_0}  \ac {b}  \ad {a} \ket{\Psi_0}
 \,, \qquad
\end{eqnarray}
which, combined with Eqs.~\eqref{eq:commAA} and~\eqref{eq:seplus}, leads to
\begin{equation}
\label{eq:sr_comm}
M^{(1)\, 11}_{ab}  = \bra{\Psi_0} \{[\ad{a},\Omega],\ac{b}\} \ket{\Psi_0} + \mu \, \delta_{ab}\,\,\,.
\end{equation}
Using the second quantized form of $T$, $N$ and $V^{\text{NN}}$, together with symmetries of interaction matrix elements, one eventually obtains the key result
\begin{subequations}
\begin{eqnarray}
M^{(1)\, 11}_{ab} &=&  T_{ab} + \sum_{cd} \bar{V}^{\text{NN}}_{adbc} \, \rho_{cd} \\
&\equiv& h^{\infty}_{ab} \label{HFfield}
 \,\,\, ,
\end{eqnarray}
\end{subequations}
where $h^{\infty}_{ab} \equiv T_{ab} +  \Sigma^{11}_{ab}$ involves the energy-{\it independent} (or static) part of the normal self-energy. Eventually, solving
\begin{eqnarray}
h^{\infty} \, \psi^{\text{ESPE}}_p &=& e^{\text{ESPE}}_{p} \, \psi^{\text{ESPE}}_p \,\,\,, \label{HFfield3}
\end{eqnarray}
provides ESPEs and associated single-particle wave functions. Defined in this way, ESPEs are manifestly independent of the single-particle basis used to compute the centroid matrix~\eqref{eq:SSDa}. They possess the meaning of an average of observable one-nucleon separation energies weighted by the probability to reach the corresponding many-body state of the $N\pm1$-body system by adding or removing a nucleon in the single-particle state $\psi^{\text{ESPE}}_p$. As such, it is however essential to understand that ESPEs are by essence non observable and display an intrinsic resolution scale dependence, just as spectroscopic factors do~\cite{Duguet:2011a}.

In spite of the breaking of particle-number symmetry leading to the coupling of additional and removal spectroscopic amplitudes via anomalous self-energies in Gorkov's equations, Eq.~\eqref{HFfield} demonstrates that the centroid matrix is equal to the normal {\it static} field $h^{\infty}$, exactly as for theories that explicitly conserve particle number~\cite{Duguet:2011a}. In other words, the centroid sum rule does not only screen out the energy-dependent part of the normal self-energy but also screens out the entire anomalous self-energy. This is an a priori non trivial, though straightforward to obtain, result. Of course, the explicit tackling of pairing correlations through anomalous propagators and self-energies does impact the results indirectly via their feedback onto the normal one-body density matrix $\rho$ entering $h^{\infty}$.

Eventually, one can demonstrate, just as for particle-number conserving theories~\cite{Duguet:2011a} that $h^{\infty}$ only involves the so-called {\it monopole part}~\cite{bansal64a,zuker69a} of the NN interaction whenever $\ket{\Psi_0}$ is a $J^{\pi}=0^+$ state.

\subsection{Natural basis}

The natural basis is the one that diagonalizes  the one-body density matrix, i.e.
\begin{eqnarray}
\rho \, \psi^{\text{nat}}_p &=& n_p^{\text{nat}} \, \psi^{\text{nat}}_p \: , \label{eq:nat_basis}
\end{eqnarray}
where basis states $\psi^{\text{nat}}_p$ are called ``natural orbitals'' and where diagonal elements $n_p^{\text{nat}}$ denote ``natural occupation numbers''. The $N$ most occupied natural orbitals define the set that better approximates the true (correlated) density matrix $\rho$ in terms of a Slater determinant wave function. Thus, the natural basis is most convenient to expand approximations of observables other than energies (e.g., radii, density distributions, etc). We stress that in general  $\psi^{\text{nat}}_p$ correspond to superpositions of orbits $\psi^{\text{ESPE}}_p$ with ESPEs both above and below the Fermi surface, chosen to optimize the density profile of the system, and therefore are a poorer approximation to energy levels. Conversely, single-particle states $\psi^{\text{ESPE}}_p$, in Eq. \eqref{HFfield3}, can be directly associated to orbits of the effective shell structure.

\subsection{One-body observables and radii}

The expectation value of a general one-body operator, $\mO$, is obtained from the normal density matrix \eqref{eq:nobdm} or, equivalently, by integration over the normal Gorkov propagator
\begin{equation}
\bra{\Psi_0}  \mO \ket{\Psi_0}  =  \text{Tr}_{{\cal H}_1} \left\{ \rho \, \mO \right\}  = \sum_{a b}  \int_{C \uparrow} \frac{d\omega}{2 \pi i}  \, G^{11}_{a b}(\omega)  o_{a b} \: , \quad
\end{equation}
where $o_{a b} \equiv (  a | \mO |  b )$ are matrix elements of the one-body operator.

For matter radii, however, one needs to sum over particle positions in the intrinsic frame. The operator for the root mean square point radius $r_{rms}$ is thus
\begin{eqnarray}
\label{eq:r_rms}
r_{rms}^2  &=& \frac{1}{\hat{N}}  \sum_i  \left( \mathbf{r}_i - \mathbf{R} \right) ^2  \nonumber \\
 &=&   \frac{1}{\hat{N}} \left( 1- \frac{1}{\hat{N}} \right ) \sum_i \mathbf{r}_i^2 \, - \, \frac{1}{\hat{N}^2} \sum_{i<j} \mathbf{r}_i \cdot \mathbf{r}_j \: , \quad
\end{eqnarray}
where $\mathbf{r}_i$ denote coordinates of nucleon $i$ in the laboratory frame and $\mathbf{R} \equiv (\sum_i \mathbf{r}_i)/\hat{N}$ is the centre-of-mass coordinate. Operator \eqref{eq:r_rms} contains a two-body correction term and depends on the number of particles. As for the centre-of-mass corrected Hamiltonian, this is the form suitable for applications in Fock space. At second order in the Gorkov self-energy, as considered in this work, there is no resummation of diagrams corresponding to correlated two-body propagators. Only the free propagation of dressed pp or hh is accounted for in the second order diagrams. Correspondingly, one can approximate the two-particle density matrix with the antisymmetrized product of correlated one-body density matrices in order to evaluate the two-body part in \eqref{eq:r_rms}. The $r_{rms}$ radius is thus calculated to first order in $\hat{N}^{-1}$ as
\begin{eqnarray}
\label{eq:r_rms_G11}
&&\langle r^2_{rms} \rangle  = \frac{1}{N} \left( 1 - \frac{1}{N}\right) \sum_{a b}  \int_{C \uparrow} \frac{d\omega}{2 \pi i} \, G^{11}_{a b} \; r^2_{a b}   \\
 &&- \frac{1}{2 N^2}  \sum_{a b c d}  \int_{C \uparrow} \frac{d\omega}{2 \pi i} \int_{C \uparrow} \frac{d\omega'}{2 \pi i} \, \langle ab | \mathbf{r}_1 \cdot \mathbf{r}_2 | cd \rangle \, G^{11}_{a c}(\omega) \,  G^{11}_{b d}(\omega')  \: ,
\nonumber
\end{eqnarray}
where $r^2_{a b}=(a|\mathbf{r}^2|b)$ and where $\langle ab | \mathbf{r}_1 \cdot \mathbf{r}_2 | cd \rangle$ are antisymmetrized two-body matrix elements while N denotes the total number of nucleons (protons plus neutrons). Eqs. \eqref{eq:r_rms} and  \eqref{eq:r_rms_G11} are valid for matter radii. Charge point radii $r_{ch}$ can be obtained from
\begin{equation}
\label{eq:r_ch}
r_{ch}^2 = \frac{1}{\hat{Z}}  \sum_p  \left( \mathbf{r}_p - \mathbf{R} \right) ^2   \: , \quad
\end{equation}
where $\hat{Z}$ is the number operator for protons and where $p$ runs only over protons while $\mathbf{R}$ is the same as in \eqref{eq:r_rms}.
Isotope shifts are calculated from differences of squared charge radii,
\begin{eqnarray}
\delta \langle r_{ch}^2 \rangle ^{N,N'} = \langle r_{ch}^2 \rangle^{N'} -  \langle r_{ch}^2 \rangle^{N} \, ,
\end{eqnarray}
where $N'$ is the number of nucleons of the system under consideration whereas $N$ characterizes a reference nucleus.

\subsection{Pairing gaps}

Experimentally, a suitable way of extracting the pairing gap goes through, e.g., the three-point mass formula
\begin{equation}
\label{eq:3pmass}
\Delta^{(3)}(N) = \frac{(-1)^N}{2} \left[ E_0^{N+1} - 2 E_0^{N} + E_0^{N-1}  \right] \: ,
\end{equation}
where $N$ is the total number of nucleons. This is motivated by the relation between the odd-even staggering of nuclear binding energies and the lack of binding of the unpaired odd nucleon, as first pointed out in Ref. \cite{Bohr:1958zz}.

To compute $\Delta^{(3)}(N)$ theoretically, one needs to perform consistent calculations of odd nuclei. In the present context, this would require to perform Gorkov calculations for a state $| \Psi_0 \rangle$ having an odd number-parity quantum number, i.e. a state such that the sum runs over odd $N$ in Eq.~\eqref{eq:psi0}. This is however beyond the scope of the present work. The next best approximation would consist in keeping an even number-parity state while accounting for the blocking of a quasi-particle within the filling approximation~\cite{perezmartin08a}. Such an approximation remains however to be formulated within the general frame of Gorkov-Green's function formalism.

In such a situation, the next best estimate to the ground-state energy of the odd system is obtained through~\cite{Duguet:2001gr, Duguet:2001gs}
\begin{equation}
\label{eq:oddE}
E_0^{N} \approx E_0^{N \, *} + \omega^{N}_{F} \: ,
\end{equation}
where $E_0^{N \, *}$ is the energy of the odd nucleus computed as it were an even one, i.e. as a fully paired vacuum with an odd number of particles on average, while $\omega^{N}_{F}$ denotes the lowest pole energy obtained from that Gorkov calculation. Obviously, for even $N$ one simply has $E_0^{N} = E_0^{N \, *}$. With such an appropriate decomposition of the energy, the three-point mass formula reads
\begin{equation}
\label{eq:3pmass-th}
\Delta^{(3)}(N) \approx \frac{(-1)^N}{2} \frac{\partial^2 E_0^{N \, *}}{\partial N^2} + \Delta_{F}(N) \: .
\end{equation}
The first contribution relates to the second derivative of the smooth part of the energy $E_0^{N \, *}$, i.e. the energy curve on which both even and odd nuclei would lie in the absence of odd-even mass staggering. Such a second derivative of $E_0^{N \, *}$ evolves very smoothly with $N$. However, the corresponding contribution to $\Delta^{(3)}(N)$ oscillates strongly around zero due to the factor $(-1)^N$ appearing in Eq.~\eqref{eq:3pmass-th}, accounting for the odd-even oscillation of experimental $\Delta^{(3)}(N)$ and having nothing to do with the pairing gap itself~\cite{Duguet:2001gr, Duguet:2001gs}. The second contribution to $\Delta^{(3)}(N)$ relates specifically to the unpaired character of the odd nucleon, and thus extracts the actual pairing gap at the Fermi energy in open-shell nuclei~\cite{Duguet:2001gr, Duguet:2001gs}
\begin{eqnarray}
\label{gapfermi}
\Delta_{F}(N)  \equiv
\left\{
\begin{tabular}{cl}
$\omega^{N}_{F}$ & \,\, for $N$ odd \\
$(\omega^{N-1}_{F}+\omega^{N+1}_{F})/2$ & \,\,  for $N$ even
\end{tabular}
\right.
\: .
\end{eqnarray}

\section{$\Phi$-derivability}
\label{phi_derivable}

Gorkov-Green's functions constitute a versatile and powerful technique that can as well be applied to time-dependent, non-equilibrium systems. When truncating the self-energy expansion, and hence approximating the solution of the many-problem, one has to pay attention to the possible violation of basic conservation laws involving e.g. particle number, total energy, total momentum, total angular momentum. For an arbitrary set of self-energy diagrams nothing assures that quantities are conserved with time even though the corresponding operator commutes with $H$.

A way to construct a class of conserving approximations was devised in the early 1960's by Baym and Kadanoff \cite{Baym:1961zz, Baym:1962sx}. Baym and Kadanoff demonstrated that if the self-energy is derived from a certain functional $\Phi$ of the one-body Green's function (and the two-body potential), previously introduced by Luttinger and Ward \cite{Luttinger:1960ua}, the resulting scheme automatically satisfies all basic conservation laws. Moreover, the resulting approximation preserves thermodynamic consistency requirements, including the Hugenhotlz-van Hove \cite{Hugenholtz:1958zz} and Luttinger \cite{Luttinger:1960} identities. Fulfilling such consistency requirements avoids ambiguities in the calculation of thermodynamic observables, i.e. different ways of computing the same quantity yield the same result.

The concept of $\Phi$-derivable approximations, introduced by Baym and Kadanoff for normal Green's functions, was generalized to Gorkov's formalism by De Dominicis and Martin \cite{DeDominicis:1964se} (see also \cite{Kita:1996aa, Kita:1996bb}).
In this case such a class of approximations relies on the existence of a closed functional $\Phi$ of the four Gorkov-Green's functions $\nG$ and the two-body interaction $V$, from which self-energy contributions are obtained via a functional derivative.
At zero temperature, such a functional is closely related to the correlation energy $\Delta E_0$, i.e. the total energy measured with respect to the unperturbed energy \cite{Nozieres:1963}
\begin{equation}
\label{eq:Phi-DE}
\Delta E_0 =   \Phi [\nG,V] + \text{Tr} \left \{ \frac{\nG^{\phantom{(0)}}}{\nG^{(0)}} -1 \right \} - \text{Tr} \left \{ \text{ln} \frac{\nG^{\phantom{(0)}}}{\nG^{(0)}} \right \}\, ,
\end{equation}
where traces have to be performed over Gorkov and  one-particle Hilbert spaces
\begin{equation}
\label{eq:Phi-tr}
\text{Tr}  \equiv \text{Tr}_{{\cal H}_1} \, \frac{1}{2} \, \text{Tr}_{\cal{G}}   \, \int \frac{d \omega}{2 \, \pi} \, .
\end{equation}

There exist two possible strategies to build a $\Phi$-derivable scheme.
Starting from a carefully chosen set of self-energy contributions, Baym-Kadanoff's functional can be formally defined through~\cite{Nozieres:1963}
\begin{eqnarray}
\label{eq:Phi-def}
\Phi [\nG,V] &\equiv&  - \frac{1}{2} \, \sum_{n=1}^{\infty} \frac{1}{n} \, \text{Tr}
\left \{ \mathbf{\Sigma}^{(n)}[\nG,V] \, \nG \right \}
\nonumber \\ \displaystyle
&\equiv& \sum_{n=1}^{\infty}  \Phi^{(n)} [\nG,V]
\, ,
\end{eqnarray}
where $\mathbf{\Sigma}^{(n)}[\nG,V]$ denotes skeleton self-energy terms of order $n$.
Compared to the standard definition, an additional $1/2$ factor appears in relation to the trace over the two-dimensional Gorkov space in Eq.~\eqref{eq:Phi-tr}. Notice that self-consistency, i.e. the use of dressed propagators in the functional, is a necessary condition for  $\Phi$-derivability.

Alternatively, one can use diagrammatic techniques to construct $\Phi$, analogously to the self-energy expansion. At order $n$ in $V$, $\Phi^{(n)}$ is given by two-fermion-line irreducible connected closed skeleton diagrams. With the obvious change from an open to a closed topology, all diagrammatic rules outlined in Appendix \ref{diag_rules} hold for the construction of the functional.
From $\Phi$ one can obtain the four self-energies by differentiating with respect to Gorkov propagators
\begin{subequations}
\label{eq:Phi-der}
\begin{equation}
\label{eq:Phi-der11}
\Sigma_{ij}^{11} (\omega) =  -  \frac{\delta \Phi [\nG,V]}{\delta G_{ji}^{11} (\omega)} \, ,
\end{equation}
\begin{equation}
\label{eq:Phi-der12}
\Sigma_{ij}^{12} (\omega) =  - 2 \, \frac{\delta \Phi [\nG,V]}{\delta G_{ji}^{21} (\omega)} \, ,
\end{equation}
\begin{equation}
\label{eq:Phi-der21}
\Sigma_{ij}^{21} (\omega) =  - 2 \, \frac{\delta \Phi [\nG,V]}{\delta G_{ji}^{12} (\omega)} \, ,
\end{equation}
\begin{equation}
\label{eq:Phi-der22}
\Sigma_{ij}^{22} (\omega) =  -  \frac{\delta \Phi [\nG,V]}{\delta G_{ji}^{22} (\omega)} \, ,
\end{equation}
\end{subequations}
as demonstrated in Appendix \ref{sec:phi-app1}.
Any subset of $\Phi$ diagrams employed to derive the self-energy via Eq. \eqref{eq:Phi-der} will generate a conserving approximation. If at a given order all terms are taken into account, the resulting self-energy will contain all possible contributions at that order.
Eventually, one should have internal consistency, i.e. if the functional is constructed from a certain set of self-energy contributions via Eq.~\eqref{eq:Phi-def}, all and only these self-energy contributions must be generated from that functional when employing Eq.~\eqref{eq:Phi-der}.

\begin{figure}[h]
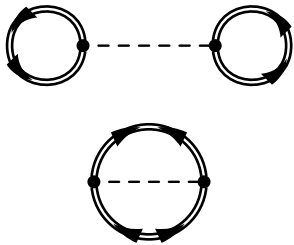

\begin{eqnarray}
\nonumber
\mbox{\phifirst}
\\ \nonumber
\\ \nonumber
\\ \nonumber
\mbox{\phifirstex}
\end{eqnarray}
\caption{Diagrams contributing to $\Phi^{(1)}$.}
\label{fig:phi1}
\end{figure}
\begin{figure}[h]
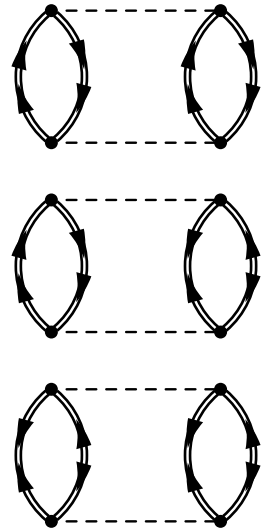

\begin{eqnarray}
\nonumber
\mbox{\phisecondnn}
\\ \nonumber
\\ \nonumber
\\ \nonumber
\mbox{\phisecondna}
\\ \nonumber
\\ \nonumber
\\ \nonumber
\mbox{\phisecondaa}
\end{eqnarray}
\caption{Diagrams contributing to $\Phi^{(2)}$.}
\label{fig:phi2}
\end{figure}

Most of the commonly used (fully) self-consistent approximations in Green's function theory are $\Phi$-derivable. It is the case of the second-order approximation used in the present work, as well as of Hartree-Fock, RPA or T-matrix approximations.
Diagrams in the $\Phi$-functional that generate first- and second-order self-energies of Figs. \ref{fig:first}, \ref{fig:secondn} and \ref{fig:seconda} are depicted respectively in Figs. \ref{fig:phi1} and \ref{fig:phi2}. Writing explicit expressions for such diagrams and applying functional derivative \eqref{eq:Phi-der}, one obtains all self-energies calculated in Appendix \ref{sec:self_eva}. As an illustration, the first order case is treated in full details in Appendix \ref{sec:phi-app2}.

\section{Conclusions}
\label{conclusions}

The aim of the present paper is to extend the reach of ab-initio nuclear structure calculations to truly open shell nuclei. This is done by implementing self-consistent Green's Function method within the general Gorkov's scheme. Such a method retains the simplicity of single-reference approaches, i.e. methods relying on a many-body expansion around a single vacuum. As for open-shell systems, this can only be done at the price of breaking $U(1)$ symmetry associated with particle-number conservation in order to tackle Cooper pair instabilities and explicitly account for pairing correlations.

The present work provides a detailed account of Gorkov's formalism and of its specification to second-order in the expansion of normal and anomalous self-energies. At the present stage, this is done in terms of two-nucleon interactions only. First numerical applications of such a scheme will be reported soon in a forthcoming publication. This constitutes the first {\it ab-initio} application of self-consistent Gorkov-Green's function method in finite nuclei using realistic two-nucleon interactions. The extension of the proposed ab-initio method to more advanced truncation schemes and to include three-nucleon forces is the aim of future works.

\section*{ Acknowledgements }

The authors would like to thank S. Baroni and P. Bo\.zek for useful discussions. This work was supported by  the United Kingdom Science and Technology Facilities Council (STFC) through travel grant No. ST/I003363. V. S. acknowledges support from Espace de Structure Nucl\'eaire Th\'eorique (ESNT).
\appendix

\section{Normalization condition}
\label{app_normalization}

In order to work out the normalization of the spectroscopic amplitudes let us consider the expansion of Gorkov's equation \eqref{eq:gorkov} around the pole $+\omega_k$. We recall that a complex function $f(z)$ can be expanded in a Laurent series around a point $c$ in the complex plane as
\begin{equation}
f(z) = \sum_{n=-\infty}^{+\infty} a_n \, (z-c)^n \: ,
\end{equation}
with
\begin{equation}
\label{eq:laurent_coeff}
 a_n \equiv \frac{1}{2 \pi i} \int_C \frac{f(z) \, dz}{(z-c)^{n+1}} \:,
\end{equation}
and where $C$ is a contour containing $c$ on which $f$ is analytic.
Here one is interested in the case in which $f(z)$ has a simple pole at $z=c$, which means that the integrand in Eq. \eqref{eq:laurent_coeff} has a pole of order $n+2$ at $z=c$ and the integral can be performed by means of the residue theorem. In particular, the $n=0$ coefficient  reads
\begin{equation}
 a_0 = \frac{1}{2 \pi i} \int_C \frac{f(z) \, dz}{(z-c)} = \lim_{z \rightarrow c} \frac{d}{dz} \left[ (z-c)^2 \frac{f(z)}{(z-c)} \right] \: .
\end{equation}
Performing a Laurent expansion of Gorkov's equation \eqref{eq:gorkov} around $\omega=+\omega_k$ and extracting the coefficients of order zero on both sides, one obtains
\begin{widetext}
\begin{eqnarray}
\lim_{\omega \rightarrow \omega_k} \frac{d}{d \omega} \,  \Big\{
\nX_{a}^{k} {\nX_{b}^{k \dagger}} + (\omega-\omega_k) \, \nB^k_{ab}(\omega) &=& \displaystyle (\omega - \omega_k)
\, \left( \omega - \Omega_U \right)^{-1}_{ab} + \sum_{cd}  \left [ \left( \omega - \Omega_U \right)^{-1}_{ac} \tilde{\nSigma}_{cd}(\omega) \,
\nX_{d}^{k} {\nX_{b}^{k \dagger}} \right.
\nonumber \\ \displaystyle
&&
\hspace{3.5cm} \left.+ (\omega - \omega_k)
\, \left( \omega - \Omega_U \right)^{-1}_{ac} \, \tilde{\nSigma}_{cd}(\omega) \, \nB^k_{db}(\omega) \right]
\Big\} \: ,
\end{eqnarray}
where the singular part in the dressed propagator has been isolated as follows
\begin{eqnarray}
\mathbf{G}_{ab} (\omega) &=&  \frac{{\nX_{a}^{k}} \, \nX_{b}^{k \dagger}}{\omega-\omega_{k} + i \eta}
+  \sum_{k' \neq k}  \frac{{\nX_{a}^{k'}} \, \nX_{b}^{k' \dagger}}{\omega-\omega_{k'} + i \eta}
+ \sum_{k'} \frac{\nY_{a}^{k'\dagger} \, {\nY_{b}^{k'}}} {\omega+\omega_{k'} - i \eta}
\nonumber \\ \displaystyle
&\equiv&
\frac{{\nX_{a}^{k}} \, \nX_{b}^{k \dagger}}{\omega-\omega_{k} + i \eta} + \nB^k_{ab}(\omega) \: .
\end{eqnarray}
Applying the derivative to all terms and dropping the ones that give zero in the limit $\omega \rightarrow \omega_k$, one has
\begin{eqnarray}
\lim_{\omega \rightarrow \omega_k}  \,  \big\{
\nB^k_{ab}(\omega)&=& \left( \omega - \Omega_U \right)^{-1}_{ab}
+ \sum_{cd}  \Big[ - \left( \omega - \Omega_U \right)^{-2}_{ac}  \tilde{\nSigma}_{cd}(\omega) \, \nX_{d}^{k} {\nX_{b}^{k \dagger}}
\nonumber \\ \displaystyle
&&  \hspace{3cm} + \left( \omega - \Omega_U \right)^{-1}_{ac}  \frac{\partial  \tilde{\nSigma}_{cd}(\omega)}{\partial \omega} \, \nX_{d}^{k} {\nX_{b}^{k \dagger}}
+ \left( \omega - \Omega_U \right)^{-1}_{ac}  \, \tilde{\nSigma}_{cd}(\omega) \, \nB^k_{db}(\omega) \Big]
\big\} \: ,
\end{eqnarray}
such that, using Eq. \eqref{eq:omu_sigma},
\begin{eqnarray}
\label{eq:norm_intyy}
\lim_{\omega \rightarrow \omega_k}  \, \big\{ \sum_{a}
\left(\omega - \Omega_U\right)^{-1}_{ea} \, \nX_{a}^{k} {\nX_{b}^{k \dagger}} + \nB^k_{eb}(\omega)
&=& \left(\omega - \Omega_U\right)^{-1}_{eb}+\sum_{cd} \Big[
 \left( \omega - \Omega_U \right)^{-1}_{ec}  \frac{\partial  \tilde{\nSigma}_{cd}(\omega)}{\partial \omega} \, \nX_{d}^{k} {\nX_{b}^{k \dagger}}
\nonumber \\ \displaystyle
&& \hspace{2.8cm}  + \left( \omega - \Omega_U \right)^{-1}_{ec}  \, \tilde{\nSigma}_{cd}(\omega) \, \nB^k_{db}(\omega) \Big]
\big\} \: .
\end{eqnarray}
\end{widetext}
Multiplying both sides of Eq. \eqref{eq:norm_intyy} by $\left( \omega - \Omega_U \right)_{fe}$ and summing over $e$ yields
\begin{equation}
\nX_{f}^{k} {\nX_{b}^{k \dagger}}
= \delta_{fb} + \sum_{d}
\left.
 \frac{\partial  \tilde{\nSigma}_{fd}(\omega)}{\partial \omega}
\right|_{\omega_k}
 \, \nX_{d}^{k} {\nX_{b}^{k \dagger}}
 \: ,
\end{equation}
where the terms involving $\nB^k$ cancel out after using the conjugate Gorkov's equation
\begin{equation}
\label{eq:gorkov_conj}
\mathbf{G}_{ab}(\omega)  \displaystyle
= \mathbf{G}^{(0)}_{ab}(\omega)
+ \sum_{cd} \mathbf{G}_{ac}(\omega) \, \tilde{\nSigma}_{cd}(\omega) \,
\mathbf{G}^{(0)}_{db}(\omega)  \: .
\end{equation}
Multiplying by ${\nX_{f}^{k \dagger}}$ from the left, summing over $f$ and renaming $(f,d)$ to $(a,b)$ one finally obtains the normalization condition
\begin{equation}
\label{eq:norm_xbis}
\sum_a  \left| \nX_{a}^{k} \right|^2
= 1 + \sum_{ab} \, \nX_{a}^{k \dagger} \,
 \left. \frac{\partial  \nSigma_{ab}(\omega)}{\partial \omega}\right|_{\omega_k}  {\nX_{b}^{k}} \: ,
\end{equation}
where only the proper self-energy appears as a result of the energy independence of the auxiliary potential. Similarly one can derive a condition for Gorkov's amplitude $\nY$
\begin{equation}
\label{eq:norm_ybis}
\sum_a  \left| \nY_{a}^{k} \right|^2 = 1 + \sum_{ab} \, {\nY_{a}^{k \dagger}} \, \left. \frac{\partial  \nSigma_{ab}(\omega)}{\partial \omega}\right|_{-\omega_k} \nY_{b}^{k} \: .
\end{equation}

\section{Diagrammatic}
\label{sec:diagrules}

\subsection{Diagrammatic rules}
\label{diag_rules}

A convenient way to express the expansion of the single-particle propagator is via diagrammatic techniques. By giving the interaction and the single-particle propagator a graphical representation and by establishing a set of rules one can generate diagrams that are in one-to-one correspondence with the terms appearing in the expansion. As it provides an immediate insight to physical processes associated with the various contributions, the diagrammatic expansion is of great help when choosing a suitable approximation. It is relevant to discuss diagrammatic rules in some details here given that there exist differences compared to rules applicable to the diagrammatic expansion involving normal contractions only.

In the present work antisymmetrized interaction matrix elements are represented by a dashed line labeled by four single-particle indices
\begin{equation}
\hspace{1.3cm}
\bar{V}_{abcd} \equiv \hspace{.8cm}
\mbox{\diagv} \hspace{-2.2cm} .
\label{eq:vgraph}
\end{equation}
Single-particle unperturbed propagators, i.e. Green's functions associated with the unperturbed Hamiltonian $\Omega_U$ introduced in Eq. \eqref{eq:g0}, are depicted as solid lines labelled by two indices and one energy flowing from the second to the first index
\begin{subequations}
\label{eq:def_enflow}
\begin{eqnarray}
\nonumber \\
\hspace{.6cm} G^{11 \, (0)}_{ab} (\omega) &\equiv& \hspace{1.2cm} 
\mbox{\diaggoozero}
\hspace{-2.7cm} \, ,
\\ \nonumber \\ \nonumber \\ \nonumber \\
G^{12 \, (0)}_{ab} (\omega) &\equiv& \hspace{1.2cm} \mbox{\diaggotzero} \hspace{-2.7cm} \, ,
\\ \nonumber \\ \nonumber \\ \nonumber \\
G^{21 \, (0)}_{ab} (\omega) &\equiv& \hspace{1.2cm} \mbox{\diaggtozero} \hspace{-2.7cm} \, ,
\\ \nonumber \\ \nonumber \\ \nonumber \\
G^{22 \, (0)}_{ab} (\omega) &\equiv& \hspace{1.2cm} \mbox{\diaggttzero} \hspace{-2.7cm} \, .
\\ \nonumber \\ \nonumber
\end{eqnarray}
\end{subequations}
One should notice that, as opposed to traditional graphical representations of Dyson's propagator, Gorkov's propagators carry two arrows specifying whether a given propagator results from the contraction of two creation operators, of two annihilation operators, or of one creation (annihilation) and one annihilation (creation) operator.

With building blocks \eqref{eq:vgraph} and \eqref{eq:def_enflow} one can construct, order by order, the (diagrammatic) perturbative expansion for each of the four Gorkov propagators \eqref{eq:gg}.
To obtain all terms of the expansion at a certain order $m$ and for one of the four Gorkov propagators, the following rules are employed:
\begin{enumerate}
\item Draw all topologically distinct connected direct diagrams with $m$ horizontal interaction lines (with 4 single-particle indices) and $2m+1$ directed propagation lines (with 2 single-particle indices each, connecting the $4m$ indices of the interaction and the 2 external ones).
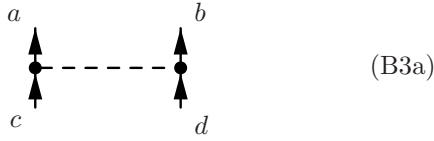
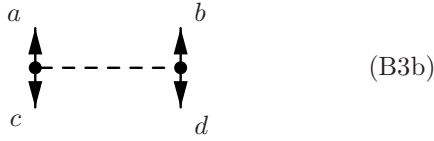
\begin{subequations}
\begin{eqnarray}
\nonumber \\
\hspace {3cm} &\mbox{\diagver}& \hspace {-2cm}  
\label{eq:allowedme} \\ \nonumber \\ \nonumber  \\ \nonumber  \\
&\mbox{\diagvew}& \label{eq:notallowedme} \hspace {-2cm} 
\\ \nonumber
\end{eqnarray}
\end{subequations}
Notice that exactly two incoming and two outgoing lines must be attached to a given interaction vertex, i.e. diagram \eqref{eq:allowedme} is allowed while diagram \eqref{eq:notallowedme} is not.

Topologically distinct diagrams cannot be transformed into each other by any translation (in the two-dimensional plane) of any of the vertices without disconnecting or reconnecting propagation lines.
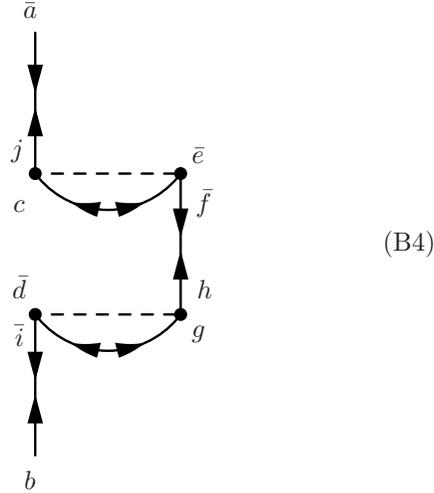
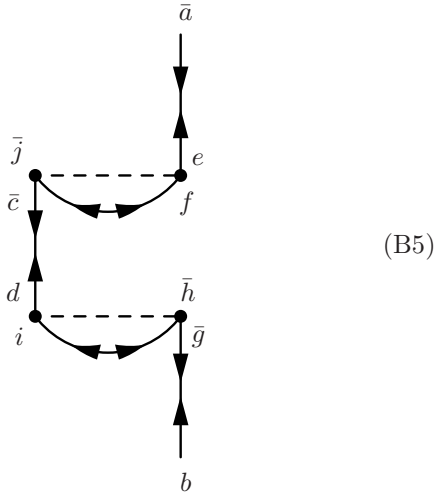
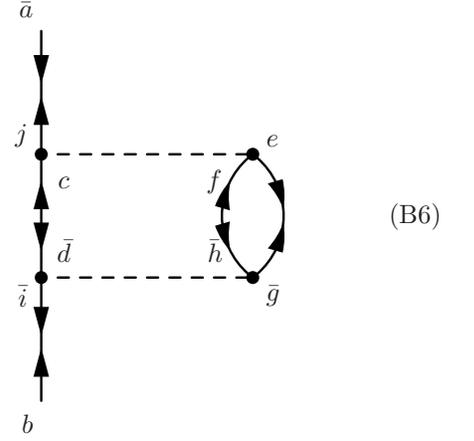
\begin{eqnarray}
\nonumber \\
\hspace {3cm} &\mbox{\topeq}& \hspace {-2cm}  
\label{eq:topeq1} \\ \nonumber \\ \nonumber 
\end{eqnarray}
\begin{eqnarray}
\hspace {3cm} &\mbox{\topeqqq}& \label{eq:topeq2} \hspace {-2cm}
\\ \nonumber \\ \nonumber 
\end{eqnarray}
\begin{eqnarray}
\hspace {3cm} &\mbox{\topneq}& \label{eq:topneq} \hspace {-2cm} 
\\ \nonumber
\end{eqnarray}
For example second-order diagrams \eqref{eq:topeq1} and \eqref{eq:topeq2} are topologically equivalent, while diagram \eqref{eq:topneq} is not.

Connected diagrams are diagrams in which it is possible to go from each interaction line to any other by moving along propagation lines.
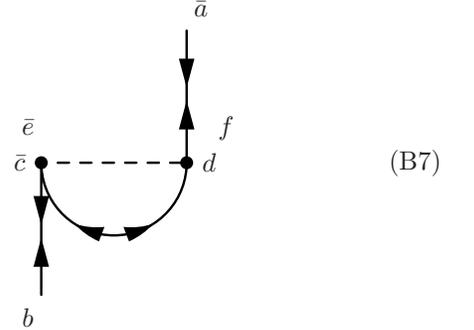
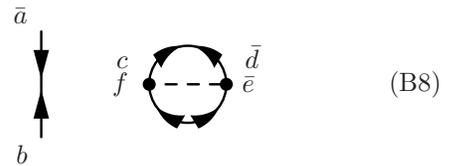
\begin{eqnarray}
\hspace {3cm} &\mbox{\diagconn}& \hspace {-2cm} 
\label{eq:connected} \\ \nonumber \\ \nonumber  \\ \nonumber  \\
&\mbox{\diagdisconn}& \label{eq:disconnected} \hspace {-2cm}
\end{eqnarray}
For example first-order diagram \eqref{eq:connected} is connected while first-order diagram \eqref{eq:disconnected} is disconnected.

For a given diagram, exchange diagrams are derived by exchanging the end points of two propagation lines coming in or out of one or more interaction vertices.
Since we are using anti symmetrized matrix elements in \eqref{eq:vgraph}, it follows that for each set of diagrams obtainable from one another by means of such exchanges, one must only retain one representative diagram, arbitrarily chosen and denoted as \textit{direct}, and discard all the other ones.
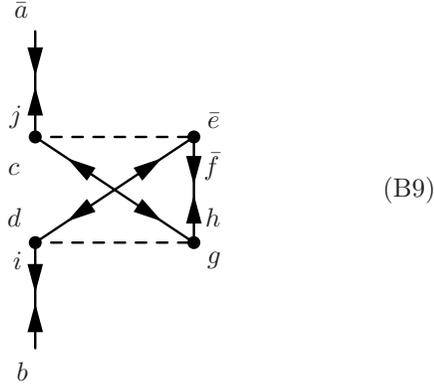
\begin{eqnarray}
\nonumber \\
\hspace {3cm} &\mbox{\diagexex}& \hspace {-2cm} 
\label{eq:dexex} \\ \nonumber
\end{eqnarray}
For example if one considers diagram \eqref{eq:topneq} as direct (the choice of the present work) it follows that one must discard diagram \eqref{eq:dexex}.

In cases where it is unclear whether diagrams are topologically distinct, one can always resort to a direct application of Wick's theorem.
\item  Assign an energy to all propagation lines such that the energy in each interaction is conserved (the energy entering a vertex must be equal to the energy exiting). As a result, a $m$-order diagram will have $m$ independent internal energies and the incoming external energy will be equal to the outgoing external one.
For each independent energy, multiply by a factor $1/2\pi$.
\item Write down a $\bar{V}$ (with corresponding s.p. indices) for each interaction line and a $G^{g_1g_2}$ (with corresponding s.p. indices and energy) for each propagation line\footnote{Any normal propagtion line can be interpreted either as $G^{11}$ or $G^{22}$. The choice of the present work consists in identifying normal lines with $G^{11}$ in the expansions of $G^{11}$, $G^{12}$ and $G^{21}$ while using $G^{22}$ in the expansion of $G^{22}$.} according to representation \eqref{eq:def_enflow}. If the energy $\omega$ flowing along the propagator has the opposite direction than in definition \eqref{eq:def_enflow}, the associated term is $G^{g_1g_2}(-\omega)$.
\item Write an overall factor $i^m$.
\item Write a factor $1/2$  for each pair of equivalent propagation lines, i.e. pairs of lines starting at the same interaction vertex and ending at the same interaction vertex and corresponding to the same type of Gorkov propagator. This factor is due to the antisymmetrization of the potential, i.e. to the fact that exchanging the incoming lines of two interactions connected by equivalent lines yields the same diagram.
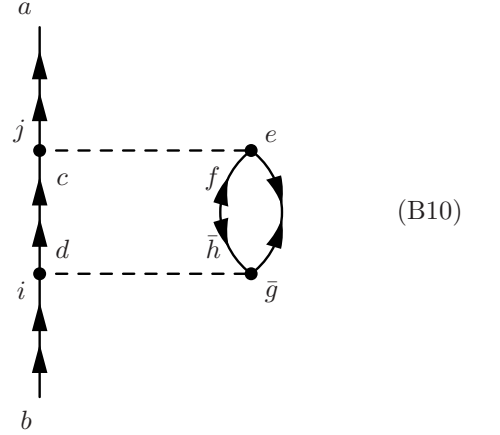
\begin{eqnarray}
\nonumber \\
\hspace {2.4cm} &\mbox{\secondnbnosdg}& \hspace {-2cm}
\label{eq:diaex_el} \\ \nonumber
\end{eqnarray}
For example diagram \eqref{eq:topneq} has a pair of equivalent lines, i.e. those labeled by $(c,\bar{d})$ and $(f,\bar{h})$, while diagram \eqref{eq:diaex_el} has none.
\item Write a factor $1/2$  for each anomalous propagator starting and ending at the same interaction vertex. This factor appears for the reason discussed in point 5 and applies, e.g., to diagram \eqref{eq:connected}.
\item Write a factor $(-1)^{N_c+N_a}$ where $N_c$ is the number of closed fermionic loops and $N_a$ is the number of anomalous contractions.
\item Interpret equal-time propagators as
\begin{subequations}
\begin{eqnarray}
\label{eq:etlim11}
\lim_{t' \rightarrow t} G^{11}_{ab}(t,t') &=& G^{11}_{ab}(0,-\eta) \: , \\
\label{eq:etlim12}
\lim_{t' \rightarrow t} G^{12}_{ab}(t,t') &=& G^{12}_{ab}(0,-\eta) \: , \\
\label{eq:etlim21}
\lim_{t' \rightarrow t} G^{21}_{ab}(t,t') &=& G^{21}_{ab}(0,-\eta) \: , \\
\label{eq:etlim22}
\lim_{t' \rightarrow t} G^{22}_{ab}(t,t') &=& G^{22}_{ab}(0,+\eta) \: ,
\end{eqnarray}
\end{subequations}
which implies that integrations over $\omega$ are performed in the complex energy plane, either by closing the contour in the upper ($C\uparrow$) or in the lower ($C\downarrow$) half plane as
\begin{subequations}
\begin{eqnarray}
\int d\omega \, G^{11}_{ab}(\omega) &\rightarrow& \int_{C\uparrow} d\omega \, G^{11}_{ab}(\omega) \: , \\
\int d\omega \, G^{12}_{ab}(\omega) &\rightarrow& \int_{C\uparrow} d\omega \, G^{12}_{ab}(\omega) \: , \\
\int d\omega \, G^{21}_{ab}(\omega) &\rightarrow& \int_{C\uparrow} d\omega \, G^{21}_{ab}(\omega) \: , \\
\int d\omega \, G^{22}_{ab}(\omega) &\rightarrow& \int_{C\downarrow} d\omega \, G^{22}_{ab}(\omega) \: .
\end{eqnarray}
\end{subequations}
When equal-time propagators appear the ordering of the annihilation and creation operators must be as in the starting Hamiltonian. Hence limits \eqref{eq:etlim11} and \eqref{eq:etlim22} must be taken in opposite ways. Since the operators in $G^{12}$ and $G^{21}$ anticommute the remaining two limits can be arbitrarily interpreted, as long as they are taken consistently.
\item Sum over all internal single-particle indices and integrate over all internal energies. External indices and energy refer to the Gorkov propagator being expanded.
\end{enumerate}

Once the expansions of the four one-body Gorkov propagators are written down, one can derive the corresponding expansions for the  self-energies by simply stripping off external propagation lines, e.g. to the term \eqref{eq:prop_strip} corresponds self-energy contribution \eqref{eq:spse}.
\begin{subequations}
\begin{eqnarray}
\label{eq:prop_strip}
 \hspace{1cm} \mbox{\diaggexa} \hspace{-1cm}
\\ \nonumber
\\ \nonumber
\\
 \hspace{1cm} \mbox{\diagseexam} \hspace{-1cm}
\label{eq:spse}
\end{eqnarray}
\end{subequations}

All self-energy contributions can be divided into two types: one-line \textit{reducible} and \textit{irreducible} self-energies.
Irreducible self-energies are constituted by diagrams that can not be separated into two parts by cutting one propagation line. For example diagram \eqref{eq:topeq1} is reducible while diagram  \eqref{eq:topneq} is irreducible. Irreducible contributions can be further divided into \textit{skeleton} and \textit{composed} diagrams. Skeleton (composed) self-energies are obtained by keeping, at a given order order, only those terms that cannot (can) be generated by successive insertions of irreducible self-energy terms of lower order. At first order, all diagrams are irreducible by definition. An example at second order is given by the two diagrams \eqref{eq:seredirred}: the first term \eqref{eq:seirred} is a skeleton diagram while the second self-energy contribution \eqref{eq:sered} can be generated by two successive insertions of the first-order term \eqref{eq:spse}.
\begin{subequations}
\label{eq:seredirred}
\begin{eqnarray}
 \nonumber \\
\label{eq:seirred}
\hspace{0.2cm} \mbox{\secondnapl} \hspace{-1cm} 
\\ \nonumber
\end{eqnarray}
\begin{eqnarray}
\hspace{-0.6cm} \mbox{\diagseexamred} \hspace{1cm}  \hspace{-0.8cm}
\label{eq:sered}
\end{eqnarray}
\end{subequations}

Once this distinction is made, one can demonstrate that the complete propagator expansion is generated by keeping irreducible skeleton self-energy diagrams only and by replacing accordingly in such diagrams all unperturbed propagators by \textit{dressed} ones. Dressed propagators are Green's functions that are solution of Gorkov's equations: their appearance in the self-energy expansion generates the self-consistency characterizing the method.

It thus follows that only irreducible skeleton self-energy diagrams with dressed propagators have to be computed. Such dressed propagators are depicted as solid double lines and are labeled by two indices as well as by an energy, just as for unperturbed ones, i.e.
\begin{subequations}
\begin{eqnarray}
\nonumber \\
\hspace{.6cm} G^{11}_{ab} (\omega) &\equiv& \hspace{1.2cm} \mbox{\diaggoo} \hspace{-2.7cm} \, ,
\\ \nonumber \\ \nonumber \\ \nonumber \\
G^{12}_{ab} (\omega) &\equiv& \hspace{1.2cm} \mbox{\diaggot} \hspace{-2.7cm} \, ,
\\ \nonumber \\ \nonumber \\ \nonumber \\
G^{21}_{ab} (\omega) &\equiv& \hspace{1.2cm} \mbox{\diaggto} \hspace{-2.7cm} \, ,
\\ \nonumber \\ \nonumber \\ \nonumber \\
G^{22}_{ab} (\omega) &\equiv& \hspace{1.2cm} \mbox{\diaggtt} \hspace{-2.7cm} \, .
\\ \nonumber \\ \nonumber
\end{eqnarray}
\label{eq:ggraphical}
\end{subequations}
\noindent Diagrammatic rules to compute irreducible self-energies are the same as for reducible ones, with the only difference that dressed propagators \eqref{eq:ggraphical} have to be used instead of unperturbed ones.

\subsection{Self-energies}
\label{sec:self_eva}

The present section addresses the derivation of first-and second-order self-energy diagrams.

\subsubsection{First order}
\label{sec:first_self_app}

The first normal contribution corresponds to the standard Hartree-Fock self-energy. It is depicted as
\begin{equation}
\hspace{.8cm}
\Sigma^{11 \, (1)}_{ab}(\omega) = \hspace{.8cm}
\mbox{\firstn} \hspace{-1.4cm} ,
\end{equation}
and reads
\begin{equation}
\label{eq:slefhf}
\Sigma^{11 \, (1)}_{ab}(\omega) =
 -i \int_{C \uparrow} \frac{d \omega'}{2 \pi} \sum_{cd} \bar{V}_{acbd} \,
G_{dc}^{11} (\omega') \, ,
\end{equation}
where the energy integral is to be performed in the upper half of the complex energy plane, according to the convention introduced in Rule 8.
Inserting the Lehmann form \eqref{eq:leh11} of the propagator one obtains
\begin{eqnarray}
\label{eq:self111}
\Sigma^{11 \, (1)}_{ab}(\omega)  &=& \displaystyle
-i \int_{C \uparrow} \frac{d \omega'}{2 \pi} \sum_{cd, k} \bar{V}_{acbd} \,
\frac{{\mU_{d}^k} \, \mU_{c}^{k *}}
{\omega'-\omega_k + i \eta}
\nonumber \\ &-& \displaystyle
i \int_{C \uparrow} \frac{d \omega'}{2 \pi} \sum_{cd, k} \bar{V}_{acbd} \,
\frac{\bar{\mV}_{d}^{k *} \, {\bar{\mV}_{c}^k}}{\omega'+\omega_k - i \eta}
\nonumber \\  &=& \displaystyle
\sum_{cd, k} \bar{V}_{acbd} \, \bar{\mV}_{d}^{k *} \, {\bar{\mV}_{c}^k} \: ,
\end{eqnarray}
where the residue theorem has been used, i.e. the first term, with $+i \eta$ in the denominator, contains no pole in the upper plane and thus cancels out. As in the standard case the Hartree-Fock self-energy is energy independent.

Similarly one computes the other normal self-energy term
\begin{equation}
\hspace{.8cm}
\Sigma^{22 \, (1)}_{ab}(\omega) = \hspace{.8cm}
\mbox{\firstnn} \hspace{-1.4cm} ,
\end{equation}
which reads
\begin{eqnarray}
\label{eq:self221}
\Sigma^{22 \, (1)}_{ab}(\omega)  &=&
-i \int_{C \downarrow} \frac{d \omega'}{2 \pi} \sum_{cd} \bar{V}_{\bar{b}\bar{d}\bar{a}\bar{c}} \,
G_{dc}^{22} (\omega')
\nonumber \\ &=& \displaystyle
-i \int_{C \downarrow} \frac{d \omega'}{2 \pi} \sum_{cd, k} \bar{V}_{\bar{b}\bar{d}\bar{a}\bar{c}} \,
\frac{{\mV_{d}^k} \, \mV_{c}^{k *}}
{\omega'-\omega_k + i \eta}
\nonumber \\ &-& \displaystyle
i \int_{C \downarrow} \frac{d \omega'}{2 \pi} \sum_{cd, k} \bar{V}_{\bar{b}\bar{d}\bar{a}\bar{c}} \,
\frac{\bar{\mU}_{d}^{k *} \, {\bar{\mU}_{c}^k}}{\omega'+\omega_k - i \eta}
\nonumber \\ &=& \displaystyle
- \sum_{cd, k} \bar{V}_{\bar{b}\bar{d}\bar{a}\bar{c}} \, {\mV_{d}^k} \, \mV_{c}^{k *}
\nonumber \\ &=& \displaystyle
- \sum_{cd, k} \bar{V}_{\bar{b}c\bar{a}d} \, {\bar{\mV}_{c}^k} \, \bar{\mV}_{d}^{k *}
\nonumber \\ &=& \displaystyle
- \Sigma^{11 \, (1)}_{\bar{b}\bar{a}}
\nonumber \\ &=& \displaystyle
- [\Sigma^{11 \, (1)}_{\bar{a}\bar{b}}]^*
\: .
\end{eqnarray}
The anomalous contributions to the self-energy at first order are
\begin{eqnarray}
\hspace{-.8cm}
&&\Sigma^{12 \, (1)}_{ab}(\omega) =  \hspace{.8cm}
\begin{tabular}{c}
\mbox{\firsta} \\ \\ \\
\end{tabular} \hspace{-1.7cm} ,
\\ \nonumber &&
\end{eqnarray}
\begin{equation}
\hspace{.8cm}
\Sigma^{21 \, (1)}_{ab}(\omega) =  \hspace{.8cm}
\begin{tabular}{c}
\\ \\
\mbox{\firstaa}
\end{tabular} \hspace{-1.7cm} ,
\end{equation}
and are written respectively as
\begin{eqnarray}
\label{eq:self121}
\Sigma_{ab}^{12 \, (1)} (\omega)&=&
 - \frac{i}{2} \int_{C \uparrow} \frac{d \omega'}{2 \pi} \sum_{cd} \bar{V}_{a\bar{b}c\bar{d}} \,
G_{cd}^{12} (\omega')
\nonumber \\ &=& \displaystyle
- \frac{i}{2} \int_{C \uparrow} \frac{d \omega'}{2 \pi} \sum_{cd, k} \bar{V}_{a\bar{b}c\bar{d}} \,
\frac{{\mU_{c}^k} \, \mV_{d}^{k *}}
{\omega'-\omega_k + i \eta}
\nonumber \\ &&- \displaystyle
 \frac{i}{2} \int_{C \uparrow} \frac{d \omega'}{2 \pi} \sum_{cd, k} \bar{V}_{a\bar{b}c\bar{d}} \,
\frac{\bar{\mV}_{c}^{k *} \, {\bar{\mU}_{d}^k}}{\omega'+\omega_k - i \eta}
\nonumber \\ &=& \displaystyle
 \frac{1}{2} \sum_{cd, k} \bar{V}_{a\bar{b}c\bar{d}} \, \bar{\mV}_{c}^{k *} \, {\bar{\mU}_{d}^k} \: ,
\end{eqnarray}
and
\begin{eqnarray}
\label{eq:self211}
\Sigma_{ab}^{21 \, (1)} (\omega)&=&
 - \frac{i}{2} \int_{C \uparrow} \frac{d \omega'}{2 \pi} \sum_{cd}  \bar{V}_{\bar{c}d\bar{a}b} \,
G_{cd}^{21} (\omega')
\nonumber \\ &=& \displaystyle
- \frac{i}{2} \int_{C \uparrow} \frac{d \omega'}{2 \pi} \sum_{cd, k} \bar{V}_{\bar{c}d\bar{a}b} \,
\frac{{\mV_{c}^k} \, \mU_{d}^{k *}}
{\omega'-\omega_k + i \eta}
\nonumber \\ &&- \displaystyle
 \frac{i}{2} \int_{C \uparrow} \frac{d \omega'}{2 \pi} \sum_{cd, k} \bar{V}_{\bar{c}d\bar{a}b} \,
\frac{\bar{\mU}_{c}^{k *} \, {\bar{\mV}_{d}^k}}{\omega'+\omega_k - i \eta}
\nonumber \\ &=& \displaystyle
 \frac{1}{2} \sum_{cd, k} \bar{V}_{\bar{c}d\bar{a}b} \, \bar{\mU}_{c}^{k *} \, {\bar{\mV}_{d}^k}
\nonumber \\ &=& \displaystyle
 \frac{1}{2} \sum_{cd, k} \bar{V}_{b\bar{a}c\bar{d}}^* \, \bar{\mU}_{d}^{k *} \, {\bar{\mV}_{c}^k}
\nonumber \\ &=& \displaystyle
[\Sigma_{ba}^{12 \, (1)}]^*
\: ,
\end{eqnarray}
where the same integration technique as in \eqref{eq:self111} has been used.

\subsubsection{Second order}
\label{sec:2oself}

Let us now proceed to the computation of the second-order contributions.
The first term is the standard second-order self-energy
\begin{eqnarray}
\hspace{.8cm}
\nonumber \\
\Sigma^{11 \, (2')}_{ab}(\omega) = \hspace{.8cm}
\mbox{\secondna} \hspace{-.2cm}
\\ \nonumber
\label{eq:fig_self1121}
\end{eqnarray}
reading
\begin{widetext}
\begin{eqnarray}
\label{eq:secfirform}
\Sigma_{ab}^{11 \, (2')} (\omega) &=& \frac{1}{2}
\int \frac{d \omega'}{2 \pi} \frac{d \omega''}{2 \pi}  d \omega'''
\sum_{cdefgh} \bar{V}_{aecf} \, \bar{V}_{dgbh} \,
G_{cd}^{11} (\omega') \, G_{fg}^{11} (\omega'') \, G_{he}^{11} (\omega''') \, \delta(\omega - \omega' -\omega'' + \omega''')
\nonumber \\ &=& \displaystyle \frac{1}{2}
\int \frac{d \omega'}{2 \pi} \frac{d \omega''}{2 \pi}
\sum_{cdefgh} \bar{V}_{aecf} \, \bar{V}_{dgbh} \,
G_{cd}^{11} (\omega') \, G_{fg}^{11} (\omega'') \, G_{he}^{11} (\omega'+\omega''-\omega) \, .
\end{eqnarray}
Notice that the minus sign coming from rule 4 is cancelled by a minus sign coming from the presence of a closed loop (rule 7).
The integrations over the two energy variables are performed in this case using two successive applications of the formula
\begin{eqnarray}
I(E) &=& \int_{-\infty}^{+\infty} \frac{d\, E'}{2\pi i} \left\{ \frac{F_1}{E' - f_1 + i\eta}
+ \frac{B_1}{E' - b_1 - i\eta} \right\}
\, \left\{ \frac{F_2}{E' -E - f_2 + i\eta}
+ \frac{B_2}{E' -E- b_2 - i\eta} \right\}
\nonumber \\ &=&
\left\{ \frac{F_1 B_2}{E - (f_1-b_2) + i\eta}
- \frac{F_2 B_1}{E+ (f_2 -b_1) - i\eta} \right\} \: .
\end{eqnarray}
The above integral, defined on the real axis, is computed by extending the integration to a large semicircle in the upper or lower complex half plane of $E'$ (this can be done since the integrand behaves as $|E'|^{-2}$ for $|E'| \rightarrow \infty$ and this branch do not contribute to the integral) and then by using the residue theorem. Of the four terms, two have poles in the same half plane and yield zero as the contour can be closed in the other half.
Applying this formula to the integral \eqref{eq:secfirform} one obtains
\begin{eqnarray}
\label{eq:self112a}
\Sigma_{ab}^{11 \, (2')} (\omega) &=& \displaystyle
-\frac{1}{2}
\int \frac{d \omega'}{2 \pi i} \frac{d \omega''}{2 \pi i}
\sum_{cdefgh,k_1k_2k_3}   \bar{V}_{aecf} \, \bar{V}_{dgbh}
\, \left\{
\frac{\mU_{c}^{k_1} \, \mU_{d}^{k_1 *}}
{\omega'-\omega_{k_1} + i \eta} + \frac{\bar{\mV}_{c}^{k_1 *} \, {\bar{\mV}_{d}^{k_1}}}{\omega'+\omega_{k_1} - i \eta}
\right\}
\nonumber \\ &\times& \displaystyle
\left\{
\frac{{\mU_{f}^{k_2}} \, \mU_{g}^{k_2 *}}
{\omega''-\omega_{k_2} + i \eta} + \frac{\bar{\mV}_{f}^{k_2 *} \, {\bar{\mV}_{g}^{k_2}}}{\omega''+\omega_{k_2} - i \eta}
\right\}
\, \left\{
\frac{{\mU_{h}^{k_3}} \, \mU_{e}^{k_3 *}}
{\omega'+\omega''-\omega-\omega_{k_3} + i \eta} +
\frac{\bar{\mV}_{h}^{k_3 *} \, {\bar{\mV}_{e}^{k_3}}}{\omega'+\omega''-\omega+\omega_{k_3} - i \eta}
\right\}
\nonumber \\ &=& \displaystyle
\frac{1}{2}
\sum_{cdefgh,k_1k_2k_3}  \bar{V}_{aecf} \, \bar{V}_{dgbh}
\, \left\{
\frac{{\mU_{c}^{k_1}} \, \mU_{d}^{k_1 *} \, {\mU_{f}^{k_2}} \, \mU_{g}^{k_2 *}
\, \bar{\mV}_{h}^{k_3 *}  \, {\bar{\mV}_{e}^{k_3}}}
{\omega-(\omega_{k_1} + \omega_{k_2} + \omega_{k_3}) + i \eta}
+ \frac{\bar{\mV}_{c}^{k_1 *} \, {\bar{\mV}_{d}^{k_1}} \,\bar{\mV}_{f}^{k_2 *} \, {\bar{\mV}_{g}^{k_2}}
\, {\mU_{h}^{k_3}} \,\mU_{e}^{k_3 *}}
{\omega+(\omega_{k_3} + \omega_{k_1} + \omega_{k_2}) - i \eta}
\right\} \: .
\end{eqnarray}
With the same technique one can evaluate all other terms contributing to the second-order self-energy. One has
\begin{eqnarray}
\label{eq:fig_self1122}
\hspace{.8cm}
\nonumber \\ \nonumber \\
\Sigma^{11 \, (2'')}_{ab}(\omega) = \hspace{.8cm}
\mbox{\secondnb} \: ,
\\ \nonumber \\ \nonumber
\end{eqnarray}
reading
\begin{eqnarray}
\label{eq:self112b}
\Sigma_{ab}^{11 \, (2'')} (\omega) &=& -
\int \frac{d \omega'}{2 \pi} \frac{d \omega''}{2 \pi}
\sum_{cdefgh} \bar{V}_{aecf} \, \bar{V}_{d\bar{g}b\bar{h}} \,
G_{cd}^{11} (\omega') \, G_{fh}^{12} (\omega'') \, G_{ge}^{21} (\omega'+\omega''-\omega) \,
 \\ &=& \displaystyle
- \sum_{cdefgh,k_1k_2k_3}  \bar{V}_{aecf} \, \bar{V}_{d\bar{g}b\bar{h}} \,
\left\{
\frac{\mU_{c}^{k_1} \, \mU_{d}^{k_1 *} \, \mU_{f}^{k_2} \, \mV_{h}^{k_2 *}
 \, \bar{\mU}_{g}^{k_3 *} \, {\bar{\mV}_{e}^{k_3}}}
{\omega-(\omega_{k_1} + \omega_{k_2} + \omega_{k_3}) + i \eta} +
\frac{\bar{\mV}_{c}^{k_1 *} \, {\bar{\mV}_{d}^{k_1}} \, \bar{\mV}_{f}^{k_2 *} \, {\bar{\mU}_{h}^{k_2}}
 \,{\mV_{g}^{k_3}} \, \mU_{e}^{k_3 *}}
{\omega+(\omega_{k_3} + \omega_{k_1} + \omega_{k_2}) - i \eta}
\right\} \nonumber \, .
\end{eqnarray}
The two diagrams of the other normal self-energy $\Sigma^{22}$ are respectively
\begin{eqnarray}
\hspace{.8cm}
\nonumber \\ \nonumber \\
\Sigma^{22 \, (2')}_{ab}(\omega) = \hspace{.8cm}
\mbox{\secondnaa} \: ,
\\ \nonumber \\ \nonumber
\end{eqnarray}
yielding
\begin{eqnarray}
\label{eq:self222a}
\Sigma_{ab}^{22 \, (2')} (\omega) &=& \frac{1}{2}
\int \frac{d \omega'}{2 \pi} \frac{d \omega''}{2 \pi}
\sum_{cdefgh} \bar{V}_{\bar{c}\bar{f}\bar{a}\bar{e}} \, \bar{V}_{\bar{b}\bar{h}\bar{d}\bar{g}} \,
G_{cd}^{22} (\omega') \, G_{fg}^{22} (\omega'') \, G_{he}^{22} (\omega'+\omega''-\omega) \,
\nonumber \\  &=& \displaystyle
\frac{1}{2}
\sum_{cdefgh,k_1k_2k_3} \bar{V}_{\bar{c}\bar{f}\bar{a}\bar{e}} \, \bar{V}_{\bar{b}\bar{h}\bar{d}\bar{g}} \, \left\{
\frac{{\mV_{c}^{k_1}} \, \mV_{d}^{k_1 *} \, {\mV_{f}^{k_2}} \, \mV_{g}^{k_2 *}
\, \bar{\mU}_{h}^{k_3 *} \, {\bar{\mU}_{e}^{k_3}}}
{\omega-(\omega_{k_1} + \omega_{k_2} + \omega_{k_3}) + i \eta}
+ \frac{\bar{\mU}_{c}^{k_1 *} \, {\bar{\mU}_{d}^{k_1}} \, \bar{\mU}_{f}^{k_2 *} \, {\bar{\mU}_{g}^{k_2}}
\, {\mV_{h}^{k_3}} \, \mV_{e}^{k_3 *}}
{\omega+(\omega_{k_3} + \omega_{k_1} + \omega_{k_2}) - i \eta}
\right\} \: ,
\end{eqnarray}
and
\begin{eqnarray}
\hspace{.8cm}
\nonumber \\ \nonumber \\
\Sigma^{22 \, (2'')}_{ab}(\omega) = \hspace{.8cm}
\mbox{\secondnbb}  \: ,
\\ \nonumber \\ \nonumber
\end{eqnarray}
reading
\begin{eqnarray}
\label{eq:self222b}
\Sigma_{ab}^{22 \, (2'')} (\omega) &=& -
\int \frac{d \omega'}{2 \pi} \frac{d \omega''}{2 \pi}
\sum_{cdefgh} \bar{V}_{\bar{c}f\bar{a}e} \, \bar{V}_{\bar{b}\bar{h}\bar{d}\bar{g}} \,
G_{cd}^{22} (\omega') \, G_{eg}^{12} (\omega'') \, G_{hf}^{21} (\omega'+\omega''-\omega) \,
 \\ &=& \displaystyle
- \sum_{cdefgh,k_1k_2k_3}  \bar{V}_{\bar{c}f\bar{a}e} \, \bar{V}_{\bar{b}\bar{h}\bar{d}\bar{g}} \,
\left\{
\frac{{\mV_{c}^{k_1}} \, \mV_{d}^{k_1 *} \, {\mU_{e}^{k_2}} \, \mV_{g}^{k_2 *}
  \, \bar{\mU}_{h}^{k_3 *} \, {\bar{\mV}_{f}^{k_3}}}
{\omega-(\omega_{k_1} + \omega_{k_2} + \omega_{k_3}) + i \eta} +
\frac{\bar{\mU}_{c}^{k_1 *} \, {\bar{\mU}_{d}^{k_1}} \, \bar{\mV}_{e}^{k_2 *} \, {\bar{\mU}_{g}^{k_2}}
 \, {\mV_{h}^{k_3}} \, \mU_{f}^{k_3 *}}
{\omega+(\omega_{k_3} + \omega_{k_1} + \omega_{k_2}) - i \eta}
\right\} \nonumber \, .
\end{eqnarray}
The first of the anomalous self-energy is
\begin{eqnarray}
\hspace{.8cm}
\nonumber \\ \nonumber \\
\Sigma^{12 \, (2')}_{ab}(\omega) = \hspace{.8cm}
\begin{tabular}{c}
\mbox{\secondaa} \\
\end{tabular}  \: ,
\\ \nonumber \\ \nonumber  \\ \nonumber
\end{eqnarray}
for what concerns the first contribution, which reads
\begin{eqnarray}
\label{eq:self122a}
\Sigma_{ab}^{12 \, (2')} (\omega) &=&
- \int \frac{d \omega'}{2 \pi} \frac{d \omega''}{2 \pi}
\sum_{cdefgh} \bar{V}_{aecf} \, \bar{V}_{h\bar{b}g\bar{d}} \,
G_{cd}^{12} (\omega') \, G_{fh}^{11} (\omega'') \, G_{ge}^{11} (\omega'+\omega''-\omega) \,
\\ &=& \displaystyle
- \sum_{cdefgh,k_1k_2k_3} \bar{V}_{aecf} \, \bar{V}_{h\bar{b}g\bar{d}} \,
\left\{
\frac{{\mU_{c}^{k_1}} \, \mV_{d}^{k_1 *} \, {\mU_{f}^{k_2}} \, \mU_{h}^{k_2 *}
\, \bar{\mV}_{g}^{k_3 *} \, {\bar{\mV}_{e}^{k_3}}}
{\omega-(\omega_{k_1} + \omega_{k_2} + \omega_{k_3}) + i \eta} +
\frac{\bar{\mV}_{c}^{k_1 *} \, {\bar{\mU}_{d}^{k_1}} \, \bar{\mV}_{f}^{k_2*} \, {\bar{\mV}_{h}^{k_2}}
\, {\mU_{g}^{k_3}} \, \mU_{e}^{k_3 *}}
{\omega+(\omega_{k_3} + \omega_{k_1} + \omega_{k_2}) - i \eta}
\right\} \, , \nonumber
\end{eqnarray}
and
\begin{eqnarray}
\hspace{.8cm}
\nonumber \\ \nonumber \\
\Sigma^{12 \, (2'')}_{ab}(\omega) = \hspace{.8cm}
\begin{tabular}{c}
\mbox{\secondab}
\\
\end{tabular}  \: ,
\\ \nonumber \\ \nonumber
\end{eqnarray}
for the second contribution yielding
\begin{eqnarray}
\label{eq:self122b}
\Sigma_{ab}^{12 \, (2'')} (\omega) &=& \frac{1}{2}
\int \frac{d \omega'}{2 \pi} \frac{d \omega''}{2 \pi}
\sum_{cdefgh} \bar{V}_{aecf} \, V_{\bar{h}\bar{b}\bar{g}\bar{d}} \,\,
G_{cd}^{12} (\omega') \, G_{fg}^{12} (\omega'') \, G_{he}^{21} (\omega'+\omega''-\omega) \,
 \\ &=& \displaystyle
\frac{1}{2}
\sum_{cdefgh,k_1k_2k_3} \bar{V}_{aecf} \, V_{\bar{h}\bar{b}\bar{g}\bar{d}} \,
\left\{
\frac{{\mU_{c}^{k_1}} \, \mV_{d}^{k_1 *} \, {\mU_{f}^{k_2}} \, \mV_{g}^{k_2 *}
\, \bar{\mU}_{h}^{k_3 *} \, {\bar{\mV}_{e}^{k_3}}}
{\omega-(\omega_{k_1} + \omega_{k_2} + \omega_{k_3}) + i \eta} +
\frac{\bar{\mV}_{c}^{k_1 *} \, {\bar{\mU}_{d}^{k_1}} \, \bar{\mV}_{f}^{k_2*} \, {\bar{\mU}_{g}^{k_2}}
\, {\mV_{h}^{k_3}} \, \mU_{e}^{k_3 *}}
{\omega+(\omega_{k_3} + \omega_{k_1} + \omega_{k_2}) - i \eta}
\right\} \, , \nonumber
\end{eqnarray}
Finally
\begin{eqnarray}
\hspace{.8cm}
\nonumber \\ \nonumber \\
\Sigma^{21 \, (2')}_{ab}(\omega) = \hspace{.8cm}
\begin{tabular}{c}
\mbox{\secondaaa}
\end{tabular} \: ,
\\ \nonumber \\ \nonumber
\end{eqnarray}
reads as
\begin{eqnarray}
\label{eq:self212a}
\Sigma_{ab}^{21 \, (2')} (\omega) &=&
- \int \frac{d \omega'}{2 \pi} \frac{d \omega''}{2 \pi}
\sum_{cdefgh} \bar{V}_{\bar{c}f\bar{a}e} \, \bar{V}_{gdhb} \,
G_{cd}^{21} (\omega') \, G_{eg}^{11} (\omega'') \, G_{hf}^{11} (\omega'+\omega''-\omega) \,
\\ &=& \displaystyle
- \sum_{cdefgh,k_1k_2k_3} \bar{V}_{\bar{c}f\bar{a}e} \, \bar{V}_{gdhb}  \,
\left\{
\frac{{\mV_{c}^{k_1}} \, \mU_{d}^{k_1 *} \, {\mU_{e}^{k_2}} \, \mU_{g}^{k_2 *}
\, \bar{\mV}_{h}^{k_3 *} \, {\bar{\mV}_{f}^{k_3}}}
{\omega-(\omega_{k_1} + \omega_{k_2} + \omega_{k_3}) + i \eta} +
\frac{\bar{\mU}_{c}^{k_1 *} \, {\bar{\mV}_{d}^{k_1}} \, \bar{\mV}_{e}^{k_2 *} \, {\bar{\mV}_{g}^{k_2}}
\, {\mU_{h}^{k_3}} \, \mU_{f}^{k_3 *}}
{\omega+(\omega_{k_3} + \omega_{k_1} + \omega_{k_2}) - i \eta}
\right\} \, . \nonumber
\end{eqnarray}
while
\begin{eqnarray}
\hspace{.8cm}
\nonumber \\
\Sigma^{21 \, (2'')}_{ab}(\omega) = \hspace{.8cm}
\begin{tabular}{c}
\mbox{\secondabb}
\end{tabular} \: ,
\\ \nonumber \\ \nonumber
\end{eqnarray}
is expressed as
\begin{eqnarray}
\label{eq:self212b}
\Sigma_{ab}^{21 \, (2'')} (\omega) &=& \frac{1}{2}
\int \frac{d \omega'}{2 \pi} \frac{d \omega''}{2 \pi}
\sum_{cdefgh} \bar{V}_{\bar{c}f\bar{a}e} \, \bar{V}_{\bar{g}d\bar{h}b} \,
G_{cd}^{21} (\omega') \, G_{eh}^{12} (\omega'') \, G_{gf}^{21} (\omega'+\omega''-\omega) \,
 \\ &=& \displaystyle
\frac{1}{2}
\sum_{cdefgh,k_1k_2k_3} \bar{V}_{\bar{c}f\bar{a}e} \, \bar{V}_{\bar{g}d\bar{h}b} \,
\left\{
\frac{{\mV_{c}^{k_1}} \, \mU_{d}^{k_1 *} \, {\mU_{e}^{k_2}} \, \mV_{h}^{k_2 *}
\, \bar{\mU}_{g}^{k_3 *} \, {\bar{\mV}_{f}^{k_3}}}
{\omega-(\omega_{k_1} + \omega_{k_2} + \omega_{k_3}) + i \eta} +
\frac{\bar{\mU}_{c}^{k_1 *} \, {\bar{\mV}_{d}^{k_1}} \, \bar{\mV}_{e}^{k_2 *} \, {\bar{\mU}_{h}^{k_2}}
\, {\mV_{g}^{k_3}} \, \mU_{f}^{k_3 *}}
{\omega+(\omega_{k_3} + \omega_{k_1} + \omega_{k_2}) - i \eta}
\right\} \nonumber \, .
\end{eqnarray}
\end{widetext}


\section{$J^{\Pi} = 0^+$ states}
\label{appli0plus}

The present section specifies the complete set of equations to $J^{\Pi} = 0^+$ states.

\subsection{Time-reversal invariant systems}
\label{sec:tris}

Let us define the time-reversal operator as
\begin{equation}
\label{eq:trop}
\mathcal{T} \equiv e^{i \pi S_y} K \, ,
\end{equation}
where $K$ is an operator that associates to a wavefunction its complex conjugate and $S_y$ is the $y$-axis spin-projection operator of the N-body system. The time-reversal operator is antiunitary (unitary and antilinear) and displays the following properties
\begin{subequations}
\label{eq:propt}
\begin{eqnarray}
\mathcal{T}^{\dagger} \mathcal{T} &=& 1 \: ,
\\
\mathcal{T}^2 &=& (-1)^N \label{eq:propt2} \: ,
\\
( a | \left. \mathcal{T} \right) | b ) &=& \left[ ( a | \left( \mathcal{T} \right. | b ) \right]^* \: ,
\\
( a | \left( \mathcal{T}^{\dagger} \right. | b ) &=& ( b | \left( \mathcal{T} \right. | a ) \: .
\end{eqnarray}
\end{subequations}
One can also introduce the time reversal operator acting in Fock space. This is done by specifying its transformation
rules on standard creation and annihilation operators defined on the tensorial product of spatial (represented by the vector $\vec{r}$), spin (represented by the spin projection $m_{\sigma}$) and isospin (represented by the isospin projection $q$) one-body Hilbert spaces
\begin{subequations}
\label{eq:propt3}
\begin{align}
\mathcal{T}^\dagger\,a^\dagger_{\vec{r}m_{\sigma} q}\,\mathcal{T}=&2 \, m_{\sigma}\,a^\dagger_{\vec{r}\tilde{m}_{\sigma}q}\,, \\
\mathcal{T}\,a^\dagger_{\vec{r} m_{\sigma} q}\,\mathcal{T}^\dagger=
&2 \, \tilde{m}_{\sigma}\,a^\dagger_{\vec{r}\tilde{m}_{\sigma}q}\,,\label{eq:phase}\\
\mathcal{T}^\dagger\,a_{\vec{r}m_{\sigma}
q}\,\mathcal{T}=&2 \, m_{\sigma}\,a_{\vec{r}\tilde{m}_{\sigma}q}\,, \\
\mathcal{T}\,a_{\vec{r}m_{\sigma} q}\,\mathcal{T}^\dagger=
&2 \, \tilde{m}_{\sigma}\,a_{\vec{r}\tilde{m}_{\sigma}q}\,,\label{eq:phase2}
\end{align}
as well as on the particle vacuum
\begin{align}
\mathcal{T}\,|0\rangle=&|0\rangle \, ,\label{eq:phase3}
\end{align}
\end{subequations}
where the notation $\tilde{m}_{\sigma}=-m_{\sigma}$ has been used. Given Eq. \eqref{eq:propt3} it is easy to prove that for any basis $\{ a_k^{\dagger} \}$ {\it closed} under time-reversal, defining
\begin{equation}
\bar{a}^\dagger_k \equiv \mathcal{T}\,a^\dagger_k\,\mathcal{T}^\dagger \, ,
\end{equation}
provides a partner basis of the type \eqref{eq:gen_aad}. Accordingly we define the action of $\mathcal{T}$ on one- and two-particle Dirac ket and bra as
\begin{subequations}
\begin{eqnarray}
\mathcal{T} \, | a ) \equiv& \eta_a \, | \tilde{a} ) \qquad  (  a | \, \mathcal{T}^{\dagger}& \equiv \eta_a \,  (  \tilde{a} | \: ,
\\
\mathcal{T} \, | ab )
\equiv& \, \eta_a \, \eta_b \, | \tilde{a} \tilde{b} ) \qquad
(  ab | \, \mathcal{T}^{\dagger}\,& \equiv \eta_a \,  \eta_b \, (  \tilde{a} \tilde{b} |
\: .
\end{eqnarray}
\end{subequations}
It follows that for kinetic energy, which fulfills $\mathcal{T}^{\dagger} \, T \, \mathcal{T} = T$,
\begin{eqnarray}
\label{eq:trikin}
T_{\bar{a}\bar{b}}
&=& (  a | \, \mathcal{T}^{\dagger}\, ) \, T \, ( \mathcal{T}  | b )
\nonumber \\ \displaystyle
&=& \left[(  a | \, (\mathcal{T}^{\dagger}\,  T \, \mathcal{T} | b )\right]^*
\nonumber \\ \displaystyle
&=& \left[(  a | \, T | b )\right]^*
\nonumber \\ \displaystyle
&=& T_{ab}^* \: ,
\end{eqnarray}
and similarly for time-reversal invariant interactions, i.e. $\mathcal{T}^{\dagger} \, V \, \mathcal{T} = V$,
\begin{eqnarray}
\label{eq:triint}
\bar{V}_{\bar{a}\bar{b}\bar{c}\bar{d}}
&=& (  ab | \, \mathcal{T}^{\dagger}\, ) V \, ( \mathcal{T}  \{\, | cd ) -  | dc )\} 
\nonumber \\ \displaystyle
&=& \left[(  ab | \, (\mathcal{T}^{\dagger}\,  V \, \mathcal{T}  \{\, | cd ) -  | dc )\} \right]^*
\nonumber \\ \displaystyle
&=& \left[(  ab | \, V \, \{ \, | cd ) -  | dc )\} \right]^*
\nonumber \\ \displaystyle
&=& \bar{V}_{abcd}^* \: .
\end{eqnarray}
Considering a time-reversal invariant system, i.e. a reference state $| \Psi_0 \rangle$ satisfying $\mathcal{T} \, | \Psi_0 \rangle = | \Psi_0 \rangle$ and $\langle \Psi_0 | \mathcal{T}^{\dagger} )= \langle \Psi_0 |$, one can prove, using property \eqref{eq:propt}, that in this particular case the anomalous density matrix \eqref{eq:aobdm} is Hermitian
\begin{eqnarray}
\label{eq:rhotildeher}
\notag\tilde{\rho}_{ab} &=& \langle \Psi_0 | \bar{a}_b \, a_a  | \Psi_0 \rangle
\nonumber \\ \nonumber
&=& \langle \Psi_0 | \, \mathcal{T}^{\dagger} \bigr) \, \bigr(\mathcal{T} \,
a_b
\, \mathcal{T}^{\dagger} \,a_a \,  | \Psi_0 \rangle \\
\notag&=& - \left[\langle \Psi_0 | \, a_b \, \bigr(\mathcal{T}^{\dagger} \, \mathcal{T}^{2} \, a_a
\,
\mathcal{T}^{\dagger} \, \mathcal{T} \, | \Psi_0 \rangle\right]^{\ast}\\
\notag&=& \langle \Psi_0 | \, \bar{a}_a \, a_{b} \, | \Psi_0 \rangle^{\ast} \\
&=&\tilde{\rho}^{\dagger}_{ab} \, .
\end{eqnarray}

\subsection{Single-particle basis}

In the remaining of the present section the many-body system under study is assumed to be in a $J^{\Pi}=0^+$ state, where the parity is $\Pi = (-1)^L$. A possible choice for labeling single-particle basis states in this context is $a~\equiv~(n,\ell,j,m,q)$, where $n$ represents the principal quantum number, $\ell$ is the orbital angular momentum, $j$ is the total angular momentum, $m$ is the projection of the total angular momentum along the $z$ axis and $q$ is the isospin projection. The spin $\sigma$ is omitted from the single-particle label because trivially $\sigma=1/2$ for all nucleons.
A choice that will appear to be more convenient below consists of labeling single-particle states according to $a~\equiv~(n,\pi,j,m,q)$, where the parity $\pi = (-1)^{\ell}$ substitutes the orbital angular momentum. Since for a given $j$, $\ell = j \pm \frac{1}{2}$ are the only possible values, there exists a one-to-one correspondence between $\ell$ and $\pi$.

Different phase conventions exist to define single-particle states in such a context. In the present work, spinors are written as
\begin{eqnarray}
\label{eq:sphericalstates}
( \, \vec{r} \,| \, a ) &=& \varphi_a (\vec{r} \, q)
\nonumber \\ \displaystyle
&=& \frac{u_{n\ell j} (r\,q)}{r} \, \sum_{m_\ell
m_{\sigma}} \, Y_{\ell}^{m_\ell}(\hat r)
  \, C^{j m}_{\ell\, m_\ell\, 1/2 \, m_{\sigma}}
  \, | m_{\sigma} )
\nonumber \\ \displaystyle
&\equiv& \frac{u_{n\ell j} (r q)}{r} \,
\Omega_{\ell j m}(\hat r)  \, ,
\end{eqnarray}
where $C$ denotes a Clebsch-Gordan coefficient according to
\begin{equation}
C^{JM}_{j_1 m_1 j_2 m_2} \equiv \langle j_1 m_1 j_2 m_2 | JM \rangle \: .
\end{equation}
The $\Omega_{j \ell m} (\hat r)$ are spherical spinors that recouple the angular
part of the wavefunction to spin-1/2 spinors. They fulfill
\begin{subequations}
\begin{eqnarray}
\vec{J}\,^2 \, \Omega_{j \ell m} (\hat{r})
& = & \hbar^2 \, j (j+1) \, \Omega_{j \ell m} (\hat{r}) \,, \\
\vec{L}\,^2 \, \Omega_{j \ell m} (\hat{r})
& = & \hbar^2 \, \ell (\ell+1) \, \Omega_{j \ell m} (\hat{r}) \, , \\
\vec{S}\,^2 \, \Omega_{j \ell m} (\hat{r}) & = & \hbar^2 \, \sigma
(\sigma+1) \, \Omega_{j \ell m} (\hat{r}) \nonumber \\ \displaystyle \label{eq:spispi}
&=& \frac{3}{4} \, \hbar^2 \, \Omega_{j \ell m} (\hat{r}) \,, \\
J_z \, \Omega_{j \ell m} (\hat{r}) & = & \hbar \, m \,
\Omega_{j \ell m} (\hat{r}) \, .
\end{eqnarray}
\end{subequations}
Spherical spinors are orthonormal, i.e.
\begin{eqnarray}
&& \int \limits_{0}^{2\pi} \! \rmd \varphi \int \limits_{0}^{\pi} \!
\rmd \vartheta \, \sin(\vartheta) \, \Omega_{j\ell m}^\dagger
(\vartheta, \varphi)  \, \Omega_{j' \ell ' m'} (\vartheta,
\varphi)
= \delta_{j j'} \, \delta_{\ell \ell '} \, \delta_{m m'} \: , \nonumber
\end{eqnarray}
and fulfill
\begin{subequations}
\begin{eqnarray}
\label{eq:sphere:coupling1} &&\sum_{m=-j}^{j} \Omega_{j \ell m}^\dagger
(\hat{r}) \, \Omega_{j \ell m} (\hat{r}) = \frac{2j+1}{4\pi} \,,
\\
\label{eq:sphere:coupling2} &&\sum_{m=-j}^{j} \Omega_{j \ell m}^\dagger
(\hat{r}) \, \vec{\sigma} \, \Omega_{j \ell m} (\hat{r}) = 0 \,.
\end{eqnarray}
\end{subequations}
As shown by Eq. \eqref{eq:spispi}, the total spin $\sigma$ is a good quantum number while its projection is
\emph{not}, since single-particle states mix $m_{\sigma} = \pm \frac{1}{2}$. If the time-reversal operator \eqref{eq:trop} is applied to a state $a$ belonging to this single-particle basis one can easily prove, using
\begin{subequations}
\begin{eqnarray}
{Y_{\ell}^{m_\ell}}^\ast(\hat r) &=&
(-1)^{m_\ell}\,Y_{\ell}^{-m_\ell}(\hat r)\,
 \\ \displaystyle
2m_{\sigma} &=&
(-1)^{m_{\sigma}-\hfb}\,,
\end{eqnarray}
\end{subequations}
as well as standard properties of Clebsh-Gordan coefficients, that
\begin{eqnarray}
\label{eq:def_eta}
(\mathcal{T} \, \varphi)_{n\ell j m} (\vec{r}\, q)
&=& \eta_{\ell j m} \, \varphi_{n\ell j -m} (\vec{r}\, q) \, ,
\end{eqnarray}
where $\eta_{\ell j m} \equiv (-1)^{\ell-j-m}$. Equation~\eqref{eq:def_eta} demonstrates that time-reversal operation connects basis state $a$, up to a phase, to state $\tilde{a} \equiv (n,\ell,j,-m,q)$ and thus constitutes an anti-unitary transformation that can be employed to define the partner basis $\{\bar{a}^{\dagger}\}$ used to introduce  Gorkov Green's functions. The creation and annihilation operators introduced in Eq. \eqref{eq:gen_aad} in this case take the form
\begin{subequations}
\begin{eqnarray}
\label{eq:gen_aad_ell}
\bar{a}_{n \ell j m q}^{\dagger} &\equiv& \eta_{\ell j m} \, a_{n \ell j -m q}^{\dagger}\, ,
\\ \displaystyle
\bar{a}_{n \ell j m q} &\equiv& \eta_{\ell j m} \, a_{n \ell j -m q} \: .
\end{eqnarray}
\end{subequations}
Equivalently, one obtains $\tilde{a} \equiv (n,\pi,j,-m,q)$ if parity is chosen to label single-particle basis (which will be our choice in practice), such that
\begin{subequations}
\begin{eqnarray}
\label{eq:gen_aad_pi}
\bar{a}_{n \pi j m q}^{\dagger} &\equiv& \eta_{\pi j m} \, a_{n \pi j -m q}^{\dagger}\, ,
\\ \displaystyle
\bar{a}_{n \pi j m q} &\equiv& \eta_{\pi j m} \, a_{n \pi j -m q} \: ,
\end{eqnarray}
\end{subequations}
with $\eta_{\pi j m} \equiv \pi \, (-1)^{j+m}$. By including $\eta_{\pi j m}$ in the definition of the building blocks of the theory (density matrices, propagators, etc ...) one can exploit symmetry properties associated with time-reversal invariance that lead to a simplification of the formalism.

\subsection{Block-diagonal structure of propagators}

It is easy to prove that the operator $a^{\dagger}_{n \ell j m q}$ ($a^{\dagger}_{n \pi j m q}$) is the $m^{\text{th}}$ component of an irreducible tensor of rank $j$ and that the corresponding annihilation operator transforms contragrediently, implying that $(-1)^{m} a_{j -m}$ is also the $m^{\text{th}}$ component of an irreducible tensor of rank $j$. Starting from such a property, one can demonstrate that, in addition to being diagonal in isospin space as only proton-proton and neutron-neutron pairing is considered here, Gorkov propagators possess a block-diagonal structure relative to quantum numbers $\pi$ ($\ell$) and $j$ and are independent on $m$, i.e.
\begin{eqnarray}
\label{eq:bdform_g_gen}
G^{g_1 g_2}_{ab}(\omega) &=& G^{g_1 g_2}_{n_a j_a \ell_a m_a q_a n_b j_b \ell_b m_b q_b}(\omega)
\nonumber \\ \displaystyle &=&
G^{g_1 g_2}_{n_a m_a \alpha n_b m_b \beta}(\omega)
\nonumber \\ \displaystyle &\equiv&
\delta_{m_am_b} \delta_{\alpha \beta} \, G^{g_1 g_2 \, [\alpha]}_{n_a n_b}(\omega) \, ,
\end{eqnarray}
where the notation $\alpha \equiv (j_a, \pi_a, q_a)$, i.e. $a = (n_a, \alpha, m_a)$, has been introduced. Similarly, one writes unperturbed propagators as
\begin{equation}
G^{g_1 g_2 \, (0)}_{ab}(\omega) \equiv
\delta_{m_am_b} \delta_{\alpha \beta} \, G^{g_1 g_2 \, [\alpha] \, (0)}_{n_a n_b}(\omega) \,  .
\end{equation}
The Lehmann representation also reflects the block-diagonal form of the propagators. In particular, there exist selection rules associated with label $k=(n_k, \kappa', m_k)$, with $\kappa'\equiv(j_k,\pi_k,Q_k)$, characterizing many-body states $| \Psi_k \rangle$ introduced in Eq. \eqref{eq:kapp}.
Considering the definition of the spectroscopic amplitudes \eqref{eq:specg} and \eqref{eq:specgb} and applying Wigner-Eckart theorem, one finds
\begin{subequations}
\label{eq:bdform_uv_all}
\begin{eqnarray}
\label{eq:bdform_u_bar}
\mU_{a}^{k} &=& \langle \Psi_0 | a_{a} | \Psi_k \rangle
\nonumber \\ \displaystyle
&=& (-1)^{m_a} \, C^{00}_{j_k m_k j -m_a} \langle \Psi_0 || a_{n_a \alpha} || \Psi_{n_k \kappa} \rangle
\nonumber \\ \displaystyle
&=& \delta_{j_k j_a} \, \delta_{m_k m_a} \frac{(-1)^{-j_a}}{\sqrt{2j_a+1}} \, \langle \Psi_0 || a_{n_a \alpha} || \Psi_{n_k \kappa} \rangle
\nonumber \\ \displaystyle
&\equiv& \delta_{\kappa \alpha} \, \delta_{m_k m_a}  \, \mU_{n_a \, [\alpha]}^{n_k} \: ,
\end{eqnarray}
where $\kappa\equiv(j_k,\pi_k,Q_k-Q_0)$, with $Q_0$ being the isospin projection of $| \Psi_0 \rangle$. It is assumed here that, analogously to angular momentum, $| \Psi_0 \rangle$ has a good isospin projection and that $Q_k$ is determined by the isospin projection of the creation/annihilation operator acting on $| \Psi_0 \rangle$, i.e. $\delta_{Q_k-Q_0, q_a}$.
From parity conservation follows $\delta_{\pi_k \pi_a}$. As the spin of $| \Psi_k \rangle$ can be different from $1/2$, the fact that $\pi_k = \pi_a$ does not imply $\ell_k = \ell_a$, hence the single-particle basis is labeled by $a=(n_a,\pi_a,j_a,m_a,q_a)$ from now on.
Similarly one has
\begin{align}
\label{eq:bdform_u}
\bar{\mU}_{a}^{k} &= \delta_{j_k j_a} \, \delta_{m_k -m_a} \, \frac{(-1)^{\ell_a+m_a}}{\sqrt{2j_a+1}} \, \langle \Psi_0 || a_{n \alpha} || \Psi_{n_k \kappa} \rangle
\nonumber \\ \displaystyle
&= \delta_{\kappa \alpha} \, \delta_{m_k -m_a} \, \eta_{a} \, \mU_{n_a \, [\alpha]}^{n_k} \: ,
\end{align}
\begin{align}
\label{eq:bdform_v_bar}
\mV_{a}^{k} &= \delta_{j_k j_a} \, \delta_{m_k m_a} \, \frac{(-1)^{\ell_a}}{\sqrt{2j_a+1}} \, \langle \Psi_0 || a_{n_a \alpha}^{\dagger} || \Psi_{n_k \kappa} \rangle
\nonumber \\ \displaystyle
&\equiv \delta_{\kappa \alpha} \, \delta_{m_k m_a} \, \mV_{n_a \, [\alpha]}^{n_k} \: , \\ \nonumber \\
\label{eq:bdform_v}
\bar{\mV}_{a}^{k} &= \delta_{j_kj_a} \, \delta_{m_k -m_a} \, \frac{(-1)^{j_a-m_a}}{\sqrt{2j_a+1}} \, \langle \Psi_0 || a_{n_a \alpha}^{\dagger} || \Psi_{n_k \kappa} \rangle
\nonumber \\ \displaystyle
&= \delta_{\kappa \alpha} \, \delta_{m_k -m_a} \, \eta_{\tilde{a}} \, \mV_{n_a \, [\alpha]}^{n_k} \: .
\end{align}
\end{subequations}
Inserting Eqs. \eqref{eq:bdform_uv_all} into Eqs. \eqref{eq:leh}, the set of Gorkov Green's functions can be written according to Eq. \eqref{eq:bdform_g_gen} as
\begin{subequations}
\label{eq:bdform_g_all}
\begin{eqnarray}
\label{eq:bdform_g11}
G^{11 \, [\alpha]}_{n n'}(\omega)  &=&
\sum_{n_k} \left\{
\frac{\mU_{n \, [\alpha]}^{n_k} \, \mU_{n' \, [\alpha]}^{n_k *}}
{\omega-\omega_{k} + i \eta} \right.
+ \left. \frac{\mV_{n \, [\alpha]}^{n_k *} \, {\mV_{n' \, [\alpha]}^{n_k}}}{\omega+\omega_{k} - i \eta} \right\} , \qquad \quad \\
\label{eq:bdform_g12}
G^{12 \, [\alpha]}_{n n'}(\omega) &=&
\sum_{n_k} \left\{
\frac{\mU_{n \, [\alpha]}^{n_k} \, \mV_{n' \, [\alpha]}^{n_k *}}
{\omega-\omega_{k} + i \eta} \right.
+ \left. \frac{\mV_{n \, [\alpha]}^{n_k *} \, {\mU_{n' \, [\alpha]}^{n_k}}}{\omega+\omega_{k} - i \eta} \right\} ,  \\
\label{eq:bdform_g21}
G^{21 \, [\alpha]}_{n n'}(\omega) &=&
\sum_{n_k} \left\{
\frac{\mV_{n \, [\alpha]}^{n_k} \, \mU_{n' \, [\alpha]}^{n_k *}}
{\omega-\omega_{k} + i \eta} \right.
+ \left. \frac{\mU_{n \, [\alpha]}^{n_k *} \, {\mV_{n' \, [\alpha]}^{n_k}}}{\omega+\omega_{k} - i \eta} \right\} ,  \\
\label{eq:bdform_g22}
G^{22 \, [\alpha]}_{n n'}(\omega)  &=&
\sum_{n_k} \left\{
\frac{\mV_{n \, [\alpha]}^{n_k} \, \mV_{n' \, [\alpha]}^{n_k *}}
{\omega-\omega_{k} + i \eta} \right.
+ \left. \frac{\mU_{n \, [\alpha]}^{n_k *} \, {\mU_{n' \, [\alpha]}^{n_k}}}{\omega+\omega_{k} - i \eta} \right\} ,
\end{eqnarray}
\end{subequations}
where only one sum over principal quantum number $n_k$ remains.

\begin{widetext}
\subsection{Matrix elements of the nuclear potential}
\label{sec:potential_struc}

Let us consider two-body interaction antisymmetrized matrix elements $\bar{V}_{abcd}$ introduced in Eq. \eqref{eq:vanti2}, which depend on angular momenta $j_a,j_b,j_c,j_d$ of the two incoming and two outgoing nucleons as well as on their third components $m_a,m_b,m_c,m_d$.
Writing all indices explicitly they read
\begin{eqnarray}
\label{eq:vanti2alli}
\bar{V}_{n_a \alpha m_a n_b \beta m_b n_c \gamma m_c n_d \delta m_d} &\equiv&
( \mbox{1:}\, (n_a \, \alpha \, m_a);\mbox{2:}\, (n_b \, \beta \, m_b) | V^{\text{NN}} |
\mbox{1:}\, (n_c \, \gamma \, m_c);\mbox{2:}\, (n_d \, \delta \, m_d) )
\nonumber \\ \displaystyle &-&
( \mbox{1:}\, (n_a \, \alpha \, m_a);\mbox{2:}\, (n_b \, \beta \, m_b) | V^{\text{NN}} |
 \mbox{1:}\, (n_d \, \delta \, m_d);\mbox{2:}\, (n_c \, \gamma \, m_c) )
\nonumber \\ \displaystyle &\equiv&
 \langle \mbox{1:}\, (n_a \, \alpha \, m_a);\mbox{2:}\, (n_b \, \beta \, m_b) | V^{\text{NN}} |
\mbox{1:}\, (n_c \, \gamma \, m_c);\mbox{2:}\, (n_d \, \delta \, m_d) \rangle \: .
\end{eqnarray}
One can go from such a representation, referred to as the $m$-scheme, to the $jj$-coupled scheme or $J$-scheme, in which incoming and outgoing two-nucleon states are labeled by total angular momenta $J$ to which individual angular momenta are recoupled.
Two-particle (non antisymmetrized) states in the two representations are connected through
\begin{subequations}
\label{eq:mmjj}
\begin{eqnarray}
\label{eq:mmjj1}
| \mbox{1:}\, (n_a \, \alpha \, m_a);\mbox{2:}\, (n_b \, \beta \, m_b) )
&=& \sum_{JM} \, C^{JM}_{j_a m_a j_b m_b}
| \mbox{1:}\, (n_a \, \alpha);\mbox{2:}\, (n_b \, \beta); \, JM ) \: ,
\\ \displaystyle
\label{eq:mmjj2}
| \mbox{1:}\, (n_a \, \alpha);\mbox{2:}\, (n_b \, \beta); \, JM )
&=& \sum_{m_a m_b} \, C^{JM}_{j_a m_a j_b m_b}
| \mbox{1:}\, (n_a \, \alpha \, m_a);\mbox{2:}\, (n_b \, \beta \, m_b) ) \: .
 \end{eqnarray}
\end{subequations}
The corresponding relations between antisymmetrized states are
\begin{subequations}
\label{eq:mmjja}
\begin{eqnarray}
\label{eq:mmjj1a}
| (n_a \, \alpha \, m_a); (n_b \, \beta \, m_b) \rangle
&=& \sqrt{1+\delta_{\alpha \beta} \, \delta_{n_a n_b}} \, \sum_{JM} \, C^{JM}_{j_a m_a j_b m_b}
| (n_a \, \alpha); (n_b \, \beta); \, JM \rangle \: ,
\\ \displaystyle
\label{eq:mmjj2a}
|  (n_a \, \alpha); (n_b \, \beta); \, JM \rangle
&=& \frac{1}{\sqrt{1+\delta_{\alpha \beta} \, \delta_{n_a n_b}}} \sum_{m_a m_b} \, C^{JM}_{j_a m_a j_b m_b}
|  (n_a \, \alpha \, m_a);(n_b \, \beta \, m_b) \rangle \: ,
 \end{eqnarray}
\end{subequations}
where the factor $\sqrt{1+\delta_{\alpha \beta} \, \delta_{n_a n_b}}$ ensures the correct normalization of the antisymmetrized state $| \mbox{1:}\, (n_a \, \alpha);\mbox{2:}\, (n_a \, \alpha); \, JM \rangle$, which is non zero for integer values of $J$.
Antisymmetrized potential matrix elements in $J$-scheme are thus related to those in $m$-scheme by means of
\begin{equation}
\bar{V}^{JMJ'M'}_{n_a \alpha n_b \beta n_c \gamma n_d \delta} =
\frac{1}{\sqrt{1+\delta_{\alpha \beta} \, \delta_{n_a n_b}}} \, \frac{1}{\sqrt{1+\delta_{\gamma \delta} \, \delta_{n_c n_d}}}
\sum_{m_a m_b m_c m_d}
C^{JM}_{j_a m_a j_b m_b} C^{J'M'}_{j_c m_c j_d m_d} \, \bar{V}_{abcd} \: ,
\end{equation}
and, conversely,
\begin{equation}
\label{eq:vmvj}
\bar{V}_{abcd} =
\sum_{JMJ'M'}
\sqrt{1+\delta_{\alpha \beta} \, \delta_{n_a n_b}} \, \sqrt{1+\delta_{\gamma \delta} \, \delta_{n_c n_d}} \,
C^{JM}_{j_a m_a j_b m_b} C^{J'M'}_{j_c m_c j_d m_d} \, \bar{V}^{JMJ'M'}_{n_a \alpha n_b \beta n_c \gamma n_d \delta} \: .
\end{equation}
Since nuclear potentials are rotationally invariant, they do not depend on $M$ or $M'$ and are non-zero only for $J=J'$, such that one can define
\begin{equation}
\bar{V}^{JMJ'M'}_{n_a \alpha n_b \beta n_c \gamma n_d \delta} \equiv \delta_{JJ'} \, \delta_{MM'} \,
\bar{V}^{J \, [\alpha \beta \gamma \delta]}_{n_a n_b n_c n_d } \: ,
\end{equation}
which allows rewriting Eq. \eqref{eq:vmvj} according to
\begin{subequations}
\label{eq:blockv_all}
\begin{equation}
\label{eq:blockv}
\bar{V}_{abcd} =  \sum_{JM} \, \sqrt{1+\delta_{\alpha \beta} \, \delta_{n_a n_b}} \, \sqrt{1+\delta_{\gamma \delta} \, \delta_{n_c n_d}}  \,
C^{JM}_{j_a m_a j_b m_b} C^{JM}_{j_c m_c j_d m_d} \, \bar{V}^{J \, [\alpha \beta \gamma \delta]}_{n_a n_b n_c n_d } \: .
\end{equation}
Similarly, one has
\begin{eqnarray}
\label{eq:blockv_2bar}
\bar{V}_{a\bar{b}c\bar{d}} &=&  \sum_{JM} \, \sqrt{1+\delta_{\alpha \beta} \, \delta_{n_a n_b}} \, \sqrt{1+\delta_{\gamma \delta} \, \delta_{n_c n_d}} \, \eta_b \eta_d \,
C^{JM}_{j_a m_a j_b -m_b} C^{JM}_{j_c m_c j_d -m_d} \, \bar{V}^{J \, [\alpha \beta \gamma \delta]}_{n_a n_b n_c n_d } \: ,
\\ \displaystyle
\nonumber \\ \displaystyle
\label{eq:blockv_4bar}
\bar{V}_{\bar{a}\bar{b}\bar{c}\bar{d}} &=& \sum_{JM} \, \sqrt{1+\delta_{\alpha \beta} \, \delta_{n_a n_b}} \, \sqrt{1+\delta_{\gamma \delta} \, \delta_{n_c n_d}}  \,
\eta_a \eta_b \eta_c \eta_d \,
C^{JM}_{j_a -m_a j_b -m_b} C^{JM}_{j_c -m_c j_d -m_d} \, \bar{V}^{J \, [\alpha \beta \gamma \delta]}_{n_a n_b n_c n_d } \: .
\end{eqnarray}
\end{subequations}

\subsection{Block-diagonal structure of self-energies}

\subsubsection{First order}

The goal of this subsection is to discuss how the block-diagonal form of the propagators and interaction matrix elements reflects in the various self-energy contributions, starting with the first-order normal self-energy $\Sigma^{11 \, (1)}$.
Substituting Eqs. \eqref{eq:blockv} and \eqref{eq:bdform_uv_all} into Eq. \eqref{eq:slefhf}, and introducing the factor
\begin{equation}
f^{n_a n_b n_c n_d}_{\alpha \beta \gamma \delta} \equiv \sqrt{1+\delta_{\alpha \beta} \, \delta_{n_a n_b}} \, \sqrt{1+\delta_{\gamma \delta} \, \delta_{n_c n_d}} \: ,
\end{equation}
one obtains
\begin{eqnarray}
\label{eq:first_bd_11}
\Sigma^{11 \, (1)}_{ab} &=&
\sum_{cd, k} \bar{V}_{acbd} \, \bar{\mV}_{d}^{k *} \, {\bar{\mV}_{c}^{k}}
\nonumber \\ \displaystyle
&=&
\sum_{n_c n_d n_k} \sum_{\gamma} \sum_{m_c} \sum_{JM} \, f^{n_a n_c n_b n_d}_{\alpha \gamma \beta \gamma}
  \, C^{JM}_{j_a m_a j_c m_c} C^{JM}_{j_b m_b j_c m_c}
\, \bar{V}^{J \, [\alpha \gamma \beta \gamma]}_{n_a n_c n_b n_d } \,
\mV_{n_d \, [\gamma]}^{n_k *} \, {\mV_{n_c \, [\gamma]}^{n_k}}
\nonumber \\ \displaystyle
&=& \delta_{\alpha \beta} \, \delta_{m_a m_b}
\sum_{n_c n_d} \sum_{\gamma} \sum_{J} \,
f^{n_a n_c n_b n_d}_{\alpha \gamma \alpha \gamma} \,
\frac{2J+1}{2j_a+1}
 \, \bar{V}^{J \, [\alpha \gamma\alpha \gamma]}_{n_a n_c n_b n_d }
\, \rho_{n_d n_c}^{[\gamma]}
\nonumber \\ \displaystyle
&\equiv& \delta_{\alpha \beta} \, \delta_{m_a m_b} \, \Sigma^{11  \, [\alpha] \, (1)}_{n_a n_b}
\nonumber \\ \displaystyle
&\equiv& \delta_{\alpha \beta} \, \delta_{m_a m_b} \, \Lambda_{n_a n_b}^{[\alpha]} \: ,
\end{eqnarray}
where the block-diagonal normal density matrix is introduced through $\rho_{ab} \equiv \delta_{\alpha \beta} \, \delta_{m_a m_b} \, \rho_{n_a n_b}^{[\alpha]}$, such that
\begin{equation}
\rho_{n_a n_b}^{[\alpha]} = \sum_{n_k} \mV_{n_b \, [\alpha]}^{n_k} \, {\mV_{n_a \, [\alpha]}^{n_k *}} \: ,
\end{equation}
and properties of Clebsch-Gordan coefficients has been used. The fact that the interaction conserves parity and charge yields $\delta_{\pi_a \pi_b}$ and $\delta_{q_a q_b}$, leading to $\delta_{\alpha \beta} = \delta_{j_a j_b} \, \delta_{\pi_a \pi_b} \, \delta_{q_a q_b}$.
Similarly for $\Sigma^{22 \, (1)}$
\begin{eqnarray}
\label{eq:first_bd_22}
\Sigma^{22 \, (1)}_{ab} &=& - \sum_{cd, k} \bar{V}_{\bar{b}c\bar{a}d} \, {\bar{\mV}_{c}^k} \, \bar{\mV}_{d}^{k *}
\nonumber \\ \displaystyle
&=& - \delta_{\alpha \beta} \, \delta_{m_a m_b} \sum_{n_c n_d} \sum_{\gamma} \sum_{J} \,
f^{n_b n_c n_a n_d}_{\alpha \gamma \alpha \gamma} \,
\frac{2J+1}{2j_a+1}
 \, \bar{V}^{J \, [\alpha \gamma \alpha \gamma]}_{n_b n_c n_a n_d } \,
\rho_{n_d n_c}^{[\gamma]}
\nonumber \\ \displaystyle
&\equiv& \delta_{\alpha \beta} \, \delta_{m_a m_b} \, \Sigma^{22 \, [\alpha] \, (1)}_{n_a n_b}
\nonumber \\ \displaystyle
&=& - \delta_{\alpha \beta} \, \delta_{m_a m_b} \, \Lambda_{n_b n_a}^{[\alpha]} \: .
\nonumber \\ \displaystyle
&=& - \delta_{\alpha \beta} \, \delta_{m_a m_b} \, [\Lambda_{n_a n_b}^{[\alpha]}]^* \: .
\end{eqnarray}
Let us consider the anomalous contributions to the first-order self-energy. Substituting Eqs. \eqref{eq:blockv_2bar}
and \eqref{eq:bdform_uv_all} into Eq. \eqref{eq:self121} one derives
\begin{eqnarray}
\label{eq:first_bd_12}
\Sigma^{12 \, (1)}_{ab} &=& \frac{1}{2}
\sum_{cd, k} \bar{V}_{a\bar{b}c\bar{d}} \, \bar{\mV}_{c}^{k* } \, {\bar{\mU}_{d}^k}
\nonumber \\ \displaystyle
&=& - \frac{1}{2}
\sum_{n_c n_d n_k} \sum_{\gamma} \sum_{m_c} \sum_{JM} \, f^{n_a n_b n_c n_d}_{\alpha \beta \gamma \gamma} \, \eta_b \eta_c \, C^{JM}_{j_a m_a j_b -m_b} C^{JM}_{j_c m_c j_c -m_c}
\, \bar{V}^{J \, [\alpha \beta \gamma \gamma]}_{n_a n_b n_c n_d } \,
\mV_{n_c \, [\gamma]}^{n_k *} \, {\mU_{n_d \, [\gamma]}^{n_k}}
\nonumber \\ \displaystyle
&=& - \frac{1}{2}
\sum_{n_c n_d} \sum_{\gamma} \sum_{m_c} \sum_{J} \, f^{n_a n_b n_c n_d}_{\alpha \beta \gamma \gamma}  \, \eta_b \eta_c \, C^{J0}_{j_a m_a j_b -m_b} C^{J0}_{j_c m_c j_c -m_c}
\, \bar{V}^{J \, [\alpha \beta \gamma \gamma]}_{n_a n_b n_c n_d } \,
\tilde{\rho}_{n_c n_d}^{[\gamma]}
\nonumber \\ \displaystyle
&=& - \frac{1}{2}
\sum_{n_c n_d} \sum_{\gamma} \, f^{n_a n_b n_c n_d}_{\alpha \beta \gamma \gamma}  \, \eta_b \pi_c (-1)^{2j_c} \, C^{00}_{j_a m_a j_b -m_b} \, \sqrt{2j_c+1}
\, \bar{V}^{0 \, [\alpha \beta \gamma \gamma]}_{n_a n_b n_c n_d } \,
\tilde{\rho}_{n_c n_d}^{[\gamma]}
\nonumber \\ \displaystyle
&=&
\delta_{\alpha \beta} \, \delta_{m_a m_b} \, \frac{1}{2} \, \sum_{n_c n_d} \sum_{\gamma} \,
f^{n_a n_b n_c n_d}_{\alpha \alpha \gamma \gamma} \, \pi_a \, \pi_c (-1)^{2j_c} \,
\frac{\sqrt{2j_c+1}}{\sqrt{2j_a+1}}
 \, \bar{V}^{0 \, [\alpha \alpha \gamma \gamma]}_{n_a n_b n_c n_d } \,
\tilde{\rho}_{n_c n_d}^{[\gamma]}
\nonumber \\ \displaystyle
&\equiv& \delta_{\alpha \beta} \, \delta_{m_a m_b} \, \Sigma^{12 \, [\alpha] \, (1)}_{n_a n_b}
\nonumber \\ \displaystyle
&\equiv& \delta_{\alpha \beta} \, \delta_{m_a m_b} \, \tilde{h}_{n_a n_b}^{[\alpha]} \: ,
\end{eqnarray}
where the block-diagonal anomalous density matrix is introduced through $\tilde{\rho}_{ab} \equiv \delta_{\alpha \beta} \, \delta_{m_a m_b} \, \tilde{\rho}_{n_a n_b}^{[\alpha]}$, such that
\begin{equation}
\tilde{\rho}_{n_a n_b}^{[\alpha]} = \sum_{n_k} \mU_{n_b \, [\alpha]}^{n_k} \, {\mV_{n_a \, [\alpha]}^{n_k *}} \: .
\end{equation}
It is interesting to note that the first-order anomalous self-energies only involve $J=0$ matrix elements as a direct result of dealing with a $J=0$ many-body state.
The other anomalous term is similarly obtained from Eq. \eqref{eq:self211} and reads
\begin{eqnarray}
\label{eq:first_bd_21}
\Sigma^{21 \, (1)}_{ab} &=& \frac{1}{2}
\sum_{cd, k} \bar{V}_{\bar{c}d\bar{a}b} \, \bar{\mU}_{c}^{k *} \, {\bar{\mV}_{d}^k}
\nonumber \\ \displaystyle
&=& - \frac{1}{2}
\sum_{n_c n_d n_k} \sum_{\gamma} \sum_{m_c} \sum_{JM} \, f^{n_a n_b n_c n_d}_{\alpha \beta \gamma \gamma}  \, \eta_a \eta_c \, C^{JM}_{j_c -m_c j_c m_c} C^{JM}_{j_a -m_a j_b m_b}
\, \bar{V}^{J \, [\gamma \gamma \alpha \beta ]}_{n_c n_d n_a n_b} \,
\mU_{n_c \, [\gamma]}^{n_k} \, \mV_{n_d \, [\gamma]}^{n_k *}
\nonumber \\ \displaystyle
&=&
\delta_{\alpha \beta} \, \delta_{m_a m_b} \,  \frac{1}{2} \, \sum_{n_c n_d n_k} \sum_{\gamma} \,
f^{n_a n_b n_c n_d}_{\alpha \alpha \gamma \gamma}  \, \pi_a \, \pi_c (-1)^{2j_c}
\frac{\sqrt{2j_c+1}}{\sqrt{2j_a+1}}
 \, \bar{V}^{0 \, [\gamma \gamma \alpha \alpha]}_{n_c n_d n_a n_b} \,
\tilde{\rho}_{n_d n_c}^{[\gamma]}
\nonumber \\ \displaystyle
&\equiv& \delta_{\alpha \beta} \, \delta_{m_a m_b}  \, \Sigma^{21 \, [\alpha] \, (1)}_{n_a n_b}
\nonumber \\ \displaystyle
&=& \delta_{\alpha \beta} \, \delta_{m_a m_b} \, \tilde{h}_{n_a n_b}^{[\alpha] \, \dagger} \: .
\end{eqnarray}

\subsubsection{Second order}

Block-diagonal forms of second-order self-energy contributions \eqref{eq:leh_self11} and \eqref{eq:leh_self} can be obtained by considering explicitly the angular momentum couplings of the three quasiparticles to their total momentum $J_{tot}$, separately for the six objects $\mM$, $\mN$, $\mP$, $\mQ$, $\mR$ and $\mS$.
One proceeds first coupling particles 1 and 2 to some momentum $J_c$, which is afterwards coupled to particle 3 to give $J_{tot}$. The recoupled $\mM$ term is computed as follows
\begin{eqnarray}
\label{eq:Mjj}
{\mM}^{k_1k_2k_3}_{a \, (J_c J_{tot})}
&=&
\sum_{m_1 m_2 m_3 M_c} C^{J_cM_c}_{j_{k_1}m_{k_1}j_{k_2}m_{k_2}} C^{J_{tot}M_{tot}}_{J_cM_cj_{k_3}m_{k_3}}  {\mM}^{k_1k_2k_3}_{a}
\nonumber \\ \displaystyle
&=&
\sum_{m_1 m_2 m_3 M_c} \sum_{rst}  C^{J_cM_c}_{j_{k_1}m_{k_1}j_{k_2}m_{k_2}} C^{J_{tot} M_{tot}}_{J_cM_cj_{k_3}m_{k_3}}
\, \bar{V}_{atrs} \, \mU_{r}^{k_1} \, \mU_{s}^{k_2} \, \bar{\mV}_{t}^{k_3}
\nonumber \\ \displaystyle
&=&
\sum_{m_1 m_2 m_3 M_c} \sum_{rst} \sum_{J_v M_v}
\delta_{\kappa_1 \rho} \, \delta_{m_{k_1} m_r} \,
\delta_{\kappa_2 \sigma} \, \delta_{m_{k_2} m_s} \,
\delta_{\kappa_3 \tau} \, \delta_{m_{k_3} -m_t} \,
(-\eta_t) \, f^{n_a n_t n_r n_s}_{\alpha \tau \rho \sigma}
\nonumber \\
&\times& \displaystyle
C^{J_cM_c}_{j_{k_1}m_{k_1}j_{k_2}m_{k_2}} C^{J_{tot}M_{tot}}_{J_cM_cj_{k_3}m_{k_3}}
C^{J_vM_v}_{j_a m_a j_t m_t} C^{J_vM_v}_{j_r m_r j_s m_s} \,
\bar{V}_{n_a n_t n_r n_s}^{J_v [\alpha \tau \rho \sigma]} \,
\mU_{n_r \, [\rho]}^{n_{k_1}} \, \mU_{n_s \, [\sigma]}^{n_{k_2}} \, \mV_{n_t \, [\tau]}^{n_{k_3}}
\nonumber \\ \displaystyle
&=&
\sum_{m_1 m_2 m_3 M_c} \sum_{n_r n_s n_t} \sum_{J_v M_v}
\eta_{k_3} \, f^{n_a n_t n_r n_s}_{\alpha \kappa_3 \kappa_1 \kappa_2}
C^{J_cM_c}_{j_{k_1} m_{k_1} j_{k_2} m_{k_2}} C^{J_{tot}M_{tot}}_{J_c M_c j_{k_3} m_{k_3}}
C^{J_vM_v}_{j_a m_a j_{k_3} -m_{k_3}} C^{J_vM_v}_{j_{k_1} m_{k_1} j_{k_2} m_{k_2}} \,
\nonumber \\
&\times& \displaystyle
\bar{V}_{n_a n_t n_r n_s}^{J_v [\alpha \kappa_3 \kappa_1 \kappa_2]} \,
\mU_{n_r \, [\kappa_1]}^{n_{k_1}} \, \mU_{n_s \, [\kappa_2]}^{n_{k_2}} \, \mV_{n_t \, [\kappa_3]}^{n_{k_3}}
\nonumber \\ \displaystyle
&=&
\sum_{m_3 M_c} \sum_{n_r n_s n_t}
\eta_{k_3} \, f^{n_a n_t n_r n_s}_{\alpha \kappa_3 \kappa_1 \kappa_2} \frac{\sqrt{2J_c+1}}{\sqrt{2j_a+1}}
\, (-1)^{J_c + 2j_{k_3} - j_a}
C^{J_{tot}M_{tot}}_{J_c M_c j_{k_3} m_{k_3}} C^{j_a m_a}_{J_c M_c j_{k_3} m_{k_3}}  \,
\nonumber \\
&\times& \displaystyle
\bar{V}_{n_a n_t n_r n_s}^{J_c [\alpha \kappa_3 \kappa_1 \kappa_2]} \,
\mU_{n_r \, [\kappa_1]}^{n_{k_1}} \, \mU_{n_s \, [\kappa_2]}^{n_{k_2}} \, \mV_{n_t \, [\kappa_3]}^{n_{k_3}}
\nonumber \\ \displaystyle
&=&
- \delta_{J_{tot} j_a} \delta_{M_{tot} m_a} \sum_{n_r n_s n_t}
\pi_{k_3} \, f^{n_a n_t n_r n_s}_{\alpha \kappa_3 \kappa_1 \kappa_2} \frac{\sqrt{2J_c+1}}{\sqrt{2j_a+1}}
\, (-1)^{J_c + j_{k_3} - j_a}
\bar{V}_{n_a n_t n_r n_s}^{J_c [\alpha \kappa_3 \kappa_1 \kappa_2]} \,
\mU_{n_r \, [\kappa_1]}^{n_{k_1}} \, \mU_{n_s \, [\kappa_2]}^{n_{k_2}} \, \mV_{n_t \, [\kappa_3]}^{n_{k_3}}
\nonumber \\ \displaystyle
&\equiv&
\delta_{J_{tot} j_a} \delta_{M_{tot} m_a}  \, {\mM_{n_a \, [\alpha \kappa_3 \kappa_1 \kappa_2] \, J_c}^{n_{k_1}n_{k_2}n_{k_3}}}
\: ,
\end{eqnarray}
where general properties of Clebsch-Gordan coefficients have been used.
Similarly one derives the $\mN$ term
\begin{eqnarray}
\label{eq:Njj}
{\mN}^{k_1k_2k_3}_{a \, (J_c J_{tot})}
&=&
\delta_{J_{tot} j_a} \delta_{M_{tot} m_a} \sum_{n_r n_s n_t}
\pi_{k_3} \, f^{n_a n_t n_r n_s}_{\alpha \kappa_3 \kappa_1 \kappa_2} \frac{\sqrt{2J_c+1}}{\sqrt{2j_a+1}}
\, (-1)^{J_c + j_{k_3} - j_a}
\bar{V}_{n_a n_t n_r n_s}^{J_c [\alpha \kappa_3 \kappa_1 \kappa_2]} \,
\mV_{n_r \, [\kappa_1]}^{n_{k_1}} \, \mV_{n_s \, [\kappa_2]}^{n_{k_2}} \, \mU_{n_t \, [\kappa_3]}^{n_{k_3}}
\nonumber \\ \displaystyle
&\equiv&
\delta_{J_{tot} j_a} \delta_{M_{tot} m_a}  \, {\mN_{n_a \, [\alpha \kappa_3 \kappa_1 \kappa_2] \, J_c}^{n_{k_1}n_{k_2}n_{k_3}}}
\: .
\end{eqnarray}
One can show that the same result is obtained by recoupling directly $\bar{\mN}$, as follows
\begin{eqnarray}
\label{eq:Njj_bar}
\bar{\mN}^{k_1k_2k_3}_{a \, (J_c J_{tot})}
&=&
\sum_{m_1 m_2 m_3 M_c} C^{J_cM_c}_{j_{k_1}m_{k_1}j_{k_2}m_{k_2}} C^{J_{tot}M_{tot}}_{J_cM_cj_{k_3}m_{k_3}}  \bar{\mN}^{k_1k_2k_3}_{a}
\nonumber \\ \displaystyle
&=&
\sum_{m_1 m_2 m_3 M_c} \sum_{rst}  C^{J_cM_c}_{j_{k_1}m_{k_1}j_{k_2}m_{k_2}} C^{J_{tot} M_{tot}}_{J_cM_cj_{k_3}m_{k_3}}
\, \bar{V}_{\bar{a}\bar{t}\bar{r}\bar{s}} \, \bar{\mV}_{r}^{k_1} \, \bar{\mV}_{s}^{k_2} \, \mU_{t}^{k_3}
\nonumber \\ \displaystyle
&=&
\sum_{m_1 m_2 m_3 M_c} \sum_{rst} \sum_{J_v M_v}
\delta_{\kappa_1 \rho} \, \delta_{m_{k_1} -m_r} \,
\delta_{\kappa_2 \sigma} \, \delta_{m_{k_2} -m_s} \,
\delta_{\kappa_3 \tau} \, \delta_{m_{k_3} m_t} \,
\eta_a \eta_t \, f^{n_a n_t n_r n_s}_{\alpha \tau \rho \sigma}
\nonumber \\
&\times& \displaystyle
C^{J_cM_c}_{j_{k_1}m_{k_1}j_{k_2}m_{k_2}} C^{J_{tot}M_{tot}}_{J_cM_cj_{k_3}m_{k_3}}
C^{J_vM_v}_{j_a -m_a j_t -m_t} C^{J_vM_v}_{j_r -m_r j_s -m_s} \,
\bar{V}_{n_a n_t n_r n_s}^{J_v [\alpha \tau \rho \sigma]} \,
\mV_{n_r \, [\rho]}^{n_{k_1}} \, \mV_{n_s \, [\sigma]}^{n_{k_2}} \, \mU_{n_t \, [\tau]}^{n_{k_3}}
\nonumber \\ \displaystyle
&=&
\sum_{m_1 m_2 m_3 M_c} \sum_{n_r n_s n_t} \sum_{J_v M_v}
\eta_a \,\eta_{k_3} \, f^{n_a n_t n_r n_s}_{\alpha \kappa_3 \kappa_1 \kappa_2}
C^{J_cM_c}_{j_{k_1} m_{k_1} j_{k_2} m_{k_2}} C^{J_{tot}M_{tot}}_{J_c M_c j_{k_3} m_{k_3}}
C^{J_vM_v}_{j_a -m_a j_{k_3} -m_{k_3}} C^{J_vM_v}_{j_{k_1} m_{k_1} j_{k_2} m_{k_2}} \,
\nonumber \\
&\times& \displaystyle
\bar{V}_{n_a n_t n_r n_s}^{J_v [\alpha \kappa_3 \kappa_1 \kappa_2]} \,
\mV_{n_r \, [\kappa_1]}^{n_{k_1}} \, \mV_{n_s \, [\kappa_2]}^{n_{k_2}} \, \mU_{n_t \, [\kappa_3]}^{n_{k_3}}
\nonumber \\ \displaystyle
&=&
\sum_{m_3 M_c} \sum_{n_r n_s n_t}
\eta_a \,\eta_{k_3} \, f^{n_a n_t n_r n_s}_{\alpha \kappa_3 \kappa_1 \kappa_2}
C^{J_{tot}M_{tot}}_{J_c M_c j_{k_3} m_{k_3}} C^{J_c M_c}_{j_a -m_a j_{k_3} -m_{k_3}}  \,
\bar{V}_{n_a n_t n_r n_s}^{J_c [\alpha \kappa_3 \kappa_1 \kappa_2]} \,
\mV_{n_r \, [\kappa_1]}^{n_{k_1}} \, \mV_{n_s \, [\kappa_2]}^{n_{k_2}} \, \mU_{n_t \, [\kappa_3]}^{n_{k_3}}
\nonumber \\ \displaystyle
&=&
\sum_{m_3 M_c} \sum_{n_r n_s n_t}
\eta_a \,\pi_{k_3} \, f^{n_a n_t n_r n_s}_{\alpha \kappa_3 \kappa_1 \kappa_2} \frac{\sqrt{2J_c+1}}{\sqrt{2j_a+1}}
\, (-1)^{j_a + j_{k_3} - J_c}
C^{J_{tot}M_{tot}}_{J_c M_c j_{k_3} m_{k_3}} C^{j_a -m_a}_{J_c M_c j_{k_3} m_{k_3}}  \,
\nonumber \\
&\times& \displaystyle
\bar{V}_{n_a n_t n_r n_s}^{J_c [\alpha \kappa_3 \kappa_1 \kappa_2]} \,
\mV_{n_r \, [\kappa_1]}^{n_{k_1}} \, \mV_{n_s \, [\kappa_2]}^{n_{k_2}} \, \mU_{n_t \, [\kappa_3]}^{n_{k_3}}
\nonumber \\ \displaystyle
&=&
\delta_{J_{tot} j_a} \delta_{M_{tot} -m_a} \sum_{n_r n_s n_t} \eta_{a} \,
\pi_{k_3} \, f^{n_a n_t n_r n_s}_{\alpha \kappa_3 \kappa_1 \kappa_2} \frac{\sqrt{2J_c+1}}{\sqrt{2j_a+1}}
\, (-1)^{J_c - j_{k_3} - j_a}
\bar{V}_{n_a n_t n_r n_s}^{J_c [\alpha \kappa_3 \kappa_1 \kappa_2]} \,
\mV_{n_r \, [\kappa_1]}^{n_{k_1}} \, \mV_{n_s \, [\kappa_2]}^{n_{k_2}} \, \mU_{n_t \, [\kappa_3]}^{n_{k_3}}
\nonumber \\ \displaystyle
&=&
- \delta_{J_{tot} j_a} \delta_{M_{tot} -m_a} \, \eta_{a} \, {\mN_{n_a \, [\alpha \kappa_3 \kappa_1 \kappa_2] \, J_c}^{n_{k_1}n_{k_2}n_{k_3}}}
\: ,
\end{eqnarray}
which recovers relation \eqref{eq:n_bar}.
The remaining quantities (see Eqs. \eqref{eq:mpr} and~\eqref{eq:mpr_bis}) are related to $\mM$ and $\mN$ by permutations of $\{ k_1, k_2, k_3 \}$ indices and can be obtained from Eqs. \eqref{eq:Mjj} and \eqref{eq:Njj} by taking into account the different recoupling of $j_{k_1}$, $j_{k_2}$ and $j_{k_3}$ to $J_{tot}$ and $J_c$ as follows
\begin{eqnarray}
\label{eq:Pjj}
{\mP}^{k_1k_2k_3}_{a \, (J_c J_{tot})}
&=& \sum_{J_d} (-1)^{J_c + J_d + j_{k_2} + j_{k_3}} \sqrt{2J_c+1} \sqrt{2J_d+1}
\left \{ \begin{array}{ccc}
j_{k_2} & j_{k_1} & J_c \\
j_{k_3} & J_{tot} & J_d
\end{array} \right \}
{\mM}^{k_1k_3k_2}_{a \, (J_d J_{tot})}
\nonumber \\ \displaystyle
&=&
- \delta_{J_{tot} j_a} \delta_{M_{tot} m_a} \sum_{n_r n_s n_t} \sum_{J_d}
\pi_{k_2} \, f^{n_a n_t n_r n_s}_{\alpha \kappa_2 \kappa_1 \kappa_3} \frac{\sqrt{2J_c+1}}{\sqrt{2j_a+1}} (2J_d+1)
\, (-1)^{J_d + j_{k_3} + j_a}
\left \{ \begin{array}{ccc}
j_{k_2} & j_{k_1} & J_c \\
j_{k_3} & J_{tot} & J_d
\end{array} \right \}
\nonumber \\ \displaystyle
&\times&
\bar{V}_{n_a n_t n_r n_s}^{J_d [\alpha \kappa_2 \kappa_1 \kappa_3]} \,
\mU_{n_r \, [\kappa_1]}^{n_{k_1}} \, \mU_{n_s \, [\kappa_3]}^{n_{k_3}} \, \mV_{n_t \, [\kappa_2]}^{n_{k_2}}
\nonumber \\ \displaystyle
&\equiv&
\delta_{J_{tot} j_a} \delta_{M_{tot} m_a}  \, {\mP_{n_a \, [\alpha \kappa_3 \kappa_1 \kappa_2] \, J_c}^{n_{k_1}n_{k_2}n_{k_3}}}
\: ,
\end{eqnarray}
\begin{eqnarray}
\label{eq:Qjj}
{\mQ}^{k_1k_2k_3}_{a \, (J_c J_{tot})}
&=& \sum_{J_d} (-1)^{J_c + J_d + j_{k_2} + j_{k_3}} \sqrt{2J_c+1} \sqrt{2J_d+1}
\left \{ \begin{array}{ccc}
j_{k_2} & j_{k_1} & J_c \\
j_{k_3} & J_{tot} & J_d
\end{array} \right \}
{\mN}^{k_1k_3k_2}_{a \, (J_d J_{tot})}
\nonumber \\ \displaystyle
&=&
 \delta_{J_{tot} j_a} \delta_{M_{tot} m_a} \sum_{n_r n_s n_t} \sum_{J_d}
\pi_{k_2} \, f^{n_a n_t n_r n_s}_{\alpha \kappa_2 \kappa_1 \kappa_3} \frac{\sqrt{2J_c+1}}{\sqrt{2j_a+1}} (2J_d+1)
\, (-1)^{J_d + j_{k_3} + j_a}
\left \{ \begin{array}{ccc}
j_{k_2} & j_{k_1} & J_c \\
j_{k_3} & J_{tot} & J_d
\end{array} \right \}
\nonumber \\ \displaystyle
&\times&
\bar{V}_{n_a n_t n_r n_s}^{J_d [\alpha \kappa_2 \kappa_1 \kappa_3]} \,
\mV_{n_r \, [\kappa_1]}^{n_{k_1}} \, \mV_{n_s \, [\kappa_3]}^{n_{k_3}} \, \mU_{n_t \, [\kappa_2]}^{n_{k_2}}
\nonumber \\ \displaystyle
&\equiv&
\delta_{J_{tot} j_a} \delta_{M_{tot} m_a}   \, {\mQ_{n_a \, [\alpha \kappa_3 \kappa_1 \kappa_2] \, J_c}^{n_{k_1}n_{k_2}n_{k_3}}}
\: ,
\end{eqnarray}
\begin{eqnarray}
\label{eq:Rjj}
{\mR}^{k_1k_2k_3}_{a \, (J_c J_{tot})}
&=& \sum_{J_d} (-1)^{2j_1 + 2J_d} \sqrt{2J_c+1} \sqrt{2J_d+1}
\left \{ \begin{array}{ccc}
j_{k_1} & j_{k_2} & J_c \\
j_{k_3} & J_{tot} & J_d
\end{array} \right \}
{\mM}^{k_3k_2k_1}_{a \, (J_d J_{tot})}
\nonumber \\ \displaystyle
&=&
- \delta_{J_{tot} j_a} \delta_{M_{tot} m_a} \sum_{n_r n_s n_t} \sum_{J_d}
\pi_{k_1} \, f^{n_a n_t n_r n_s}_{\alpha \kappa_1 \kappa_3 \kappa_2} \frac{\sqrt{2J_c+1}}{\sqrt{2j_a+1}} (2J_d+1)
\, (-1)^{J_d + j_{k_1} + j_a}
\left \{ \begin{array}{ccc}
j_{k_1} & j_{k_2} & J_c \\
j_{k_3} & J_{tot} & J_d
\end{array} \right \}
\nonumber \\ \displaystyle
&\times&
\bar{V}_{n_a n_t n_r n_s}^{J_d [\alpha \kappa_1 \kappa_3 \kappa_2]} \,
\mU_{n_r \, [\kappa_3]}^{n_{k_3}} \, \mU_{n_s \, [\kappa_2]}^{n_{k_2}} \, \mV_{n_t \, [\kappa_1]}^{n_{k_1}}
\nonumber \\ \displaystyle
&\equiv&
\delta_{J_{tot} j_a} \delta_{M_{tot} m_a}  \, {\mR_{n_a \, [\alpha \kappa_3 \kappa_1 \kappa_2] \, J_c}^{n_{k_1}n_{k_2}n_{k_3}}}
\: ,
\end{eqnarray}
\begin{eqnarray}
\label{eq:Sjj}
{\mS}^{k_1k_2k_3}_{a \, (J_c J_{tot})}
&=& \sum_{J_d} (-1)^{2j_1 + 2J_d} \sqrt{2J_c+1} \sqrt{2J_d+1}
\left \{ \begin{array}{ccc}
j_{k_1} & j_{k_2} & J_c \\
j_{k_3} & J_{tot} & J_d
\end{array} \right \}
{\mN}^{k_3k_2k_1}_{a \, (J_d J_{tot})}
\nonumber \\ \displaystyle
&=&
\delta_{J_{tot} j_a} \delta_{M_{tot} m_a} \sum_{n_r n_s n_t} \sum_{J_d}
\, \pi_{k_1} \, f^{n_a n_t n_r n_s}_{\alpha \kappa_1 \kappa_3 \kappa_2} \frac{\sqrt{2J_c+1}}{\sqrt{2j_a+1}} (2J_d+1)
\, (-1)^{J_d + j_{k_1} + j_a}
\left \{ \begin{array}{ccc}
j_{k_1} & j_{k_2} & J_c \\
j_{k_3} & J_{tot} & J_d
\end{array} \right \}
\nonumber \\ \displaystyle
&\times&
\bar{V}_{n_a n_t n_r n_s}^{J_d [\alpha \kappa_1 \kappa_3 \kappa_2]} \,
\mV_{n_r \, [\kappa_3]}^{n_{k_3}} \, \mV_{n_s \, [\kappa_2]}^{n_{k_2}} \, \mU_{n_t \, [\kappa_1]}^{n_{k_1}}
\nonumber \\ \displaystyle
&\equiv&
\delta_{J_{tot} j_a} \delta_{M_{tot} m_a} \,  {\mS_{n_a \, [\alpha \kappa_3 \kappa_1 \kappa_2] \, J_c}^{n_{k_1}n_{k_2}n_{k_3}}}
\: .
\end{eqnarray}
These terms are finally put together to form the different contributions to second-order self-energies. Let us consider $\Sigma^{11 \, (2')}_{ab}$  as an example (see Eq. \eqref{eq:selftwoa}). By inserting Eqs. \eqref{eq:Mjj} and \eqref{eq:Njj} and summing over all possible total and intermediate angular momenta one has
\begin{eqnarray}
\Sigma^{11 \, (2')}_{ab}
&=&
\displaystyle \frac{1}{2}
\sum_{J_{tot} M_{tot} J_c} \sum_{k_1k_2k_3}
\left\{
\frac{{\mM}^{k_1k_2k_3}_{a \, (J_c J_{tot})}
\left({\mM}^{k_1k_2k_3}_{b \, (J_c J_{tot})}\right)^*}
{\omega-(\omega_{k_1} + \omega_{k_2} + \omega_{k_3}) + i \eta}
+ \frac{{\mN}^{k_1k_2k_3}_{a \, (J_c J_{tot})}
\left({\mN}^{k_1k_2k_3}_{b \, (J_c J_{tot})}\right)^*}
{\omega+(\omega_{k_3} + \omega_{k_1} + \omega_{k_2}) - i \eta}
\right\}
\nonumber \\ \displaystyle
&=&
\displaystyle
\delta_{\alpha \beta} \, \delta_{m_a m_b} \,
\frac{1}{2}
 \sum_{J}  \sum_{n_{k_1} n_{k_2} n_{k_3}} \sum_{\kappa_1 \kappa_2 \kappa_3}
\left\{
\frac{{\mM_{n_a \, [\alpha \kappa_3 \kappa_1 \kappa_2] \, J_c}^{n_{k_1}n_{k_2}n_{k_3}}}
\left({\mM_{n_b \, [\alpha \kappa_3 \kappa_1 \kappa_2] \, J_c}^{n_{k_1}n_{k_2}n_{k_3}}}\right)^*}
{\omega-(\omega_{k_1} + \omega_{k_2} + \omega_{k_3}) + i \eta}
+ \frac{{\mN_{n_a \, [\alpha \kappa_3 \kappa_1 \kappa_2] \, J_c}^{n_{k_1}n_{k_2}n_{k_3}}}
\left({\mN_{n_b \, [\alpha \kappa_3 \kappa_1 \kappa_2] \, J_c}^{n_{k_1}n_{k_2}n_{k_3}}}\right)^*}
{\omega+(\omega_{k_3} + \omega_{k_1} + \omega_{k_2}) - i \eta}
\right\}
\nonumber \\ \displaystyle
&\equiv&
\delta_{\alpha \beta} \, \delta_{m_a m_b} \, \Sigma^{11  \, [\alpha] \, (2')}_{n_a n_b}
\: .
\end{eqnarray}
Proceeding similarly for the other terms and defining
\begin{subequations}
\label{eq:CDme}
\begin{eqnarray}
\mC_{n_a \, [\alpha \kappa_3 \kappa_1 \kappa_2] \, J_c}^{n_{k_1}n_{k_2}n_{k_3}} &\equiv& \frac{1}{\sqrt{6}} \left [
\mM_{n_a \, [\alpha \kappa_3 \kappa_1 \kappa_2]\, J_c}^{n_{k_1}n_{k_2}n_{k_3}} -
\mP_{n_a \, [\alpha \kappa_3 \kappa_1 \kappa_2] \, J_c}^{n_{k_1}n_{k_2}n_{k_3}} -
\mR_{n_a \, [\alpha \kappa_3 \kappa_1 \kappa_2] \, J_c}^{n_{k_1}n_{k_2}n_{k_3}}
 \right ] \: ,
\\
\mD_{n_a \, [\alpha \kappa_3 \kappa_1 \kappa_2] \, J_c}^{n_{k_1}n_{k_2}n_{k_3}} &\equiv& \frac{1}{\sqrt{6}} \left [
\mN_{n_a \, [\alpha \kappa_3 \kappa_1 \kappa_2] \, J_c}^{n_{k_1}n_{k_2}n_{k_3}} -
\mQ_{n_a \, [\alpha \kappa_3 \kappa_1 \kappa_2] \, J_c}^{n_{k_1}n_{k_2}n_{k_3}} -
\mS_{n_a \, [\alpha \kappa_3 \kappa_1 \kappa_2] \, J_c}^{n_{k_1}n_{k_2}n_{k_3}}
 \right ] \: ,
\end{eqnarray}
\end{subequations}
one finally writes
\begin{subequations}
\label{eq:second_bd}
\begin{eqnarray}
\Sigma^{11  \, [\alpha] \, (2)}_{n_a n_b} &=& \displaystyle
\sum_{n_{k_1} n_{k_2} n_{k_3}} \sum_{J_c} \sum_{\kappa_1 \kappa_2 \kappa_3}
\left\{
\frac{{\mC_{n_a \, [\alpha \kappa_3 \kappa_1 \kappa_2] \, J_c}^{n_{k_1}n_{k_2}n_{k_3}}} \,
\left(\mC_{n_b \, [\alpha \kappa_3 \kappa_1 \kappa_2] \, J_c}^{n_{k_1}n_{k_2}n_{k_3}}\right)^*}
{\omega-(\omega_{k_1} + \omega_{k_2} + \omega_{k_3}) + i \eta}
+ \frac{\left(\mD_{n_a \, [\alpha \kappa_3 \kappa_1 \kappa_2] \, J_c}^{n_{k_1}n_{k_2}n_{k_3}}\right)^* \,
{\mD_{n_b \, [\alpha \kappa_3 \kappa_1 \kappa_2] \, J_c}^{n_{k_1}n_{k_2}n_{k_3}}} \, }
{\omega+(\omega_{k_3} + \omega_{k_1} + \omega_{k_2}) - i \eta}
\right\} \: , \: \: \: \: \:\: \: \: \: \: 
\\
\Sigma^{12  \, [\alpha] \, (2)}_{n_a n_b} &=& \displaystyle
\sum_{n_{k_1} n_{k_2} n_{k_3}} \sum_{J_c} \sum_{\kappa_1 \kappa_2 \kappa_3}
\left\{
\frac{{\mC_{n_a \, [\alpha \kappa_3 \kappa_1 \kappa_2] \, J_c}^{n_{k_1}n_{k_2}n_{k_3}}} \,
\left(\mD_{n_b \, [\alpha \kappa_3 \kappa_1 \kappa_2] \, J_c}^{n_{k_1}n_{k_2}n_{k_3}}\right)^*}
{\omega-(\omega_{k_1} + \omega_{k_2} + \omega_{k_3}) + i \eta}
+ \frac{\left(\mD_{n_a \, [\alpha \kappa_3 \kappa_1 \kappa_2] \, J_c}^{n_{k_1}n_{k_2}n_{k_3}}\right)^* \,
{\mC_{n_b \, [\alpha \kappa_3 \kappa_1 \kappa_2] \, J_c}^{n_{k_1}n_{k_2}n_{k_3}}} \, }
{\omega+(\omega_{k_3} + \omega_{k_1} + \omega_{k_2}) - i \eta}
\right\} \: , \: \: \: \: \:\: \: \: \: \: 
\\
\Sigma^{21  \, [\alpha] \, (2)}_{n_a n_b} &=& \displaystyle
\sum_{n_{k_1} n_{k_2} n_{k_3}} \sum_{J_c} \sum_{\kappa_1 \kappa_2 \kappa_3}
\left\{
\frac{{\mD_{n_a \, [\alpha \kappa_3 \kappa_1 \kappa_2] \, J_c}^{n_{k_1}n_{k_2}n_{k_3}}} \,
\left(\mC_{n_b \, [\alpha \kappa_3 \kappa_1 \kappa_2] \, J_c}^{n_{k_1}n_{k_2}n_{k_3}}\right)^*}
{\omega-(\omega_{k_1} + \omega_{k_2} + \omega_{k_3}) + i \eta}
+ \frac{\left(\mC_{n_a \, [\alpha \kappa_3 \kappa_1 \kappa_2] \, J_c}^{n_{k_1}n_{k_2}n_{k_3}}\right)^* \,
{\mD_{n_b \, [\alpha \kappa_3 \kappa_1 \kappa_2] \, J_c}^{n_{k_1}n_{k_2}n_{k_3}}} \, }
{\omega+(\omega_{k_3} + \omega_{k_1} + \omega_{k_2}) - i \eta}
\right\} \: , \: \: \: \: \:\: \: \: \: \: 
\\
\Sigma^{22  \, [\alpha] \, (2)}_{n_a n_b} &=& \displaystyle
\sum_{n_{k_1} n_{k_2} n_{k_3}} \sum_{J_c} \sum_{\kappa_1 \kappa_2 \kappa_3}
\left\{
\frac{{\mD_{n_a \, [\alpha \kappa_3 \kappa_1 \kappa_2] \, J_c}^{n_{k_1}n_{k_2}n_{k_3}}} \,
\left(\mD_{n_b \, [\alpha \kappa_3 \kappa_1 \kappa_2] \, J_c}^{n_{k_1}n_{k_2}n_{k_3}}\right)^*}
{\omega-(\omega_{k_1} + \omega_{k_2} + \omega_{k_3}) + i \eta}
+ \frac{\left(\mC_{n_a \, [\alpha \kappa_3 \kappa_1 \kappa_2] \, J_c}^{n_{k_1}n_{k_2}n_{k_3}}\right)^* \,
{\mC_{n_b \, [\alpha \kappa_3 \kappa_1 \kappa_2] \, J_c}^{n_{k_1}n_{k_2}n_{k_3}}} \, }
{\omega+(\omega_{k_3} + \omega_{k_1} + \omega_{k_2}) - i \eta}
\right\} \: . \: \: \: \: \:\: \: \: \: \: 
\end{eqnarray}
\end{subequations}

\subsection{Block-diagonal structure of Gorkov's equations}

In the previous subsections it has been proven that all single-particle Green's functions and all self-energy contributions entering Gorkov's equations display the same block-diagonal structure if the systems is in a $0^+$ state.
Defining
\begin{equation}
T_{ab} - \mu \, \delta_{ab} \equiv \delta_{\alpha \beta} \, \delta_{m_a m_b} \, \left[ T_{n_a n_b}^{[\alpha]} - \mu^{[q_a]} \, \delta_{n_a n_b} \right] \: ,
\end{equation}
introducing block-diagonal forms for amplitudes $\mW$ and $\mZ$ through
\begin{subequations}
\label{eq:wz_bd_def}
\begin{eqnarray}
\mW^{{k_1}{k_2}{k_3}}_{k \, (J_c J_{tot})} &\equiv&
\delta_{J_{tot} j_k} \delta_{M_{tot} m_k}  \, \mW^{n_{k_1}n_{k_2}n_{k_3}}_{n_k \, [\kappa_3 \kappa_1 \kappa_2] \, J_c}  \: ,
\\
\mZ^{{k_1}{k_2}{k_3}}_{k \, (J_c J_{tot})} &\equiv&
-\delta_{J_{tot} j_k} \delta_{M_{tot} -m_k}  \, \eta_k \, \mZ^{n_{k_1}n_{k_2}n_{k_3}}_{n_k \, [\kappa_3 \kappa_1 \kappa_2] \, J_c}  \: ,
\end{eqnarray}
\end{subequations}
with
\begin{subequations}
\label{eq:wz_bd}
\begin{eqnarray}
(\omega_k-E_{k_1 k_2 k_3}) \, \mW^{n_{k_1}n_{k_2}n_{k_3}}_{n_k \, [\kappa_3 \kappa_1 \kappa_2] \, J_c} &\equiv&
\sum_{n_a \alpha}
\left[ \left(\mC_{n_a \, [\alpha \kappa_3 \kappa_1 \kappa_2] \, J_c}^{n_{k_1}n_{k_2}n_{k_3}}\right)^* \,
\mU_{n_a \, [\alpha]}^{n_{k}} +
\left(\mD_{n_a \, [\alpha \kappa_3 \kappa_1 \kappa_2] \, J_c}^{n_{k_1}n_{k_2}n_{k_3}}\right)^* \,
\mV_{n_a \, [\alpha]}^{n_{k}} \right] \: ,
\\
(\omega_k+E_{k_1 k_2 k_3}) \, \mZ^{n_{k_1}n_{k_2}n_{k_3}}_{n_k \, [\kappa_3 \kappa_1 \kappa_2] \, J_c} &\equiv&
\sum_{n_a \alpha}  \left[
{\mD_{n_a \, [\alpha \kappa_3 \kappa_1 \kappa_2] \, J_c}^{n_{k_1}n_{k_2}n_{k_3}}} \,
\mU_{n_a \, [\alpha]}^{n_{k}} +
{\mC_{n_a \, [\alpha \kappa_3 \kappa_1 \kappa_2] \, J_c}^{n_{k_1}n_{k_2}n_{k_3}}} \,
\mV_{n_a \, [\alpha]}^{n_{k}}
\right] \: ,
\end{eqnarray}
\end{subequations}
and using Eqs. \eqref{eq:first_bd_11}, \eqref{eq:first_bd_22}, \eqref{eq:first_bd_12}, \eqref{eq:first_bd_21}, \eqref{eq:second_bd},
one finally writes Eqs. \eqref{eq:gorkov_premat} as
\begin{subequations}
\label{eq:gorkov_bd_final}
\begin{eqnarray}
\label{eq:gorkov_premat_bd}
\omega_k \, \mU_{n_a \, [\alpha]}^{n_{k}} &=&
\sum_{n_b}
\left [
(T_{n_a n_b}^{[\alpha]} - \mu^{[q_a]} \, \delta_{n_a n_b} + \Lambda_{n_a n_b}^{[\alpha]}) \, \mU_{n_b \, [\alpha]}^{n_{k}} + \tilde{h}_{n_a n_b}^{[\alpha]} \, \mV_{n_b \, [\alpha]}^{n_{k}}
\right]
\nonumber \\ \displaystyle &+&
\sum_{n_{k_1} n_{k_2} n_{k_3}}  \sum_{\kappa_1 \kappa_2 \kappa_3} \sum_{J_c}
\left [
\mC_{n_a \, [\alpha \kappa_3 \kappa_1 \kappa_2] \, J_c}^{n_{k_1}n_{k_2}n_{k_3}} \,
\mW^{n_{k_1}n_{k_2}n_{k_3}}_{n_k \, [\kappa_3 \kappa_1 \kappa_2] \, J_c}
+ \left(\mD_{n_a \, [\alpha \kappa_3 \kappa_1 \kappa_2] \, J_c}^{n_{k_1}n_{k_2}n_{k_3}}\right)^* \,
\mZ^{n_{k_1}n_{k_2}n_{k_3}}_{n_k \, [\kappa_3 \kappa_1 \kappa_2] \, J_c}
\right]  \: ,
\\ \displaystyle
\omega_k \, \mV_{n_a \, [\alpha]}^{n_{k}} &=&
\sum_{n_b}
\left [
-(T_{n_a n_b}^{[\alpha]} - \mu^{[q_a]} \, \delta_{n_a n_b} + \Lambda_{n_a n_b}^{[\alpha] \, *}) \, \mV_{n_b \, [\alpha]}^{n_{k}} + \tilde{h}_{n_a n_b}^{[\alpha] \, \dagger} \, \mU_{n_b \, [\alpha]}^{n_{k}}
\right]
\nonumber \\ \displaystyle &+&
\sum_{n_{k_1} n_{k_2} n_{k_3}}  \sum_{\kappa_1 \kappa_2 \kappa_3} \sum_{J_c}
\left [
{\mD_{n_a \, [\alpha \kappa_3 \kappa_1 \kappa_2] \, J_c}^{n_{k_1}n_{k_2}n_{k_3}}} \,
\mW^{n_{k_1}n_{k_2}n_{k_3}}_{n_k \, [\kappa_3 \kappa_1 \kappa_2] \, J_c}
+ \left(\mC_{n_a \, [\alpha \kappa_3 \kappa_1 \kappa_2] \, J_c}^{n_{k_1}n_{k_2}n_{k_3}}\right)^* \,
\mZ^{n_{k_1}n_{k_2}n_{k_3}}_{n_k \, [\kappa_3 \kappa_1 \kappa_2] \, J_c}
\right]  \: .
\end{eqnarray}
\end{subequations}
The latter four equations constitute the block-diagonal form of Gorkov's equations.
Note that pole energies $\omega_k$ only depend on $n_k$ and $\kappa$, i.e. they display a degeneracy with respect to the magnetic quantum number $m_k$.

\section{$\Phi$-functional}
\label{sec:phi-app}

\subsection{Connection between $\Phi$ and self-energies}
\label{sec:phi-app1}

Performing the trace over Gorkov space, the $n$-th order $\Phi$-functional defined in Eq.~\eqref{eq:Phi-def} reads
\begin{eqnarray}
\label{eq:Phi-TrG}
\Phi^{(n)} [\nG,V]  &=&
- \frac{1}{4n} \, \text{Tr}_{{\cal H}_1, \omega} \left \{
\Sigma^{11 \, (n)} \, G^{11} +
 \Sigma^{12 \, (n)} \,  G^{21} +
 \Sigma^{21 \, (n)} \,  G^{12} +
 \Sigma^{22 \, (n)} \, G^{22}
\right \}
\nonumber \\ \displaystyle
&=&
- \frac{1}{4n} \, \text{Tr}_{{\cal H}_1, \omega} \left \{
2 \, \Sigma^{11 \, (n)} \, G^{11} +
 \Sigma^{12 \, (n)} \,  G^{21} +
 \Sigma^{21 \, (n)} \,  G^{12}
\right \}
\, , \hspace{.5cm}
\end{eqnarray}
where Eqs.~\eqref{eq:symG11} and \eqref{eq:symS11} have been used to express $G^{22}$ and $\Sigma^{22}$ in terms of $G^{11}$ and $\Sigma^{11}$.

Let us differentiate expression \eqref{eq:Phi-TrG} with respect to the normal propagator $G^{11}$. One finds
\begin{eqnarray}
\label{eq:Phi-Gkv11}
\frac{\delta \Phi^{(n)} [\nG,V]}{\delta G^{11}_{ji}(\omega)}  &=&
- \frac{1}{4n} \left \{ \sum_{ab} \int \frac{d \omega'}{2 \, \pi}  \left [
2 \frac{\delta \Sigma^{11 \, (n)}_{ab}(\omega')}{\delta G^{11}_{ji}(\omega)} G^{11}_{ba}(\omega') +
 \frac{\delta \Sigma^{12 \, (n)}_{ab}(\omega')}{\delta G^{11}_{ji}(\omega)} G^{21}_{ba}(\omega') +
 \frac{\delta \Sigma^{21 \, (n)}_{ab}(\omega')}{\delta G^{11}_{ji}(\omega)} G^{12}_{ba}(\omega') \right ]  +
2 \, \Sigma^{11 \, (n)}_{ij}(\omega) \right \}
\nonumber \\ \displaystyle
&=& - \frac{1}{4n} \,   \left \{
(4n-2) \, \Sigma^{11 \, (n)}_{ij}(\omega) +
2 \, \Sigma^{11 \, (n)}_{ij}(\omega) \right \}
\nonumber \\ \displaystyle
&=& -
 \Sigma^{11 \, (n)}_{ij}(\omega)
\, ,
\end{eqnarray}
The factor $(4n-2)$ comes from all possible ways of cutting one normal propagation line $G^{11}$ in an $n$-th order self-energy diagram $\Sigma^{g_1 g_2 \, (n)}$ and reconstructing $\Sigma^{11 \, (n)}$ by performing the convolution with the multiplied propagator $G^{g_2 g_1}$. Notice that such a result is not a straightforward generalization of the standard proof of Ref. \cite{Nozieres:1963} as the reconstructed self-energy $\Sigma^{11 \, (n)}$ (second line in Eq. \eqref{eq:Phi-Gkv11}) has contributions from all self-energy types (first line in Eq. \eqref{eq:Phi-Gkv11}), each containing a different number of $G^{11}$ lines.

Similarly, one can work out the derivative with respect to $G^{12}$, obtaining
\begin{eqnarray}
\label{eq:Phi-Gkv12}
\frac{\delta \Phi^{(n)} [\nG,V]}{\delta G^{12}_{ji}(\omega)}  &=&
- \frac{1}{4n} \left \{ \sum_{ab} \int \frac{d \omega'}{2 \, \pi}  \left [
2 \frac{\delta \Sigma^{11 \, (n)}_{ab}(\omega')}{\delta G^{12}_{ji}(\omega)} G^{11}_{ba}(\omega') +
 \frac{\delta \Sigma^{12 \, (n)}_{ab}(\omega')}{\delta G^{12}_{ji}(\omega)} G^{21}_{ba}(\omega') +
 \frac{\delta \Sigma^{21 \, (n)}_{ab}(\omega')}{\delta G^{12}_{ji}(\omega)} G^{12}_{ba}(\omega') \right ]  +
 \Sigma^{21 \, (n)}_{ij}(\omega) \right \}
\nonumber \\ \displaystyle
&=& - \frac{1}{4n} \,   \left \{
(2n-1) \, \Sigma^{11 \, (n)}_{ij}(\omega) +
\Sigma^{21 \, (n)}_{ij}(\omega) \right \}
\nonumber \\ \displaystyle
&=& -  \frac{1}{2} \,
 \Sigma^{21 \, (n)}_{ij}(\omega)
\, .
\end{eqnarray}
The factor 2 between the normal and the anomalous case can be intuitively understood as follows. In a closed diagram, whenever a $G^{12}$ is present, a corresponding $G^{21}$ must appear. To a $G^{11}$, on the other hand, always corresponds another $G^{11}$, yielding twice as many possibilities of cutting such a line.

Summing over $n$ on both sides of Eqs.~\eqref{eq:Phi-Gkv11} and \eqref{eq:Phi-Gkv12}, together with the analogous ones for $G^{21}$ and $G^{22}$, one recovers Eqs. \eqref{eq:Phi-der}.

\subsection{Derivation of $\mathbf{\Sigma}^{(1)}$ from $\Phi^{(1)}$}
\label{sec:phi-app2}

At first order, the $\Phi$-functional is the sum of two diagrams
\begin{eqnarray}
\nonumber \\ \displaystyle
\hspace{.8cm}
\Phi^{(1)}  [\nG,V] &=& \hspace{2cm}
\mbox{\philabn} \hspace{-.8cm} + \hspace{.2cm} \mbox{\philaba} \hspace{-.8cm} \: .  \hspace{1.6cm}
\\ \displaystyle \nonumber
\\ \displaystyle \nonumber
\:
\end{eqnarray}
Using diagrammatic rules outlined in Appendix \ref{diag_rules} one can write the corresponding expression
\begin{eqnarray}
\label{eq:phi_1}
\Phi^{(1)}  [\nG,V] &=& \frac{i}{2} \int \frac{d \omega'}{2 \pi} \frac{d \omega''}{2 \pi} \sum_{abcd}
\left[ \bar{V}_{acbd} \, G_{dc}^{11} (\omega') \, G^{11}_{ba} (\omega'')  \right]
+  \frac{i}{4} \int \frac{d \omega'}{2 \pi}  \sum_{abcd}
\left[ \bar{V}_{a\bar{b}c\bar{d}} \, G_{cd}^{12} (\omega') \, G^{21}_{ba} (\omega')  \right]
\: .
\end{eqnarray}
Applying Eqs. \eqref{eq:Phi-der} and employing Eq. \eqref{eq:symG11} one can recover first-order self-energy terms computed in Appendix \ref{sec:first_self_app}
\begin{subequations}
\begin{eqnarray}
\Sigma^{11 \, (1)}_{ij} (\omega) &=&
-  \frac{\delta \Phi^{(1)} [\nG,V]}{\delta G^{11}_{ji}(\omega)}
\nonumber \\ \displaystyle &=&
- \frac{i}{2}  \int \frac{d \omega'}{2 \pi}
\left \{
\sum_{cd} \left[ \bar{V}_{jcid} \, G_{dc}^{11} (\omega')  \right]
+ \sum_{ab} \left[ \bar{V}_{ajbi} \, G^{11}_{ba} (\omega')  \right]
\right \}
\nonumber \\ \displaystyle &=&
- i \int \frac{d \omega'}{2 \pi}
\sum_{ab}  \bar{V}_{ajbi} \, G^{11}_{ba} (\omega')
\label{eq:phi_11_self}
\:\: ,
\\ \displaystyle
\nonumber \\ \displaystyle
\Sigma^{12 \, (1)}_{ij} (\omega) &=&
- 2 \,  \frac{\delta \Phi^{(1)} [\nG,V]}{\delta G^{21}_{ji}(\omega)}
\nonumber \\ \displaystyle &=&
- \frac{i}{2}  \int \frac{d \omega'}{2 \pi}
\sum_{cd}  \bar{V}_{i\bar{j}c\bar{d}} \, G_{dc}^{12} (\omega')
\label{eq:phi_12_self}
\:\:\: ,
\\ \displaystyle
\nonumber \\ \displaystyle
\Sigma^{21 \, (1)}_{ij} (\omega) &=&
- 2 \,  \frac{\delta \Phi^{(1)} [\nG,V]}{\delta G^{12}_{ji}(\omega)}
\nonumber \\ \displaystyle &=&
- \frac{i}{2}  \int \frac{d \omega'}{2 \pi}
\sum_{cd}  \bar{V}_{a\bar{b}j\bar{i}} \, G_{ba}^{21} (\omega')
\label{eq:phi_21_self}
\\ \displaystyle
\nonumber \\ \displaystyle
\Sigma^{22 \, (1)}_{ij} (\omega) &=&
-  \frac{\delta \Phi^{(1)} [\nG,V]}{\delta G^{22}_{ji}(\omega)}
\nonumber \\ \displaystyle &=&
- \frac{i}{2}  \frac{\delta}{\delta G^{22}_{ji}(\omega)} \left \{  \int \frac{d \omega'}{2 \pi} \frac{d \omega''}{2 \pi} \sum_{abcd}
\left[ \bar{V}_{acbd} \, G_{dc}^{11} (\omega') \, G^{11}_{ba} (\omega'')  \right]
\right \}
\nonumber \\ \displaystyle &=&
- \frac{i}{2}  \frac{\delta}{\delta G^{22}_{ji}(\omega)} \left \{  \int \frac{d \omega'}{2 \pi} \frac{d \omega''}{2 \pi} \sum_{abcd}
\left[ \bar{V}_{acbd} \, G_{\bar{c}\bar{d}}^{22} (-\omega') \, G^{22}_{\bar{a}\bar{b}} (-\omega'')  \right]
\right \}
\nonumber \\ \displaystyle &=&
- i \int \frac{d \omega'}{2 \pi}
\sum_{ab}  \bar{V}_{\bar{a}\bar{j}\bar{b}\bar{i}} \, G^{22}_{ab} (\omega')
\label{eq:phi_22_self}
\:\: .
\end{eqnarray}
\end{subequations}
\end{widetext}

\end{fmffile}


\bibliography{gorkovbib}

\end{document}